# Timing and Synchronization

*A.Gallo*
INFN-LNF, Frascati (Italy)

**Abstract**
Several modern accelerator facilities require the synchronization of equipment, which is distributed over large distances, down to the femto-second scale. This document describes the resulting problems, gives a basic description of concepts for the solution, shows several solution presently in use and finishes with a linear model to compute the resulting phase-noise of a synchronization system.

## 1 Introduction

The terms **timing and synchronization** in accelerators denote the particular branch of the particle accelerator physics devoted to the **tight control of the temporal sequence of the events** that are crucial to reach the design performances and **produce the required outcomes** (such as sub-atomic particles in colliders, photons in FELs or Compton sources, accelerated particle bunches in Plasma driven boosters, …) **at the desired rate** in a regular and reproducible way.
In general every accelerator is designed and built to produce some specific physical processes, and **one necessary condition** for an effective and stable operation is that some **events** have to happen at the **same time** (simultaneously for an observer in the laboratory frame) or in a **rigidly defined temporal sequence**, within a maximum allowed time error budget. If the simultaneity or the time separation of the events fluctuates beyond the specifications, the performances of the machine will be spoiled, and the quantity and quality of the accelerator products will be compromised. Clearly, the tolerances on the time fluctuations are different for different kind of accelerators. The smaller the tolerances, the tighter the level of synchronization required.
In the past, let's say for accelerators built before the turn of the century, the level of the required synchronization among different machine sub-systems was in the order of the pico-seconds (or looser). This was sufficient, for instance, to secure the longitudinal position of the interaction points in colliders, to preserve the injection/extraction efficiency in storage/damping rings or to maintain the bunch footprint in the longitudinal phase space within acceptable limits in linear accelerators. These tasks were mainly accomplished by proper controlling and stabilizing both phase and amplitude of the RF fields in the accelerating sections, so that synchronization was essentially an additional function embedded in the RF distribution and Low-level RF control systems.
In the In the last 2 decades a **new generation of accelerator** projects such as FEL radiation sources or plasma wave based boosters has pushed the level of the synchronization specifications **down to the fs scale**. The increasing presence and role of **short laser pulse systems** in modern facilities has driven both a step forward of the **performance demand** and a transition in the **technology employed** (from µwave electronics to electro-optics and fully optics). In this context, timing and synchronization have evolved to a new, well-identified branch of the accelerator physics and technology, involving concepts and competences from various fields including Electronics, RF, Laser, Optics, Control, Diagnostics, Beam dynamics [1] [2] [3].

This introductory lecture to the subject is structured as follow:

- an introduction, including a description of the typical synchronization requirements for modern facilities;
- a section 2, covering **basic definitions, concepts and glossary**;
- a section 3, describing the **architecture and the performances** of the synchronization systems, including the **beam arrival time diagnostics**;
- a section 4, devoted to the **beam arrival time fluctuation**, providing **a linear model** to compute expected performances starting from the phase noise spectra of the various machine sub-systems impacting the beam longitudinal dynamics.

## 1.1 Synchronization needs for FEL radiation sources

As already mentioned, the **advent of Free Electron Laser** (FEL) radiation sources has been the main driver that has pushed the synchronization requirements down by about **3 orders of magnitude** from the ps domain to the fs domain. Referring to this pretty recent (less than 2 decades old) kind of accelerator based facilities, it is worth analyzing what are the physics process more critically sensitive to the relative synchronization among the faster time-varying sub systems (the RF fields and the laser pulses) and the beam bunches. This helps to build a model where the **main relevant subjects** are identified and recognized therefore as the most important "**clients**" of a **central synchronization system** of the facility. The global task of this system is to keep **tightly under control** the "**heartbeats**" of the clients to constrain them to **follow a facility common clock** and be coherent to it to the highest possible extent.

The simplest FEL regime is the **SASE** (Self-Amplified Spontaneous Emission) and requires **high-brightness bunches**. The beam brightness B is proportional to the bunch current $I_{bunch}$ and inversely proportional to the transverse emittance $\epsilon_{\perp}$ squared, i.e. $B \div I_{bunch}/\epsilon_{\perp}^2$.

A high-brightness electron bunch travelling in the gap of an undulator emits spontaneous synchrotron radiation which acts back on the bunch itself affecting the intra-bunch longitudinal dynamics and producing a **micro-bunch sub-structure** in the **charge distribution**[4]. The beam micro-bunching enhances the radiation emission further, in a positive feedback mechanism that leads to an exponential growth of the radiation intensity until a saturation level is reached. High brightness corresponds to large peak current $I_{bunch}$ and small transverse emittance $\epsilon_{\perp}$. Large bunch currents are typically obtained by **short** (of the order of 1 ps duration) **laser pulses** illuminating a **photo-cathode** embedded in an **RF Gun accelerating structure**, and furtherly increased with **bunch compression** techniques.

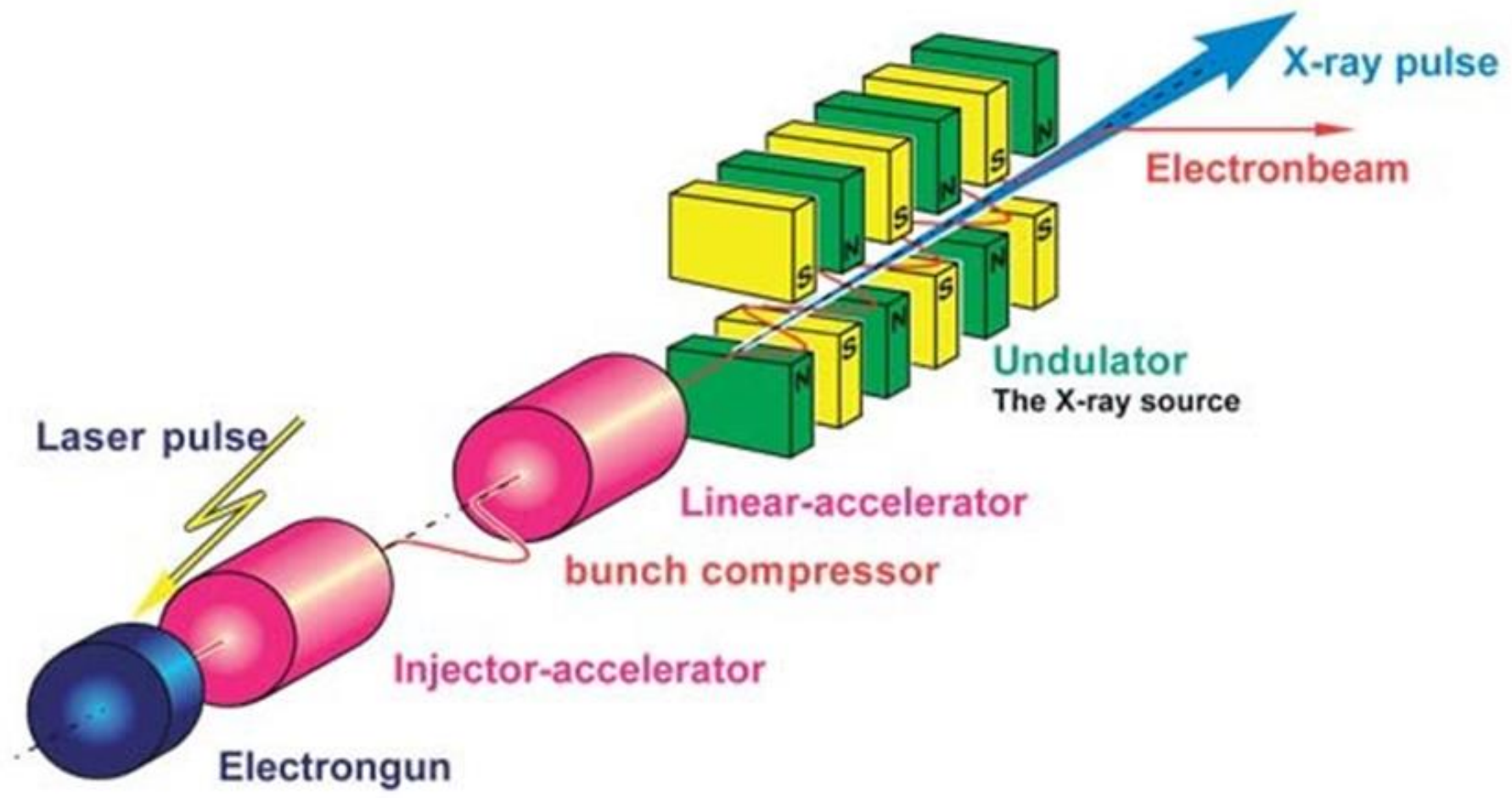

Fig. 1: Schematics of a Self-Amplified Spontaneous Emission radiation facility

The RF Gun working frequency is typically $> 1$ GHz (L-band and S-band mostly used), and beam characteristics reproducibility at the RF Gun exit requires the photo-cathode laser shots extracting the photo-electrons at a **precisely defined accelerating RF field phase**, with a maximum tolerable error well **below $\mathbf{1\ deg}$ RF**. Converted in time units, this corresponds to a relative time fluctuation between the arrival time of the laser at the photocathode position and the RF fields in the resonant cavity already **below $\mathbf{1\ ps}$**.
The bunch length is furtherly reduced by large factors (typically 10 or more) by **compressor stages** [5]. The magnetic compressor scheme is based on a non-isochronous transfer line (a magnetic chicane) exploiting a longitudinal energy chirp imprinted on the bunch distribution by accelerating the bunch off-crest in the linac portion upstream the chicane. The bunch result to be compressed since the **tail** is **more energetic** than the **head**, and travels a **short distance** in the chicane catching-up the head. Clearly the nominal chirp value matching the characteristics of the magnetic chicane corresponds to a precise value of the RF accelerating field phase on the beam. This condition sets a **limit** of the acceptable **relative fluctuation** between the **arrival time** of the **beam** in the linac ahead the chicane and the **phase** of the linac **RF fields**. Excessive fluctuations would result in shot-to-shot bunch length variations, as well as **fluctuations** of the **bunch arrival time** downstream the chicane. Again, **stability** of the **bunch compression** process typically requires **synchronization errors** between bunch arrival time and linac RF at level of $\mathbf{\leq 100 \div 300\ fs}$.
Small transverse emittances $\epsilon_{\perp}$ is obtained with tight control of the global machine working point, including amplitude and phase of the RF fields, magnetic focusing and laser characteristics.

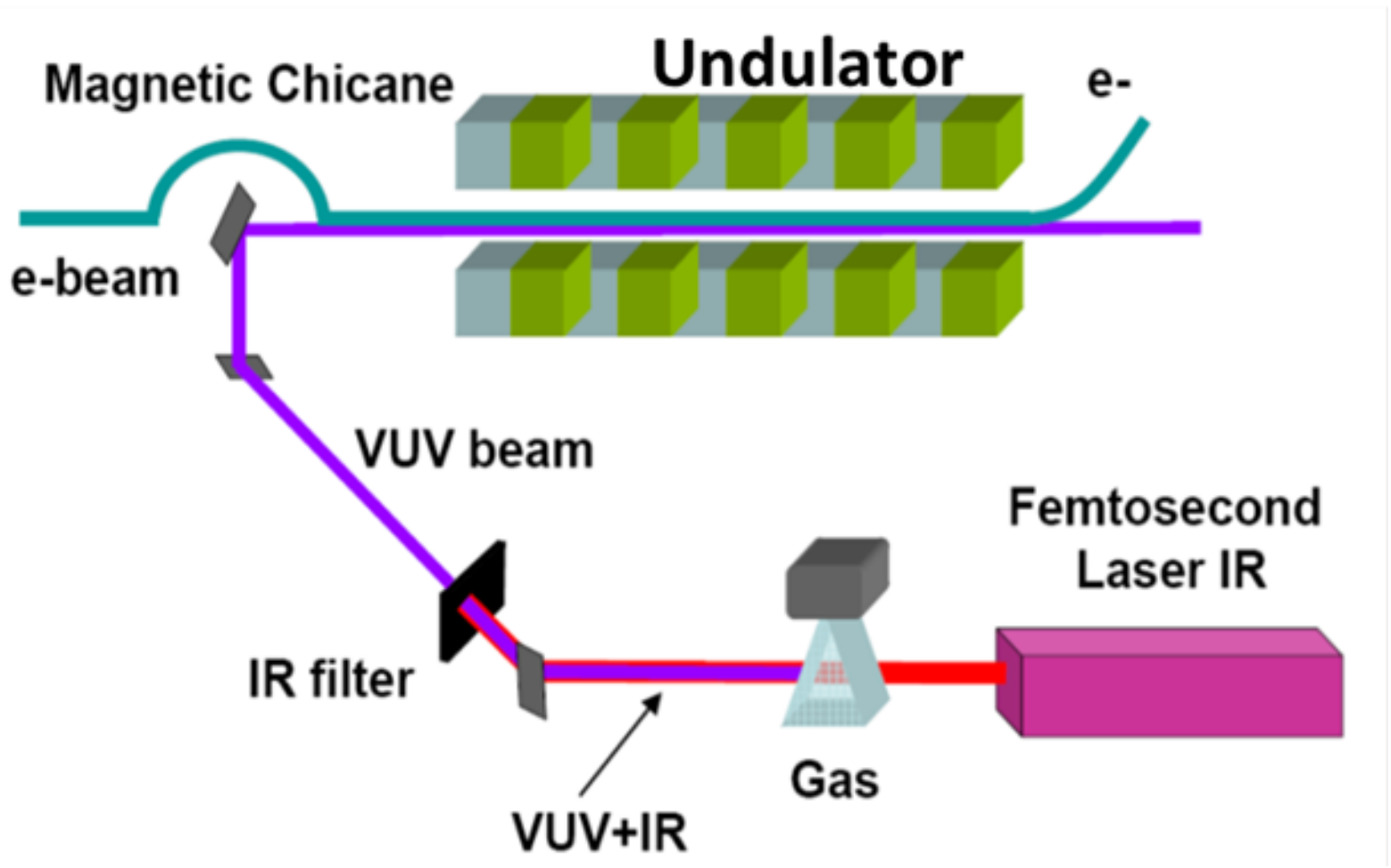

Fig. 2: Schematics of a seeded FEL radiation process implementation

In a simple SASE configuration the micro-bunching process, which is the base of the FEL radiation production, starts from noise. Characteristics such as radiation intensity and envelope profile can vary considerably from shot to shot. A better control of the radiation properties resulting in **more uniform** and reproducible shot to shot **pulse characteristics** can be achieved in the **"seeded" FEL** configuration [6].

To "trigger" and guide the avalanche process generating the exponentially-growing radiation intensity, the high brightness bunch is made to interact with a VUV short and intense pulse obtained by HHG (High Harmonic Generation) in gas driven by an infrared pulse generated by a dedicated high power laser system (typically TiSa). The presence of the external radiation since the beginning of the micro-bunching process inside the magnetic undulators seeds and drives the FEL radiation growth in a steady, repeatable configuration. The electron bunch and the VUV pulse, both very short, must constantly overlap in space and time shot to shot. This condition pushes the **synchronization constraints** furtherly down to $\leq \mathbf{100\ fs}$ between the **bunch** and the **IR laser pulse** which generates the seed.

FEL radiation facility serve a huge community of users coming from many different fields. There is a family of experiments that relies on **pump-probe techniques** [7]. This methodology consists in initializing some physical/chemical processes by means of an ultra-short laser pulse, and then probing the system status with the FEL radiation. The dynamics of the process under study is captured and stored in a "snapshots" record by sweeping shot-to-shot the relative delay between pump and probe pulses. Therefore **pump laser** and **FEL pulses** need to be **synchronized** at level of the **time-resolution** required by the experiments (often down to $\approx \mathbf{10\ fs}$).

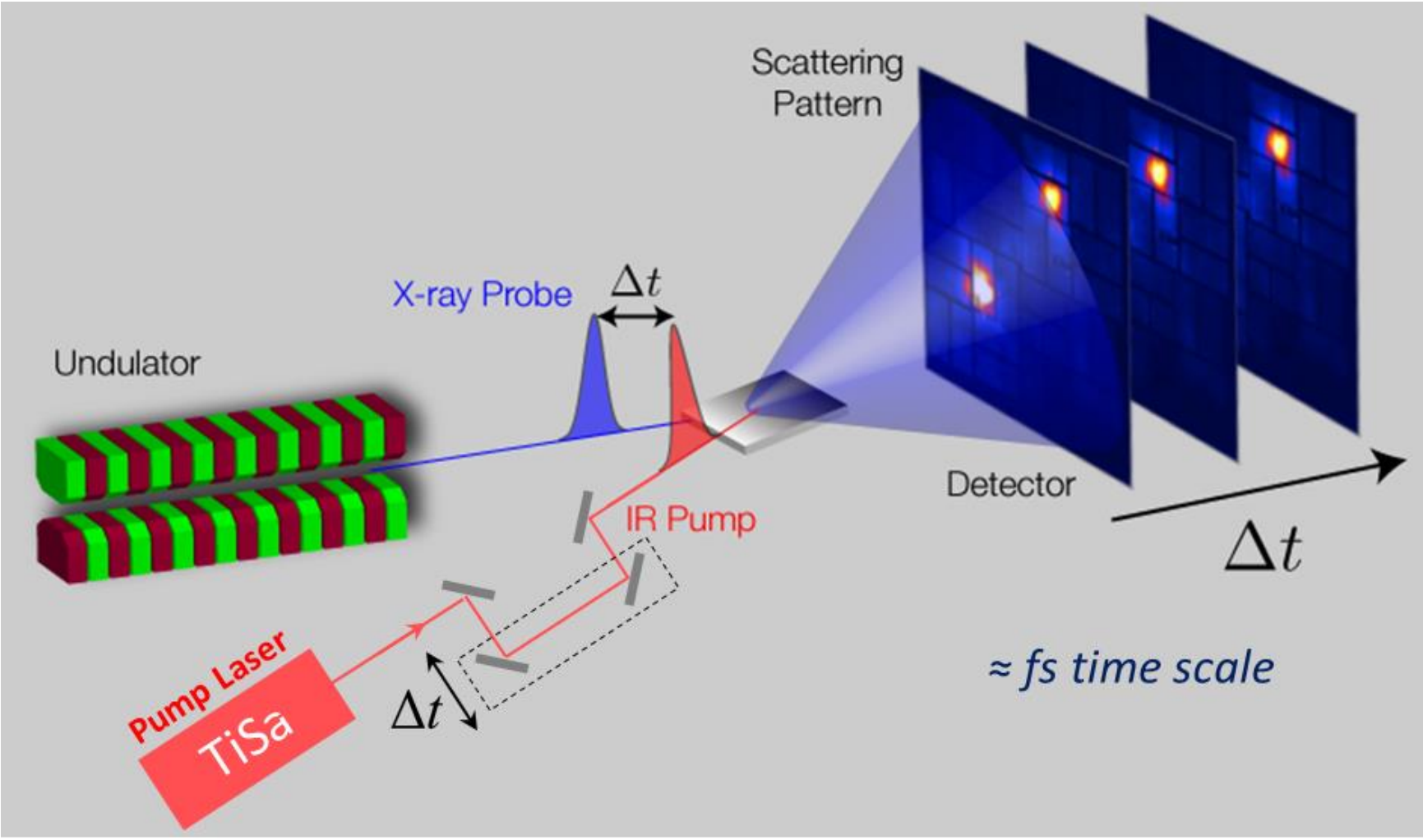

Fig. 3: Schematics of a pump-probe experiment using FEL radiation

Plasma acceleration is presently one of the most exciting frontiers in accelerator physics, which promises to overcome the gradient limits of the RF technology in the way to more compact facilities. There is a huge effort in the accelerator community to pass from a proof-of-principle experimental stage to a technology mature enough to be adopted  as the ground base of a user facility. The Design Study EuPRAXIA ("European Plasma Research Accelerator with eXcellence  In Applications"), funded by EU and started in 2017, will produce the CDR for the worldwide first high energy plasma-based accelerator that can provide beam to FEL radiation users.

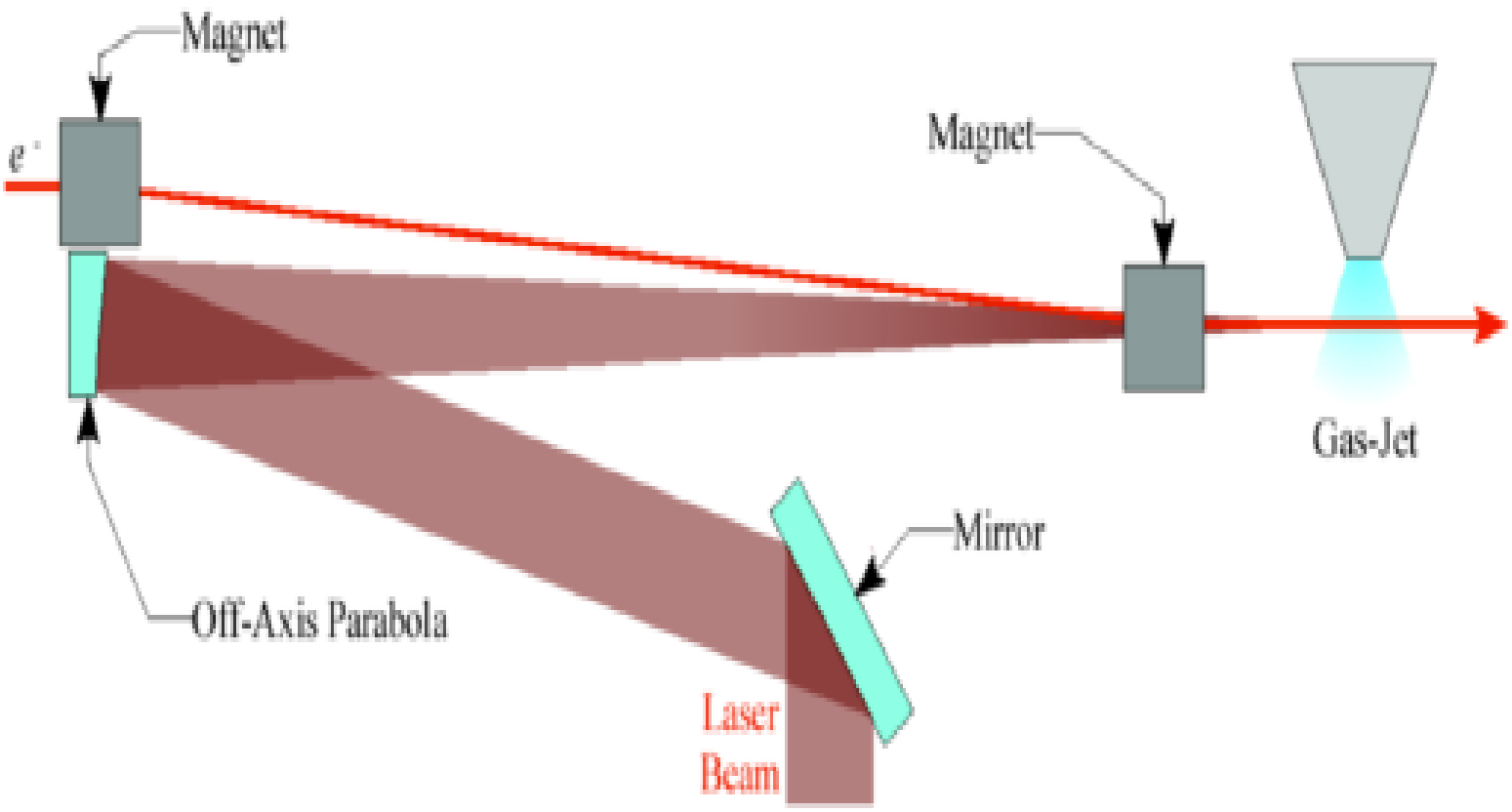

Fig. 4a: Schematics of external injection in the plasma wave

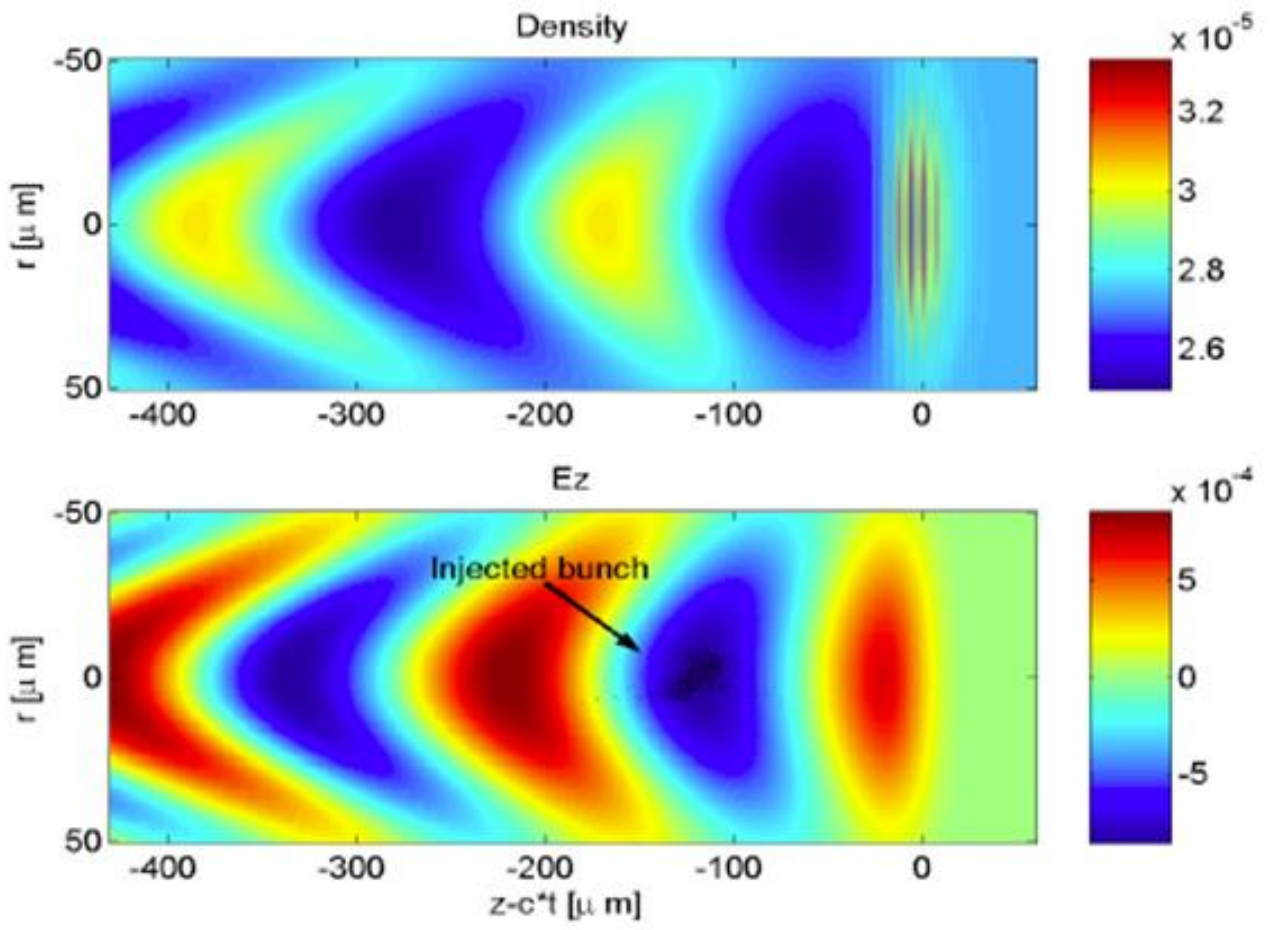

Fig. 4b: Accelerating field in a plasma wave

Plasma acceleration can be implemented in various configurations. In the **Wakefield Laser-Plasma Acceleration** (WLPA) technique [8] an extremely intense laser pulse is injected through a gas jet or in a capillary to **generate a plasma wave** carrying large accelerating gradients (many GV/m). Then an **external pre-accelerated bunch** has to be injected **in the plasma wave**, whose "accelerating buckets" are typically few 100 µm long. The injected bunch has to be very short to limit the energy spread after acceleration, and ideally needs to be injected constantly in the same position of the plasma wave to avoid shot-to-shot energy fluctuations [9]. This requires **synchronization** at the level of a **small fraction** of the **plasma wave period**, which means errors $< \mathbf{10\ fs}$ between the arrival times of the driving laser pulse and the trailing electron bunch centroids.
Clearly, the scientific interest for plasma accelerators goes much beyond the FEL user facilities and it includes all applications requiring high gradients, primarily the high energy physics.
FEL radiation user facilities are good and instructive examples of large accelerator complex systems showing a variety of synchronization problems, needs and specifications, and in this perspective they will be used as a reference all throughout this lecture. Obviously, all the material presented (concepts, schematics, layouts, …) can be applied in general to any kind of accelerator facility, or even outside the accelerator field for any big installation requiring fine temporal alignment between different parts down to the fs scale.

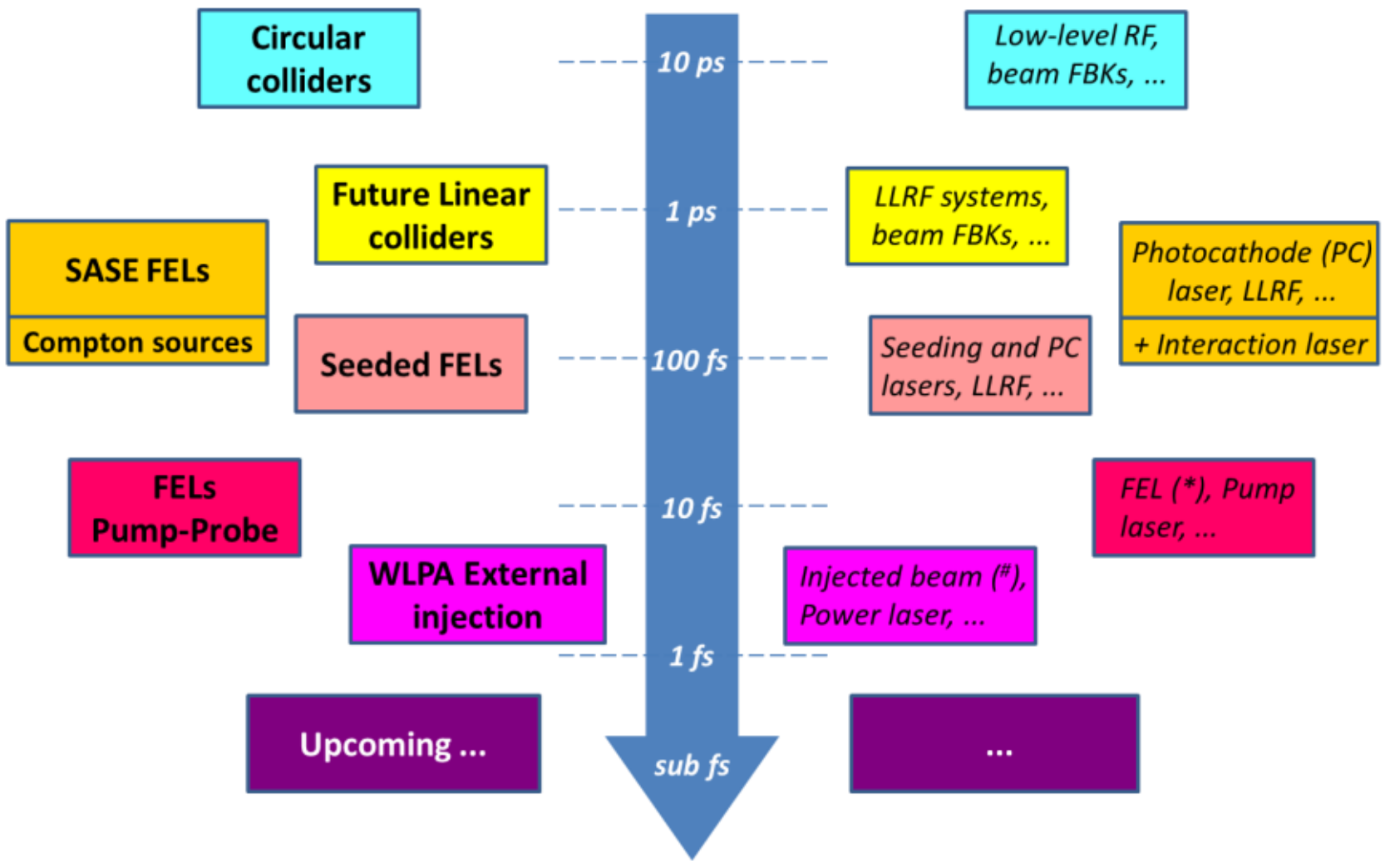

Fig. 5: Typical synchronization specs for different kind of accelerators and applications

# 2 Definitions, Basics and Glossary

## 2.1 Definitions and basic schematics

Every accelerator is built to produce some specific physical processes (shots of bullet particles, nuclear and sub-nuclear reactions, synchrotron radiation, FEL radiation, Compton photons, ...). It turns out that **one necessary condition** for an efficient and reproducible event production is the **relative temporal alignment** of all the accelerator sub-systems involved in the process (such as RF fields, PC laser system, ...), and of the beam bunches with any other system they have to interact with during and after the acceleration (such as RF fields, seeding lasers, pump lasers, interaction lasers, ...).
The **synchronization system** is the **complex** including all the hardware, the feedback processes and the control algorithms required to keep **time-aligned** the **beam bunches** and all the **machine critical sub-systems** within the facility specifications.

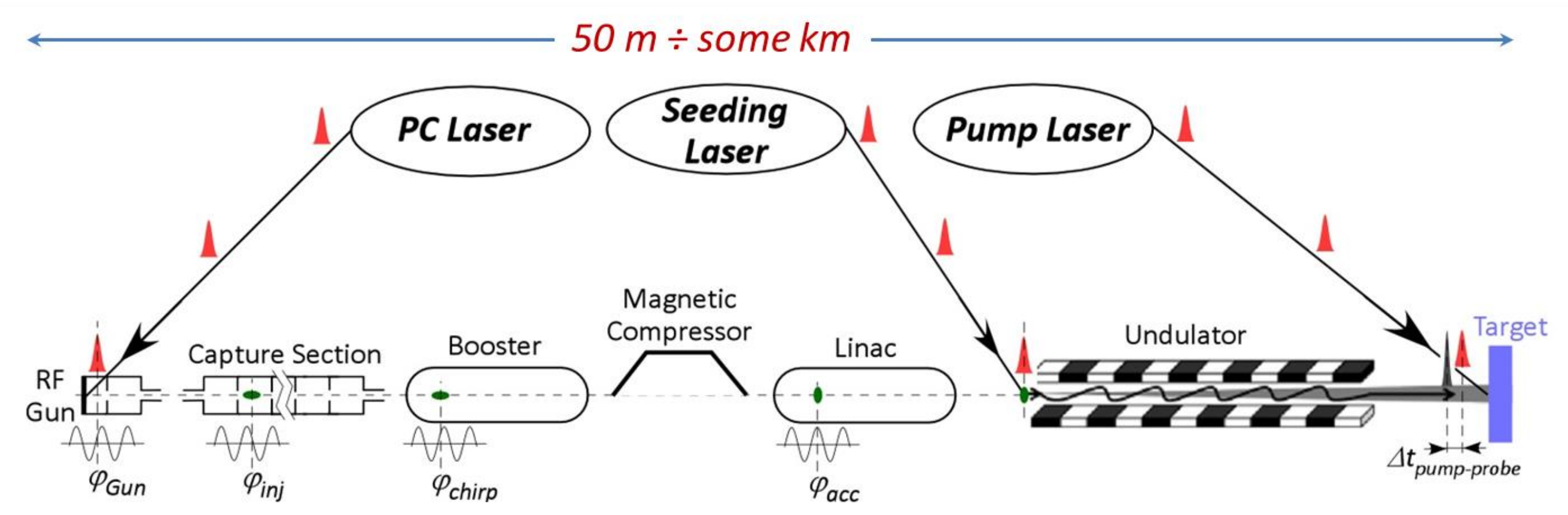

Fig. 6: temporal alignment of critical sub-systems in an accelerator facility

**Master Oscillator:**

In general each sub-system requiring temporal alignment (RF power plants connected to accelerating devices, mode-locked laser systems with Chirped Pulse Amplification, …) has its own fundamental repetition frequency, given by the oscillator in the core of the system itself. Clearly, the physical nature of the various oscillators can be different (optical cavities for laser systems, μ-wave VCOs for RF systems), and they all need to be forced to be coherent since free-run oscillators always drift one respect to the other over long time scale, even if they were as precise as atomic clocks. So the pace of every oscillator in the facility need to be continuously correct and re-synchronized by comparison with a **common master clock** that has to be distributed to the all "clients" (i.e. to all the sub-systems requiring time alignment) spread over the facility site with a “star” network architecture.

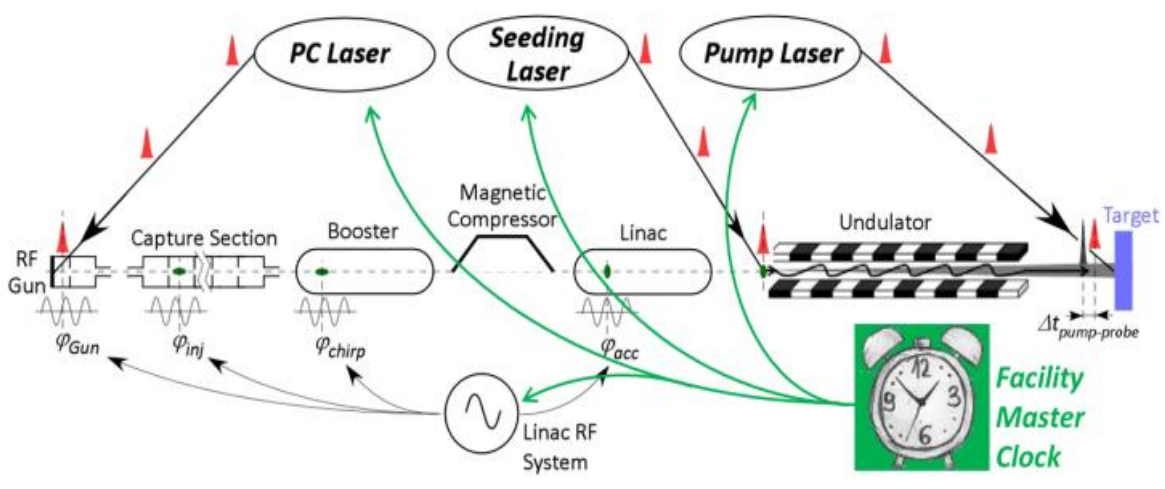

Fig. 7a: the facility master clock

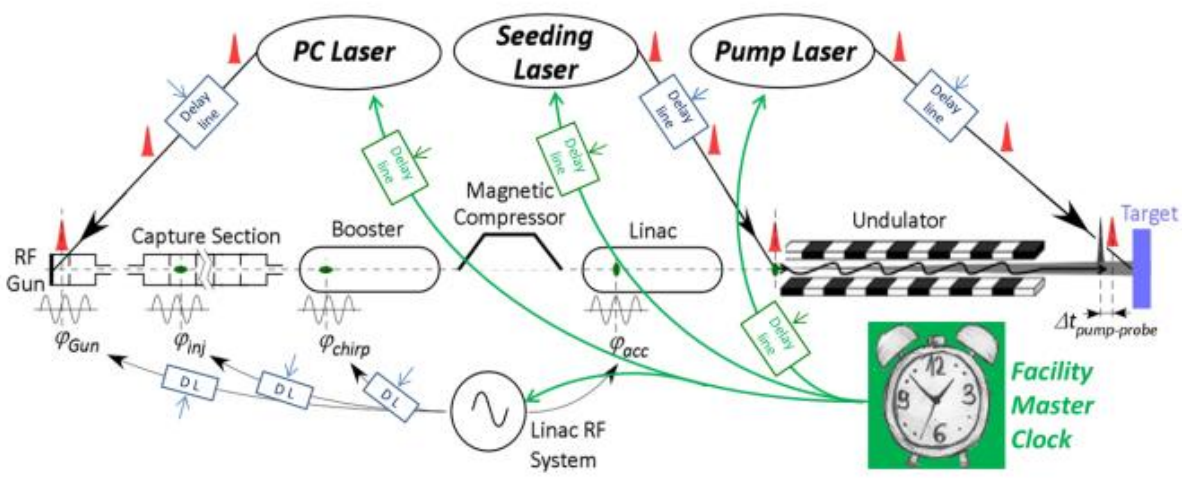

Fig 7b: delay lines for synchronization adjustment

Once the **local oscillators** have been locked to the reference, they can be shifted in time by means of **delay lines** of various types, such as translation stages with mirrors for lasers, and trombone-lines or electrical phase shifters for RF signals. This allows setting, correcting, optimizing and changing the working point of the facility synchronization. Delay lines can be placed either downstream the oscillators or on the reference signal on its path to the client oscillator. The function accomplished is exactly the same.

For simplicity, in most of the following sketches the presence of the delay lines will be omitted, but the reader should keep in mind that in a real system they will be surely present in any location where the operation requires a tunable delay.

The common master clock of a particle accelerator based facility is called **Master Oscillator** and is typically a high spectral purity, low phase noise μ-wave generator acting as timing reference for the machine sub-systems. It is often indicated as the **RMO** (RF Master Oscillator or Reference Master Oscillator). This kind of sources are available on the market for various purposes, and their performances are improving pushed by the user requirements. The physics and technology of the low noise sources is beyond the scope of this lecture, interested readers can find an introductory presentation in [10] [11]. How the intrinsic phase noise of the reference source can affect the behaviour and ultimate performances of a whole synchronization system will be discussed later on this paper.

The timing reference signal can be distributed as a pure sine-wave voltage through coaxial cables, or through optical-fiber links after being encoded in the repetition rate of a pulsed (mode-locked) laser or, sometimes, in the amplitude modulation of a CW laser. The laser oscillator distributing the timing reference signal is called **OMO** (Optical Master Oscillator) [12][13]. Optical transmission of the timing reference provides lower signal attenuation and larger bandwidth, so optical technology is definitely preferred for synchronization reference distribution, at least for large facilities. However, fiber links are dispersive and have to be compensated by adding proper negative dispersion patches to avoid excessive pulse broadening at destination.

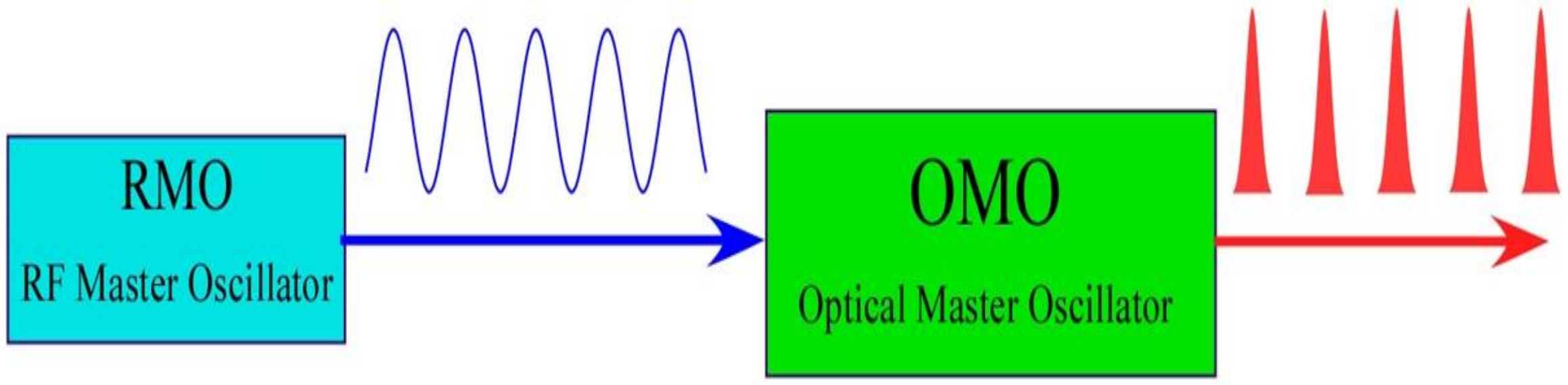

Fig. 8: RF and Optical Master Oscillators

A mode-locked laser [14] consists in an **optical cavity** hosting an active amplifying medium capable of sustaining a **large number** of longitudinal **modes** with frequencies $\nu_k = k\nu_0 = kc/L$ within the bandwidth of the active medium, being $L$ the cavity round trip length and $k$ any integer. If the modes are forced to **oscillate in phase** and the medium emission bandwidth is wide enough, a **very short pulse** ($\approx 100$ fs) travels forth and back in the cavity and a sample is coupled out through a leaking mirror.

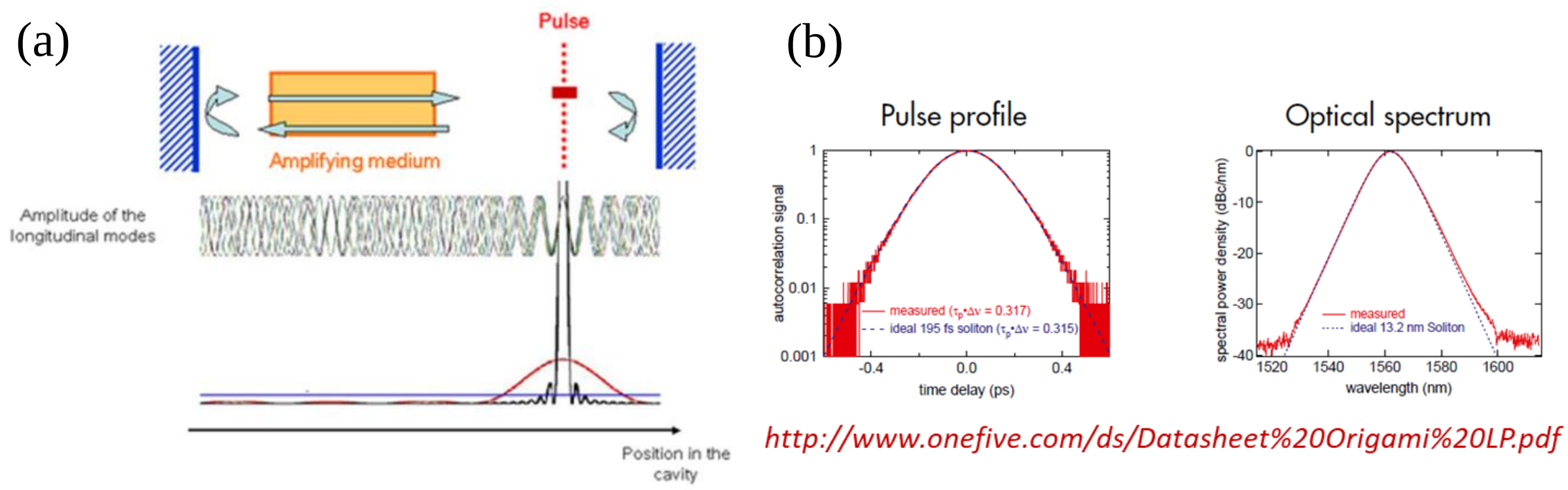

Fig. 9: Schematics (a) and pulse characteristics (b) of a wideband mode locked laser cavity

As it will be described in more details later on, the **total length** L of the cavity can be **tuned** and continuously readjusted to lock the laser pulse **repetition frequency** to the **RMO reference** by moving the longitudinal position of a piezo-controlled mirror placed along the optical beam path, i.e. by implementing an electro-mechanical Phase Locked Loop (PLL) configuration.

Jitter and Drift:

The synchronization error of a client with respect to the reference is categorized as **jitter** or **drift** depending on the time scale of the fluctuation. In general the term "**jitter**" indicates **fast variations**, caused by inherent residual lack of coherency between oscillators, even if they are locked at the best. The term "**drift**" instead is used for **slow variations**, mainly caused by modifications of the environment conditions, primarily the temperature but also the humidity and the aging of materials and components. The boundary between the two definitions is somehow arbitrary since it relies on a qualitative classification of the phenomena into the categories fast/slow. For instance, synchronization errors due to mechanical vibrations can be classified in either category: the acoustic waves are mainly considered "jitter", the infra-sounds are often included in the "drift" budget.

In pulsed accelerators, where the beam is produced in the form of a sequence of bunch trains with a certain repetition rate (10 Hz ÷ 120 Hz typically), the repetition rate value itself can be assumed as a reasonable boundary between jitters and drifts. In this respect, **drifts** are phenomena substantially **slower** than **repetition rate** and will produced effects on the beam that can be monitored and corrected pulse-to-pulse. The adjective "slow" in this case means "slow enough to be corrected by a feedback system". On the contrary, **jitters** phenomena are **faster** than **repetition rate** and will result in a pulse-to-pulse chaotic scatter of the beam characteristics that has to be minimized by adopting proper measures and precautions but that cannot be actively corrected.

## 2.2 General Architecture of a Facility Synchronization System:

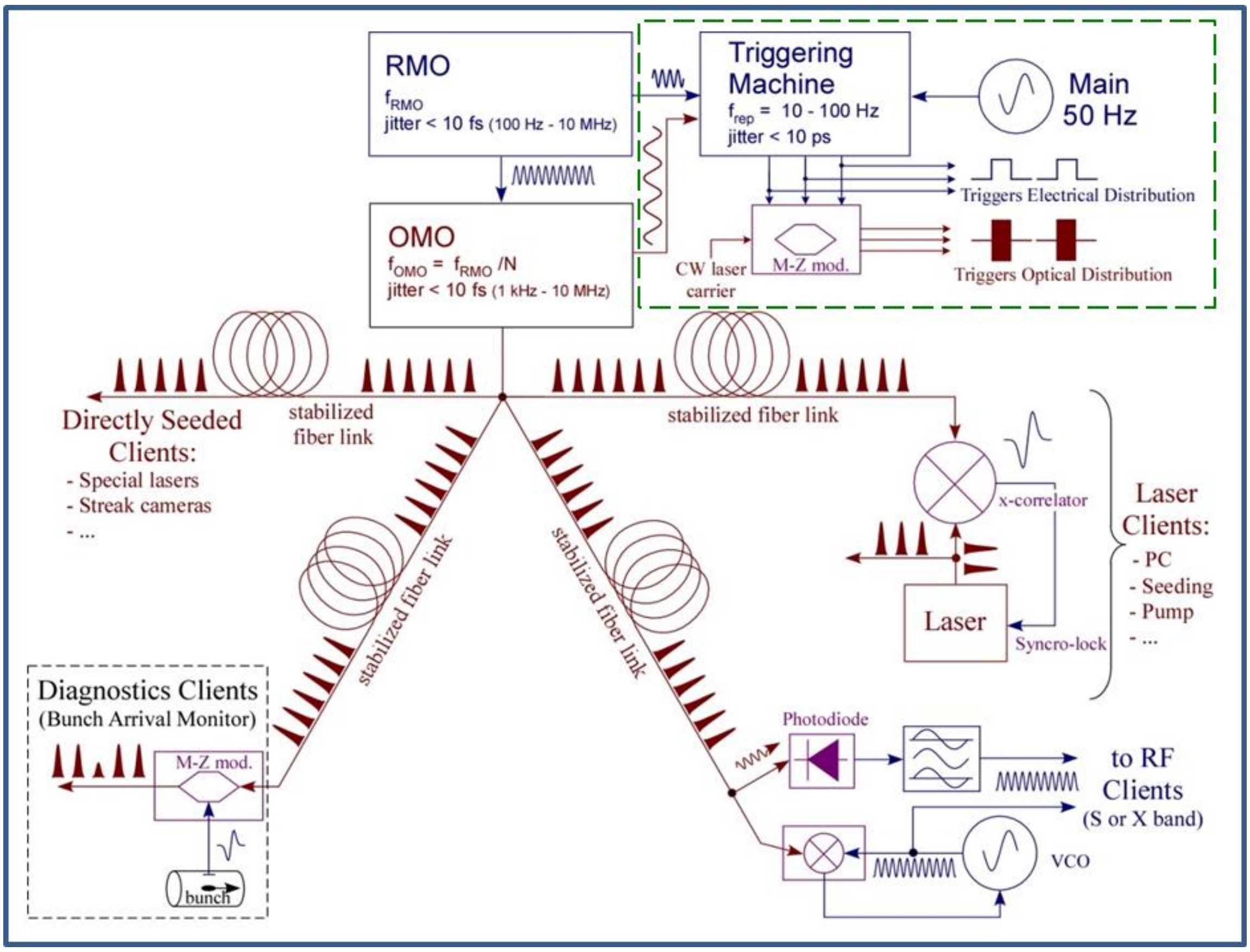

Fig. 10: schematic sketch of a facility synchronization system

A general schematics of a facility synchronization system is shown in Fig. 10. The main tasks of such a system are:

- Generate and transport the reference signal to any client local position with constant delay and minimal drifts;
- Lock the client (laser, RF, ...) fundamental frequencies to the reference with minimal residual jitter;
- Monitor clients and beam, and apply delay corrections to compensate residual (out-of-loop) drifts.

In the top-right part of the sketch (within the dashed rectangle area) the **trigger generation** section is also represented.

Triggers:

Triggers are **digital signals** with **proper relative delays** to start (or enable, gate, etc. …) a number of fundamental processes in the accelerator operation such as: firing injection/extraction kickers, RF pulse forming, switch on RF klystron HV, open/close Pockels cells in laser system, start acquisition in digitizer boards, start image acquisition with gated cameras, and many others. Generation, distribution and delay fine adjustment of trigger signals are tasks still belonging to the **facility timing business**, but the required precision is orders of magnitude less demanding than what is needed for the main client oscillators. Time resolution and stability of trigger signals is therefore way more relaxed (< 1 ns often sufficient, ≈10 ps more than adequate). Sometimes the trigger managing systems is still called "timing system", to be distinguished from the "synchronization system" which is the one properly devoted to the finest temporal alignment of the clients down to the fs scale. Other times the two systems, which are strongly integrated and correlated as shown in Fig. 10, are described as a whole, and denoted as "timing and synchronization system". Trigger managing is certainly an important aspect of the accelerator operation, but it will not be extensively covered in this lecture that will be completely focused on the fs synchronization issues.

### 1.2) Phase noise power spectrum

This paragraph is devoted to the introduction of the **phase noise** concept as a specific random variable describing the **deviation** of the argument of a real oscillation from a **pure ideal** sine wave [10], [15]. Let us start summarizing some general property of random variables. Let us consider a generic random variable $\mathrm{x(t)}$ representing a physical observable quantity.

a) The process is defined **stationary** when its statistical properties are invariant for any $\mathrm{t'}$ time shift $\mathrm{x(t) \rightarrow x(t+t')}$:

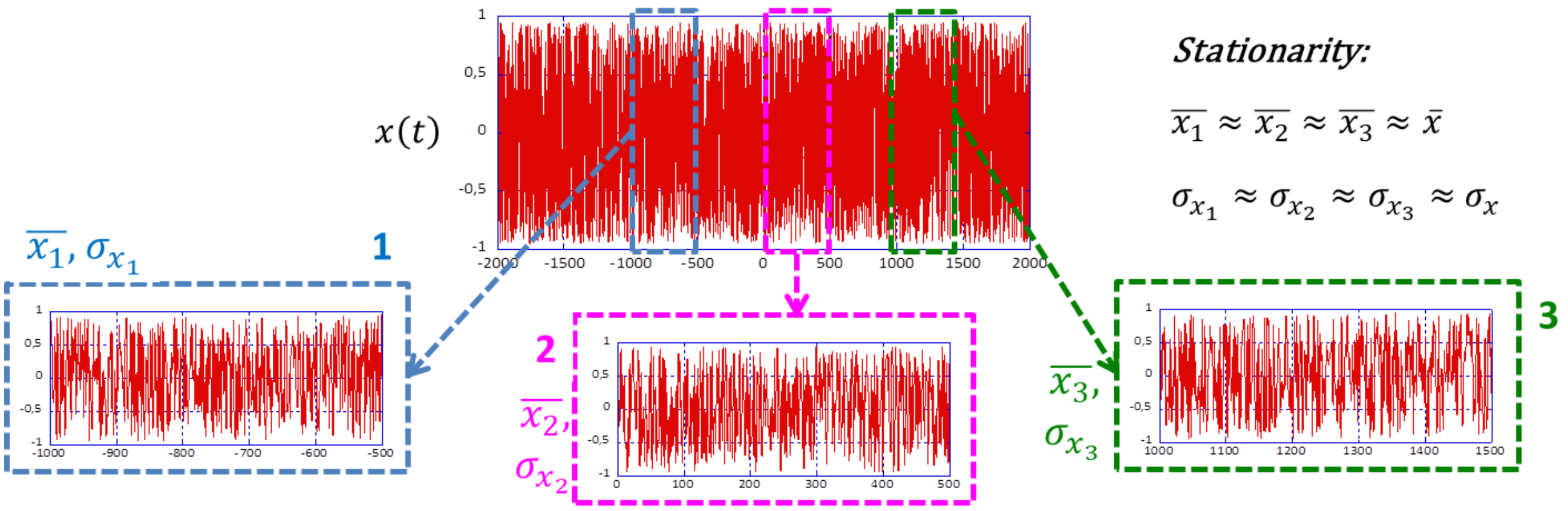

Fig. 11: in a stationary process the statistical property are independent on time

b) The process is defined **ergodic** when the statistical properties can be estimated by a single process realization. For example if we consider a set of resistors of the same value in the same environment, the characteristics of their thermal noise can be extrapolated from a measurement taken over a single sample.

c) Two random variables $\mathrm{x(t)}$ and $\mathrm{y(t)}$ are **uncorrelated** when they are statistically independent. Under this assumption the following relations hold:

$$\overline{x(t)\cdot y(t)} = \overline{x(t)}\cdot\overline{y(t)}$$

$$z(t) \stackrel{\text{def}}{=} a\cdot x(t) + b\cdot y(t) \rightarrow \sigma_z^2 = a^2\sigma_x^2 + b^2\sigma_y^2 \tag{1}$$

$$with\ \sigma_z^2 \stackrel{\text{def}}{=} \overline{z^2(t)} - \overline{z(t)}^2\ ,\quad \sigma_z = standard\ deviation$$

In the following of the lecture, referring to the phase noise in oscillators, we will deal in general with random processes that are **both stationary** and **ergodic**. The phase noises of different sources will be considered **uncorrelated** whenever they will be found to be fully **independent**, while they will be more or less **tightly correlated** when locked to a common **reference**.

It is often very useful and convenient to perform a spectral decomposition of the statistical properties of a random variable $x(t)$, especially of its rms value $x_{rms}$ and standard deviation $\sigma_x$. But since $x_{rms} \neq 0$, a real random variable $x(t)$ is in general not directly Fourier transformable. To work around this inconsistency, the random variable $x(t)$ can be observed only for a finite time $\Delta T$ and truncated outside the interval $[-\Delta T/2, \Delta T/2]$ to remove any possible limitation in the function transformability. The truncated function $x_{\Delta T}(t)$ defined as:

$$x_{\Delta T}(t) = \begin{cases} x(t) & -\Delta T/2 \leq t \leq \Delta T/2 \\ 0 & \text{elsewhere} \end{cases} \tag{2}$$

is therefore Fourier transformable. Let $X_{\Delta T}(f)$ be its Fourier transform. We have:

$$x_{rms}^2 = \lim_{\Delta T \to \infty} x_{\Delta T_{rms}}^2 = \lim_{\Delta T \to \infty} \frac{1}{\Delta T} \int_{-\infty}^{+\infty} x_{\Delta T}^2(t)\, dt = \lim_{\Delta T \to \infty} \frac{1}{\Delta T} \int_{-\infty}^{+\infty} |X_{\Delta T}(f)|^2\, df \stackrel{\text{def}}{=} \int_0^{+\infty} S_x(f)\, df$$

$$\text{with } S_x(f) \stackrel{\text{def}}{=} \lim_{\Delta T \to \infty} 2 \cdot \frac{|X_{\Delta T}(f)|^2}{\Delta T} \tag{3}$$

where the **Parseval's theorem** [15] has been used to pass from the time integral to the frequency integral. By using this approach the rms value of a random variable is decomposed in frequency components expressed by the function $S_x(f)$ which is called "power spectrum" or "power spectral density" of the random variable $x(t)$. The power spectrum $S_x(f)$ is the square module of the truncated function Fourier transform normalized to the observation interval $\Delta T$. Because of the normalization the function can remain limited or integrable while $\Delta T \to \infty$, differently from the unnormalized $|X_{\Delta T}(f)|^2$ function. Please also notice that the duration of the time observation $\Delta T$ sets the minimum frequency $f_{min} \approx 1/\Delta T$ containing meaningful information in the spectrum of $x_{\Delta T}(t)$.

Let's now compute the response of a linear, time-invariant (LTI) network characterized by its Green's function $h(t)$ when excited by a random variable $x(t)$. The signal $x_{out}(t)$ emerging from the network is given by the convolution product between the input variable $x(t)$ and the Green's function $h(t)$:

$$x_{out}(t) = \int_{-\infty}^{t} x(\tau) \cdot h(t-\tau)\, d\tau \stackrel{\text{def}}{=} x(t) * h(t) \tag{4}$$

The Green's function $h(t)$ corresponds to the network response to the elementary stimulus represented by a Dirac's delta $\delta(t)$. In Fourier and Laplace domains, the convolution product is transformed in a standard algebraic product, according to:

$$\begin{aligned} X_{out}(\omega) &= X(\omega) \cdot H(\omega) && \textit{Fourier domain} \\ X_{out}(s) &= X(s) \cdot H(s) && \textit{Laplace domain} \end{aligned} \tag{5}$$

where $H(\omega)$ and $H(s)$ are the $LTI$ transfer functions given by the Fourier and Laplace transforms of the Green's function $h(t)$. Whenever the input variable $x(t)$ is not Fourier transformable, the first equation in (5) is intended for any truncated version $x_{\Delta T}(t)$. According to the definition (3), the power spectral noise $S_{out}(\omega)$ emerging from the $LTI$ network is given by:

$$S_{out}(\omega) = |H(\omega)|^2 \cdot S_{in}(\omega) \tag{6}$$

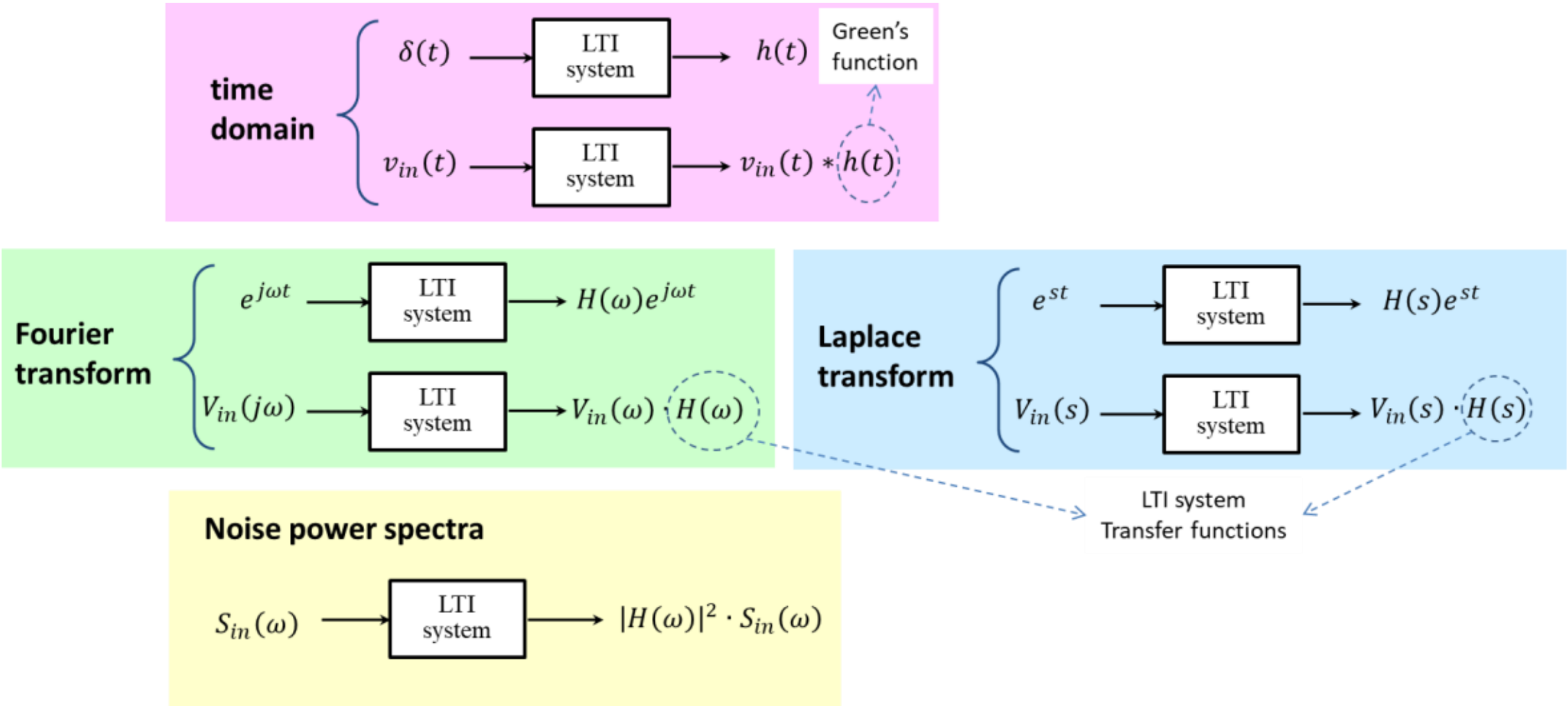

Fig. 12: Signal and noise transformation through a Linear Time-Invariant system

The fundamental task of a Synchronization system is to **lock firmly** the **phase** of each client to the **reference** oscillator in order to minimize the residual jitter. The **clients** are basically **VCOs** (Voltage Controlled Oscillators), i.e. local oscillators (electrical for RF systems, optical for laser systems) whose fundamental frequency can be changed by applying a voltage to a control port.

Before discussing the lock schematics and performances, it is worth introducing some **basic concepts** on **phase noise** in real oscillators. An **ideal oscillator** is a physical system capable of generating a physical output which is a **pure harmonic oscillation**. An ideal μ-wave oscillator, for instance, will generate a voltage $v(t)$ given by:

$$v(t) = V_0 \cdot \cos(\omega_0 t + \varphi_0) \tag{7}$$

whose amplitude $V_0$, frequency $\omega_0$ and phase $\varphi_0$ do not change with time. In the real world this never happen, and a **real oscillator** is better described by the following expression:

$$v(t) = V_0 \cdot [1 + \alpha(t)] \cdot \cos[\omega_0 t + \varphi(t)] \tag{8}$$

where $\alpha(t)$ and $\varphi(t)$ account for the unavoidable **amplitude** and **phase fluctuations**. However, two minimum conditions have to be satisfied in order to model the voltage $v(t)$ as an imperfect oscillation, namely $|\alpha(t)| \ll 1$ and $|d\varphi/dt| \ll \omega_0$ .

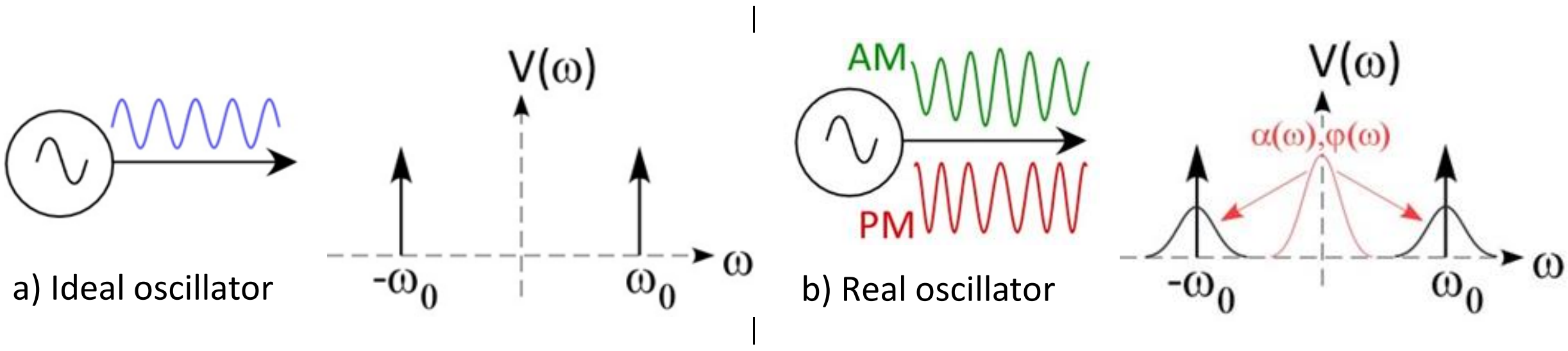

Fig. 13: Ideal (a) vs. Real (b) oscillator signals in time and frequency domains

The spectrum of the ideal signal is just a line at $\pm\omega_0$ frequency, while the **spectrum** of the real signal also includes the **up-conversion** of the **baseband** phase and amplitude **spectra** around the $\pm\omega_0$ carrier. According to the theory of frequency modulated signals [15], the up-conversion of the baseband phase spectrum is linear as far as $|\varphi(t)| \ll 1$.

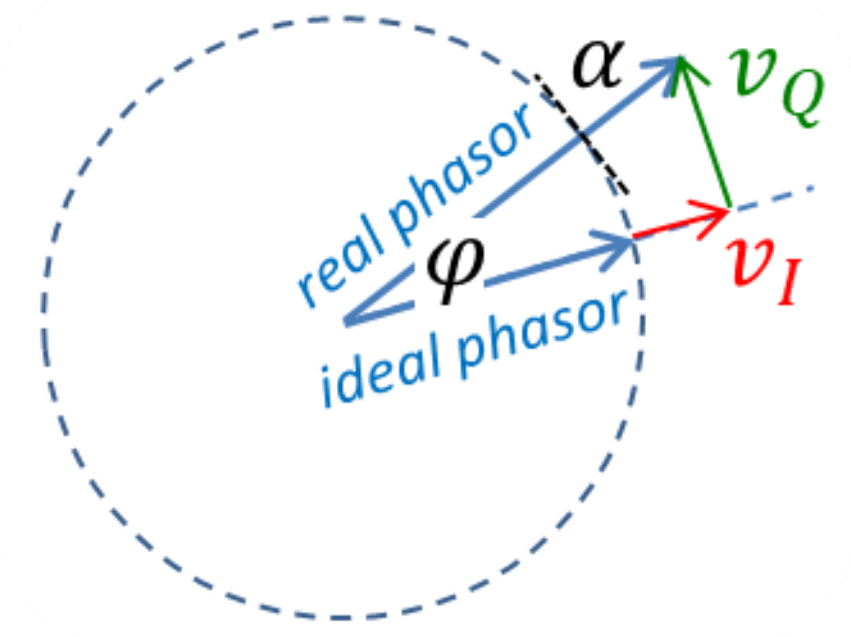

Fig. 13c: In-phase and out-of-phase voltages

Thinking at $\alpha(t)$ and $\varphi(t)$ as the perturbations of the carrier phasor expressed in polar components, we can alternatively project them in Cartesian components $v_I(t)$ and $v_Q(t)$, the in-phase and out-of-phase voltage perturbations, as shown in Fig. 13c. As far as $|\varphi(t)| \ll 1$ the coordinate transformation $(\alpha, \varphi) \rightarrow (v_I, v_Q)$ is straightforward:

$$v(t) = V_0 \cdot \cos(\omega_0 t) + v_I(t) \cdot \cos(\omega_0 t) - v_Q(t) \cdot \sin(\omega_0 t)$$

$$\begin{cases} \alpha(t) = v_I(t)/V_0 \\ \varphi(t) = v_Q(t)/V_0 \end{cases} \quad \text{if } v_I(t), v_Q(t) \ll V_0 \tag{9}$$

Real oscillator outputs are amplitude (AM) and phase (PM) modulated carrier signals. In general it turns out that **close to the carrier** frequency the contribution of the **PM noise** to the signal spectrum **dominates** the contribution of the **AM noise**, and phase noise is the main object of this lecture. However, amplitude noise in RF systems directly reflects in energy modulation of the bunches, that may cause bunch arrival time jitter when beam travels through dispersive and bended paths (i.e. when the transport matrix element $R_{56} \neq 0$ as in magnetic chicanes).

Let's consider a real oscillator neglecting the AM component:

$$v(t) = V_0 \cdot \cos[\omega_0 t + \varphi(t)] = V_0 \cdot \cos\big[\omega_0\big(t + \tau(t)\big)\big] \quad \text{with} \quad \tau(t) \equiv \varphi(t)/\omega_0 \tag{10}$$

The **statistical properties** of $\varphi(t)$, or equivalently those of $\tau(t)$, qualify the **oscillator performances**, primarily the values of the standard deviations $\sigma_\varphi$ and $\sigma_\tau$ (or equivalently $\varphi_{rms}$ and $\tau_{rms}$ since we may assume a zero average value for the random variables). As for every noise phenomena they can be computed through the **phase noise power spectral density** $S_\varphi(f)$ of the random variable $\varphi(t)$. Again, for practical reasons, we limit our observations of the random variable $\varphi(t)$ to a finite time slot $\Delta T$. So we may truncate again the function outside the interval $[-\Delta T/2, \Delta T/2]$ to recover the Fourier transformability.

$$\varphi_{\Delta T}(t) = \begin{cases} \varphi(t) & -\Delta T/2 \le t \le \Delta T/2 \\ 0 & elsewhere \end{cases} \tag{11}$$

Let $\Phi_{\Delta T}(f)$ be the Fourier transform of the truncated function $\varphi_{\Delta T}(t)$. We have:

$$(\varphi_{rms}^2)_{\Delta T} = \int_{f_{min}}^{+\infty} S_\varphi(f)\, df \qquad with \qquad S_\varphi(f) \stackrel{\text{def}}{=} 2\frac{|\Phi_{\Delta T}(f)|^2}{\Delta T} \tag{12}$$

where $S_\varphi(f)$ is defined as the **phase noise power spectral density**, whose dimensions are $1/\mathrm{Hz}$, sometimes indicated as $\mathrm{rad}^2/\mathrm{Hz}$. Please notice that in principle we could again extend the definition to $\Delta T \rightarrow \infty$ according to:

$$\varphi_{rms}^2 = \lim_{\Delta T\rightarrow\infty} (\varphi_{rms}^2)_{\Delta T} = \int_0^{+\infty} \left(2 \cdot \lim_{\Delta T\rightarrow\infty} \frac{|\Phi_{\Delta T}(f)|^2}{\Delta T}\right) df = \int_0^{+\infty} S_\varphi(f)\, df \qquad (13)$$

but we must be **aware** that in this case $\varphi_{rms}$ is **likely to diverge**. This is physically possible since the power in the carrier does only depend on the amplitude and not on the phase. That's why in general the **phase noise rms value** in real networks is specified for a **given frequency range** of integration $[f_1, f_2]$, which corresponds to a certain limited observation time $\Delta T$.

Sometimes, instead of the standard $S_\varphi(f)$ , the function "**Single Sideband Power Spectral Density**" $\mathcal{L}(f)$ is used (also called "script-L"), defined according to the IIIE standard 1139-1999 as [10]:

$$\mathcal{L}(f) = \frac{1}{2} S_\varphi(f) = \frac{|V_{\Delta T}(f_c + f)|^2/\Delta T}{V_0^2/2} = \frac{\text{power in 1 Hz phase modulation single sideband}}{\text{total signal power}} \qquad (14)$$

where $V_{\Delta T}(f)$ is the Fourier transform of the oscillator voltage $v(t)$ defined in (10) over an observation of time duration $\Delta T$ long enough to allow for a precise measurement of the noise spectral density around f ($f \cdot \Delta T \gg 1$). The numerator of eq. (14) is actually the single sideband power density of the oscillator signal, that can be eventually directly measured by modern Real Time Spectrum Analyzers which perform high resolution FFT of strings of time domain samples. The more traditional Swept Spectrum Analyzers, instead, measure the signal power emerging form a selectable filter placed across the measurement frequency which define the measurement resolution bandwidth $f_{RBW}$. The eq. (14) numerator is given in this case by the measured power normalized to the selected resolution bandwidth $P(f_c + f)/\, f_{RBW}$. Once the Single Sideband Power Spectral Density $\mathcal{L}(f)$ is introduced, eq. 12 becomes:

$$\varphi_{rms}^2\,|_{\Delta T} = 2 \cdot \int_{f_{min}}^{+\infty} \mathcal{L}(f)\, df \qquad \text{with} \qquad \mathcal{L}(f) = \begin{cases} \dfrac{|\Phi_{\Delta T}(f)|^2}{\Delta T} & f \geq 0 \\ 0 & f < 0 \end{cases} \qquad (15)$$

Definition (14) allows in principle to extract the baseband phase noise power spectrum from a spectral analysis of the oscillator signal in the close vicinity of the carrier. However, the phase baseband spectrum is linearly translated in the carrier sidebands according to eq. (9) expansion only under the condition $|\varphi(t)| = |v_Q(t)/V_0| \ll 1$. This is typically true when $\Delta T$ is short enough, since the peak values of the phase fluctuation are likely to grow with observation time.

The units of the **single sideband power spectral density $\mathcal{L}(f)$** are just the same as those of the **phase noise power spectral density $S_\varphi(f)$**, namely $\mathrm{Hz}^{-1}$, sometimes indicated as $\mathrm{rad}^2/\mathrm{Hz}$. However, the logarithmic scale $10 \cdot \mathrm{Log}_{10}[\mathcal{L}(f)]$ is often used, and the units are indicated as $\mathrm{dB_c/Hz}$. The $\mathrm{dB_c}$ unit has the meaning of "**dB normalized to the carrier**" and is a direct consequence of eq. (14) definition as far as a logarithmic scale is considered.

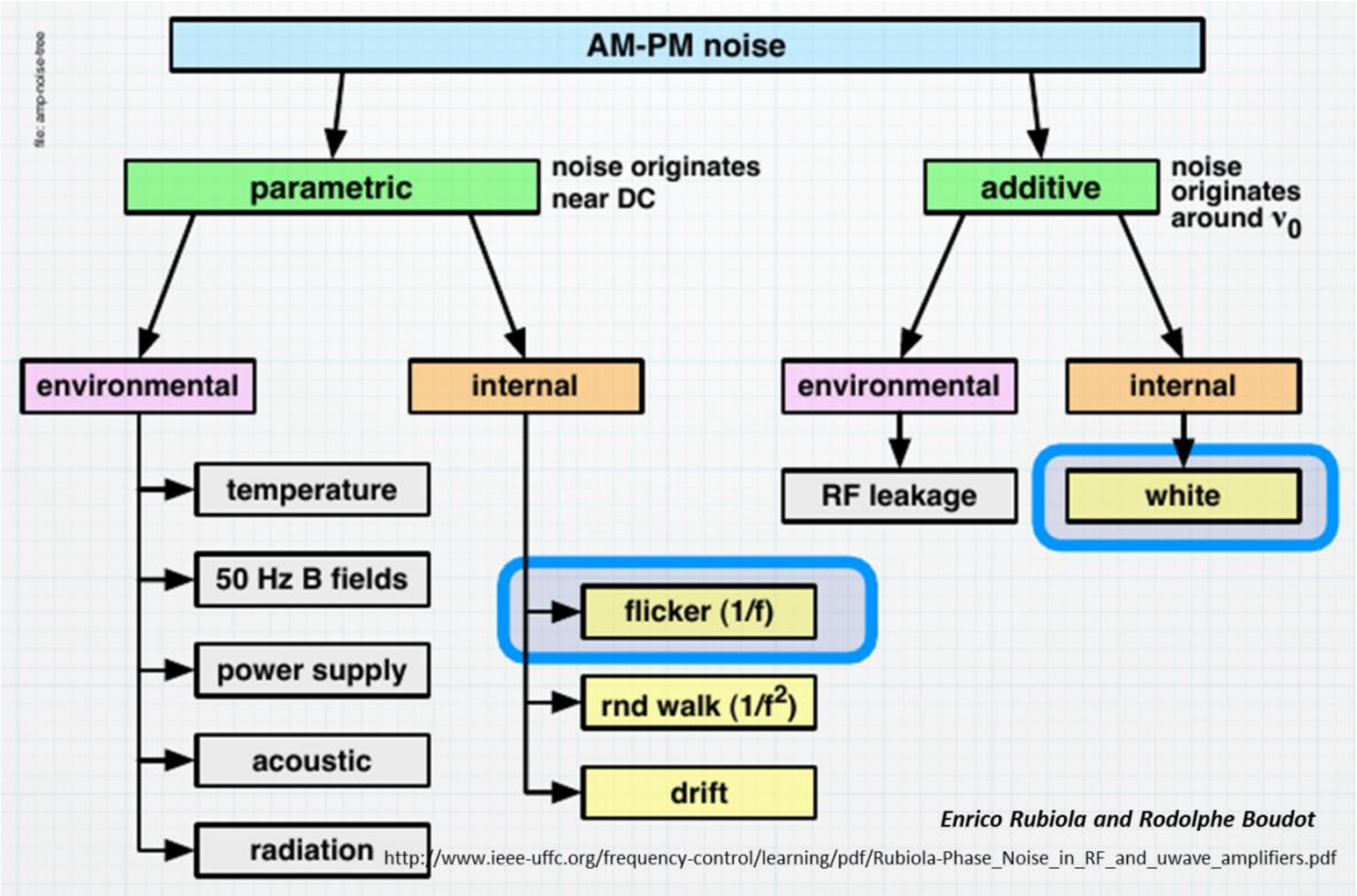

Fig. 14: AM and PM noise generation mechanisms

The physical nature of the AM and PM noise in oscillators and, more generally, in electronics circuits is a wide and complex topic. The noise sources can directly affect the signal as ***additive contributions***, or can be up-converted by modulating the original signal (***parametric contributions***). In literature the ***"close-in phase noise"***, i.e. the phase noise power spectrum close to the carrier, is generally expanded in ***powers of*** $\boldsymbol{1/f}$ according to:

$$S_\varphi(f) = \sum \frac{b_{-k}}{f^k} \qquad k = 0,1,2,3, \ \dots \tag{16}$$

where the terms of the expansion have different names and correspond to different physical mechanisms. The physics dimension of the expansion coefficients is $[b_{-k}] = rad^2 Hz^{k-1}$. When a certain mechanism directly modulates the carrier frequency $f_c$ more than its phase, then according to eq. 6) and taking into account the frequency-to-phase transfer function:

$$\Delta f_c(t) = \frac{d\varphi}{dt} \quad \rightarrow \quad \Delta \mathrm{F}_c(f) = f \cdot \Phi(f) \tag{17}$$

it turns out that frequency noise terms $1/f^k$ are converted into phase noise terms of the $1/f^{k+2}$ order.

| | **Type** | **Origin** | $S_\varphi(f)$ |
|---|---|---|---|
| $f^0$ | White | *Thermal noise of resistors* | $F \cdot kT/P_0$<br>$F = noise\ figure$ |
| | Shot | *Current quantization* | $2q\bar{\imath}R/P_0$ |
| $f^{-1}$ | Flicker | *Flicking PM* | $b_{-1}/f$ |
| $f^{-2}$ | White FM | *Thermal FM noise* | $b_0^{FM} \cdot \frac{1}{f^2}$ |
| | Random walk | *Brownian motion* | $b_{-2}/f^2$ |
| $f^{-3}$ | Flicker FM | *Flicking FM* | $\frac{b_{-1}^{FM}}{f} \cdot \frac{1}{f^2}$ |
| $f^{-4}$ | Random walk FM | *Brownian motion FM* | $\frac{b_{-2}^{FM}}{f^2} \cdot \frac{1}{f^2}$ |
| $f^{-n}$ | ... | *high orders ...* | |

Table I: Terms of the $1/f$ expansion of the "close-in" phase noise

Clearly, terms of high order dominates at low frequency while low order ones dominates at high frequency. Contributions of terms of different order to the total noise spectrum cross at some specific frequencies called "corner frequencies".

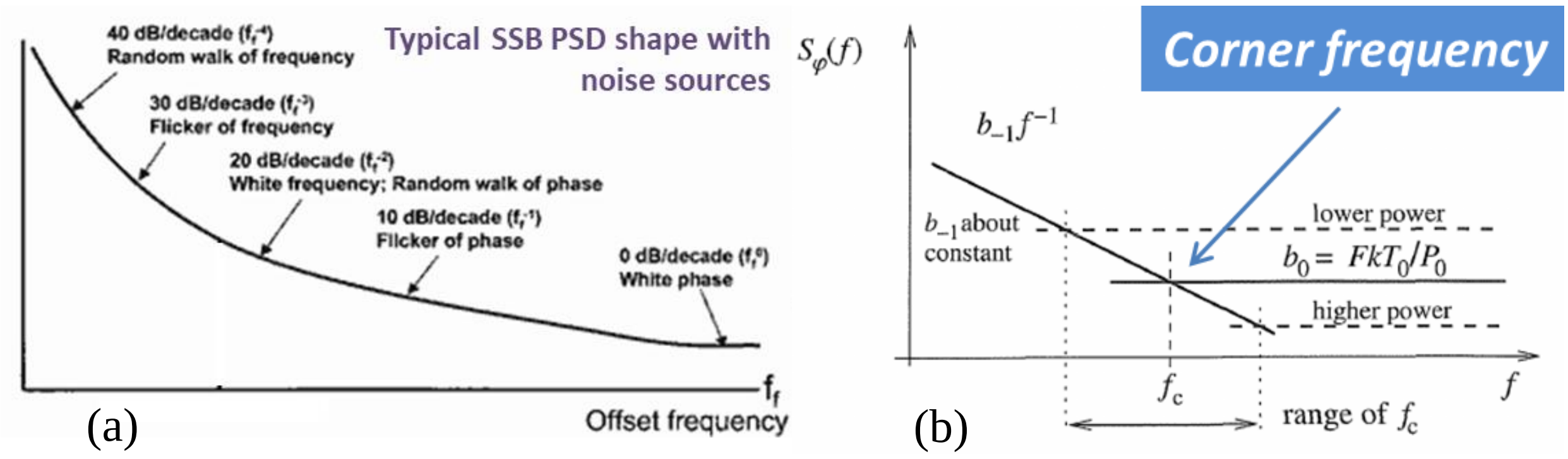

Fig. 15: Close-in noise phase terms (a) and corner frequency (b)

A typical "close-in" phase noise spectrum showing contributions of all $1/f$ expansion terms is reported in Fig. 15 (a), while the corner frequency between white and flicker noise terms is shown in Fig. 15 (b).

Phase jitter $\sigma_\varphi$ and time jitter $\sigma_t$ of a source are simply related by:

$$\sigma_t^2 = \frac{\sigma_\varphi^2}{\omega_c^2} = \frac{1}{\omega_c^2}\int_{f_{min}}^{+\infty} S_\varphi(f)\, df \tag{18}$$

In a frequency synthesizer the time jitter is defined by the quality of the reference generator and it is, in first approximation, independent on the actually selected carrier frequency. According to eq. (18), this means that observed phase noise spectrum $S_\varphi(f)$ will grow as the square of the carrier frequency, as shown in Fig. 16 (b). This scaling has to be properly taken into account to a fair comparison of phase noise spectra of oscillators tuned at different frequencies.

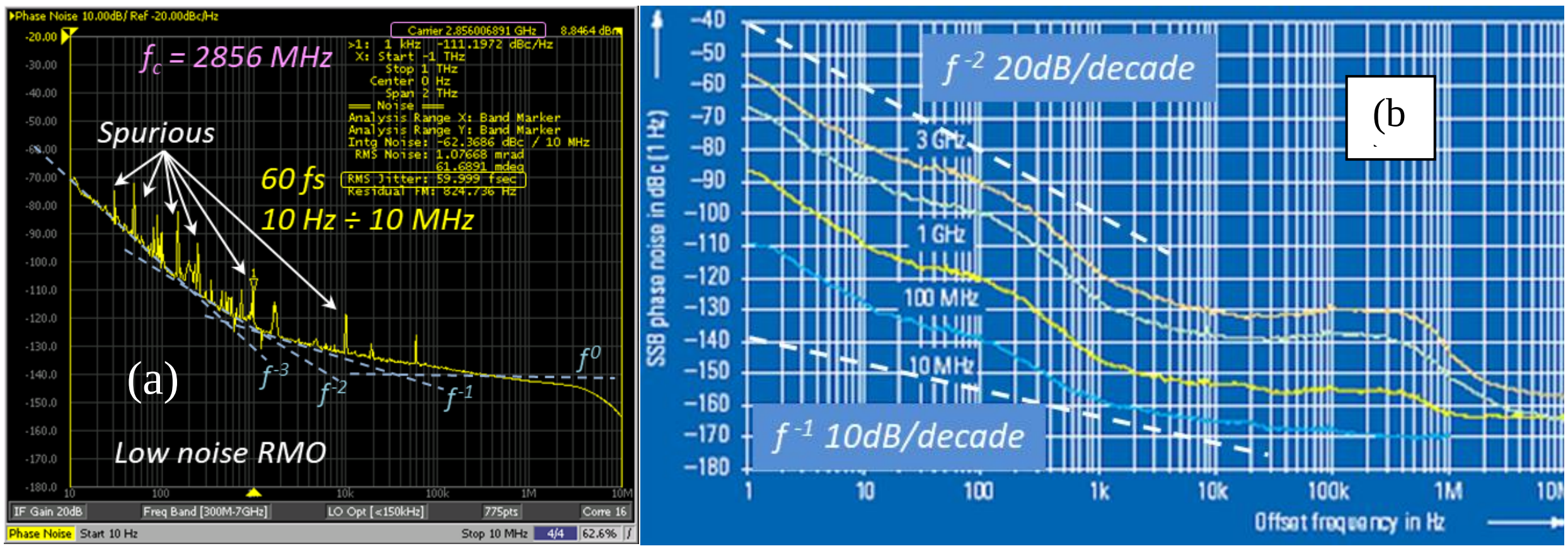

Fig. 16: Measured phase noise spectra of a 2856 MHz reference oscillator (a) and of a commercial frequency synthesizer (b)

## 2.3 Phase Locked Loops (PLLs) and phase noise

The **Phase Locked Loop** (PLL) is a **feedback** system controlling the **phase** of an oscillator, and represents the very fundamental circuit architecture concerning **synchronization** purposes [16] [17]. PLLs are a very general subject in RF electronics, being used for a large variety of tasks, obviously including the synchronization of different oscillators to a common reference and the carrier extraction from modulated signals (clock recovery). In our context PLLs are used to precisely phase-lock the **clients** of the synchronization system to the **master clock** (RMO or OMO). A schematic sketch of a PLL system is shown in Fig. 17 (a). In the most general form, it allows locking the output frequency $\omega_{out}$ to the frequency of a reference oscillator $\omega_{ref}$ provided that the ratio of the two frequencies is a rationale number ($\omega_{out}/\omega_{ref} = D/N$ with $D, N$ integer numbers). In the simplest case the two frequencies are equals ($D/N = 1$). The main building blocks of a PLL system, as shown in the sketch, are:

- A Voltage Controlled Oscillator (VCO), whose frequency range includes $(D/N)\omega_{ref}$ ;
- A phase detector, to compare the scaled VCO phase to the reference;
- A loop filter, which sets the lock bandwidth;
- A prescaler or synthesizer (*N/D* frequency multiplier), if different frequencies are required.

The feedback architecture works in a very standard way. The rescaled **VCO** output is **phase compared** to the **reference** oscillator by using a dedicated phase detector. In a fully electronic PLL the phase detector can be a simple Doubled Balanced Mixers, while for electro-optical or fully optical systems more sensitive devices would be better used, as it will be shown in a next chapter. The **phase error** signal is then processed by a **loop filter/amplifier** whose transfer function is tailored to optimize the PLL global frequency response. The output signal is fed into the **VCO control port** to correct the output phase and **minimize** the residual **error** as a result of the negative feedback action.

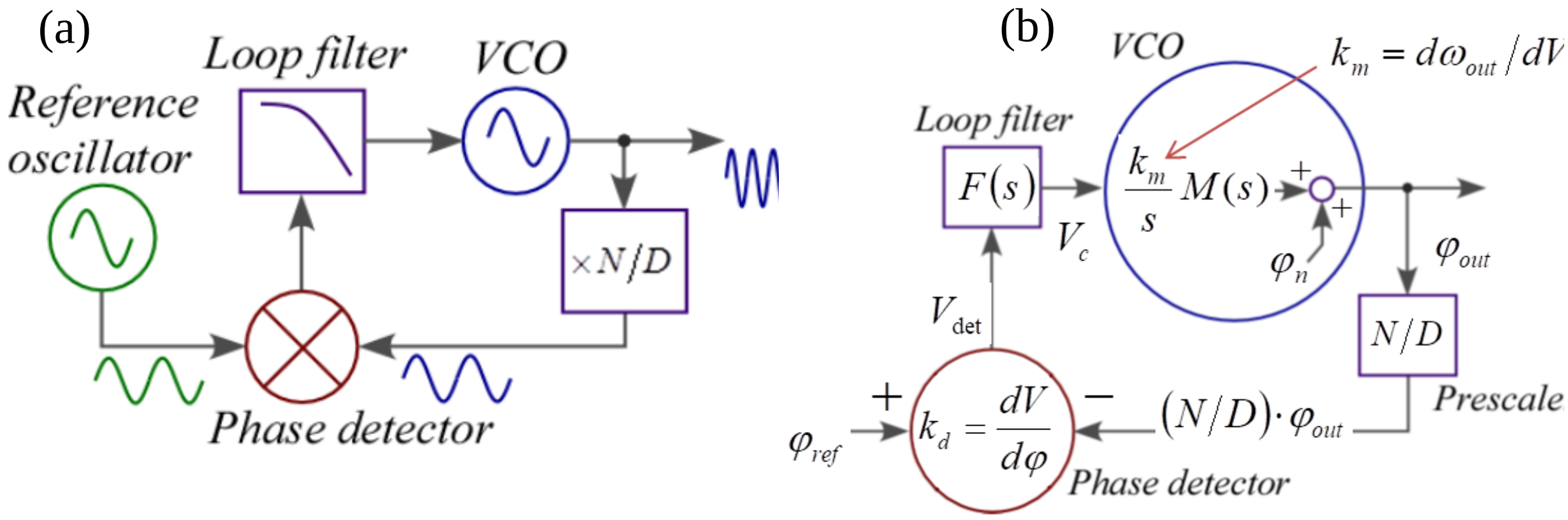

Fig. 17: PLL schematics (a) and linear model (b)

The PLL behaviour can be more formally described referring to the linear model shown in Fig. 17 (b). The phase detector is modelled by its sensitivity $k_d$ [V/rad], while the VCO output phase is related to the control voltage $V_c$ by the transfer function $k_m M(s)/s$, where $k_m$ [rad $s^{-1}V^{-1}$] is the VCO frequency modulation sensitivity, $M(s)$ accounts for the frequency response (or the bandwidth limitation) of the control port, and the term $1/s$ correspond to the frequency-to-phase transfer function (the integration operator in time domain). Finally, $F(s)$ is the loop filter/amplifier transfer function, which is properly designed to increase gain and/or stability of the PLL closed loop transfer function. By solving the Fig. 17 (b) block diagram one gets in the end:

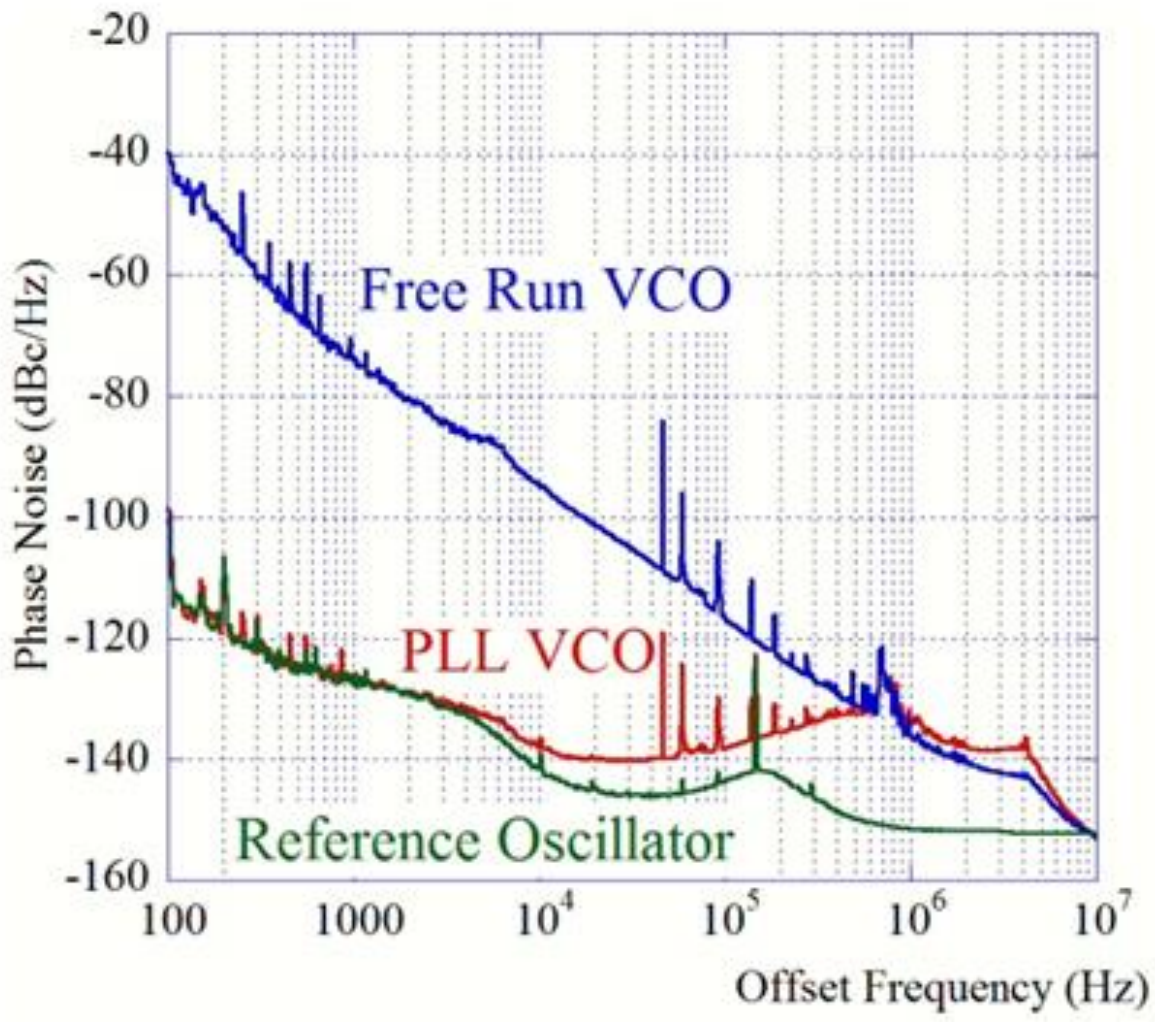

Fig. 18: Phase noise spectrum of a flat loop filter PLL

$$\Phi_{out}(s) = \frac{D}{N}\frac{H(s)}{1+H(s)}\Phi_{ref}(s) + \frac{1}{1+H(s)}\Phi_n(s) \quad with \quad H(s) = \frac{N}{D}\frac{k_d k_m}{s}F(s)M(s) \tag{19}$$

where $\Phi_n(s)$ is the Laplace transform of the VCO free-run phase noise and $H(s)$ is the PLL global open loop transfer function. The **output phase** spectrum is well **locked** to the **reference** one at frequencies where $|H(j\omega)| \gg 1$, while it remains similar to the **free run** one whenever $|H(j\omega)| \ll 1$.

**Loop filters** provide PLL stability, tailoring the frequency response, and set the **loop gain** and the **cut-off frequency**. Loop filter optimization is a key point to improve the PLL performances. A flat loop filter ($\mathrm{F(s) = F_0 = konst.}$) corresponds to pure integrator loop transfer function thanks to the pole at $\mathrm{f = 0}$ provided by the frequency-to-phase conversion operated by the cascade of the VCO and phase detector. The measured phase noise spectrum of a PLL of this type is shown in Fig. 18.

Although this configuration already provides an extremely high (namely infinite) dc loop gain, a residual **dc phase error** $\Delta\varphi_e = \varphi_{out} - \varphi_{ref}$ is necessary to **drive** the VCO to the **reference oscillator frequency** $\omega_{ref}$ according to:

$$\Delta\varphi_e \approx -\,\omega_{ref}/(k_d k_m F_0) \tag{20}$$

This result can be deducted by the basic PLL working principle, but it could be also demonstrated more rigorously. Let's consider to switch on a PLL at $t = 0$. Then the following relations in time and Laplace domains hold:

$$\begin{cases} \omega_{ref}(t) = \omega_{ref} \cdot 1(t) \\ \varphi_{ref}(t) = \omega_{ref} \cdot t \cdot 1(t) \end{cases} \xrightarrow{\text{Laplace transform}} \begin{cases} \Omega_{ref}(s) = \omega_{ref}/s \\ \Phi_{ref}(s) = \omega_{ref}/s^2 \end{cases} \tag{21}$$

Combining Eqs. (19) and (21) and assuming $D/N = 1$ we got:

$$\Delta\Phi_e(s) = \Phi_{out}(s) - \Phi_{ref}(s) = -\frac{1}{1+H(s)}\Phi_{ref}(s) = -\frac{s}{s + k_d k_m F_0} \cdot \frac{\omega_{ref}}{s^2} \tag{22}$$

The regime value of the phase error variable can be computed using the Laplace transform limit theorem, according to:

$$\lim_{t\to\infty} \Delta\varphi_e(t) = \lim_{s\to 0} s \cdot \Delta\Phi_e(s) = -\,\omega_{ref}/(k_d k_m F_0) \tag{23}$$

which is again Eq. (20).

Non-zero PLL phase offset is in general unwanted since it can result in some detrimental effects such as:

- phase variations following the VCO characteristics drifts
- AM-to-PM mixing because of the off-quadrature signals at the phase detector inputs.

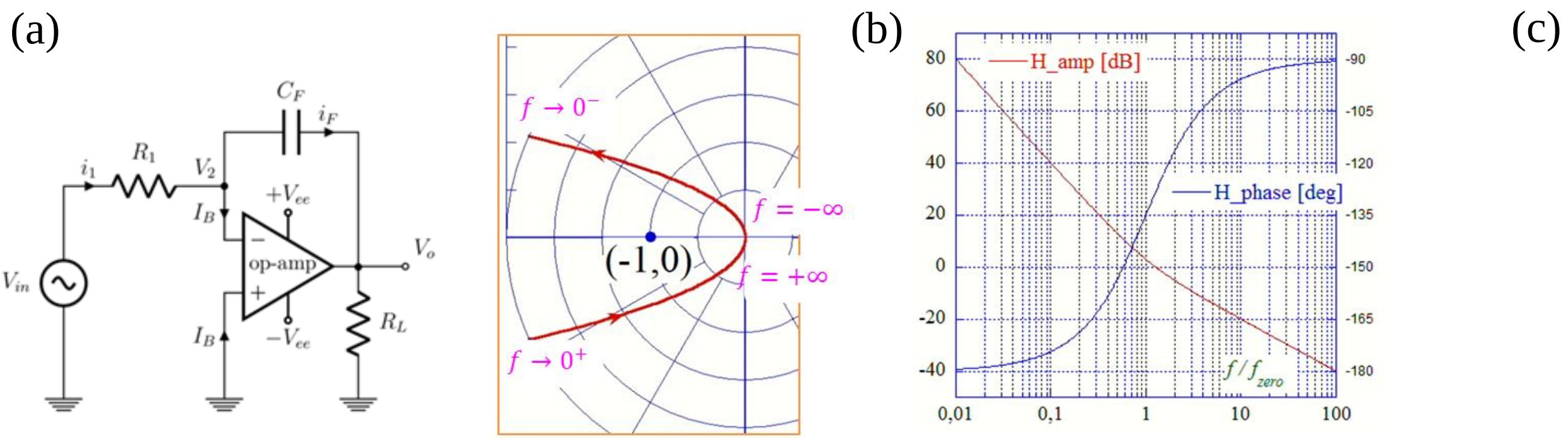

Fig. 19: Loop filter circuit featuring an extra dc pole (a), Nyquist Locus (b) and Bode plot (c) of the PLL open loop transfer function $H(j\omega)$

Residual phase offset $\Delta\varphi_e$ can be compressed to zero by adding one (or more) dc pole ($s = 0$) and one (or more) compensating zero at a certain $s = \omega_z$ in the loop filter transfer function, according to:

$$F(s) = \frac{\omega_0}{s}(1 - s/\omega_z) \tag{24}$$

The required transfer function can be easily implemented by using an operational amplifier circuit as represented in Fig. 19 (a). Provided that the frequency of the zero $\omega_z$ is properly tuned such that $|H(j\omega_z)| \approx 1$, it can be simply demonstrated that the PLL system is solidly stable as shown by the Nyquist locus of Fig. 19 (b), and it features a steep -40 dB/decade frequency response at low frequency as shown in Fig. 19 (c). By using the limit theorem of Eq. (23) with the modified loop transfer function of Eq. (24) one finally gets $\Delta\varphi_e \approx 0$ .

Loop filters can also incorporate a global equalization of the open loop transfer function to improve the PLL performances. This could be required, for instance, to compensate and suppress a peaking frequency response of a VCO modulation port, as shown in the Bode plot of Fig. 20, which allows increasing the loop gain.

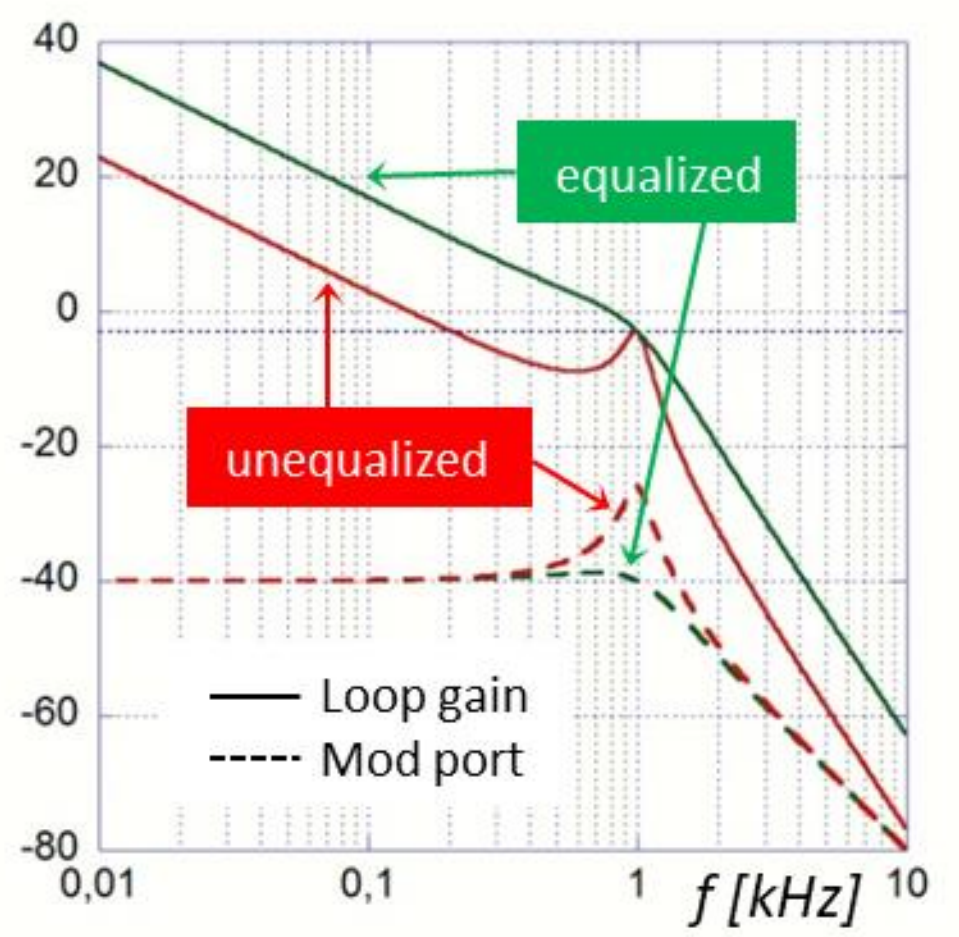

Fig. 20: PLL loop gain with (green) and without (red) equalization.

## 2.4 PLLs for synchronization system clients

The PLL topology is widely used also to synchronize (lock) the client oscillators to the master oscillator of a particle accelerator facility synchronization system. However, the **devices used** to implement each block represented in the Fig. (17) schematics can be very much **different** depending on the **nature** and characteristics of the **signals** involved. As matter of fact both the reference and the client oscillator can be either RF VCOs or laser cavities, and proper phase detectors are chosen consequently. Besides the classic RF vs. RF detectors (such as balanced mixers), there are **more sensitive** dedicated **detectors** capable to **directly compare** RF vs. laser pulse trains (electro-optical detector), or even different pulse trains one vs. the other (fully optical detectors). These kind of devices will be briefly illustrated in the following.

To allow for locking to an external reference, laser oscillators are made to behave as VCOs by trimming the **optical cavity length** through a **piezo controlled mirror**. This technique works fine for the scope, but it anyway presents **limited modulation bandwidth** (few kHz typical) because of the piezo frequency response, and **limited dynamic range** ($\Delta f/f \approx 10^{-6}$) because of the μm range of the piezo induced deformation. This last limitation is typically overcome by adding motorized translational stages to enlarge the mirror positioning range.

The optical cavity PLL bandwidth is set by the piezo frequency response in the 1 ÷ 10 kHz range, while RF VCOs can be locked in a **much wider bandwidth** exceeding the MHz range. However, as matter of fact **mode-locked laser** oscillators exhibit excellent **low-phase noise** spectrum at frequencies beyond the PLL bandwidth, which makes the PLL bandwidth limitation irrelevant in most cases, and provides acceptable levels of residual phase noise.

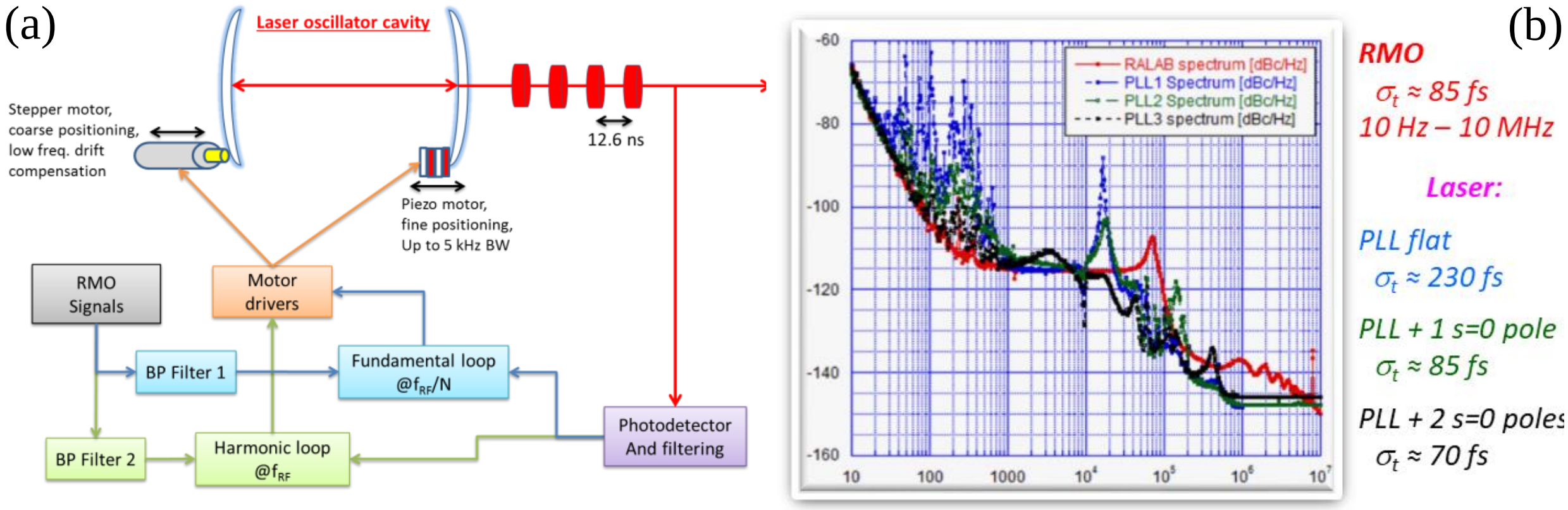

Fig. 21: laser oscillator PLL: schematics (a) and residual phase noise spectra (b) for different loop filters.

## 2.5 Precision phase noise measurements

Phase noise of RF sources can be measured in various ways. The most commonly used technique is the **PLL with Reference Source**. The phase of the Source Under Test (SUT) signal is measured with respect to a tunable reference Local Oscillator locked to the source. The **baseband signal** used to drive the PLL is also **acquired** and processed to extract the **relative phase** error information.

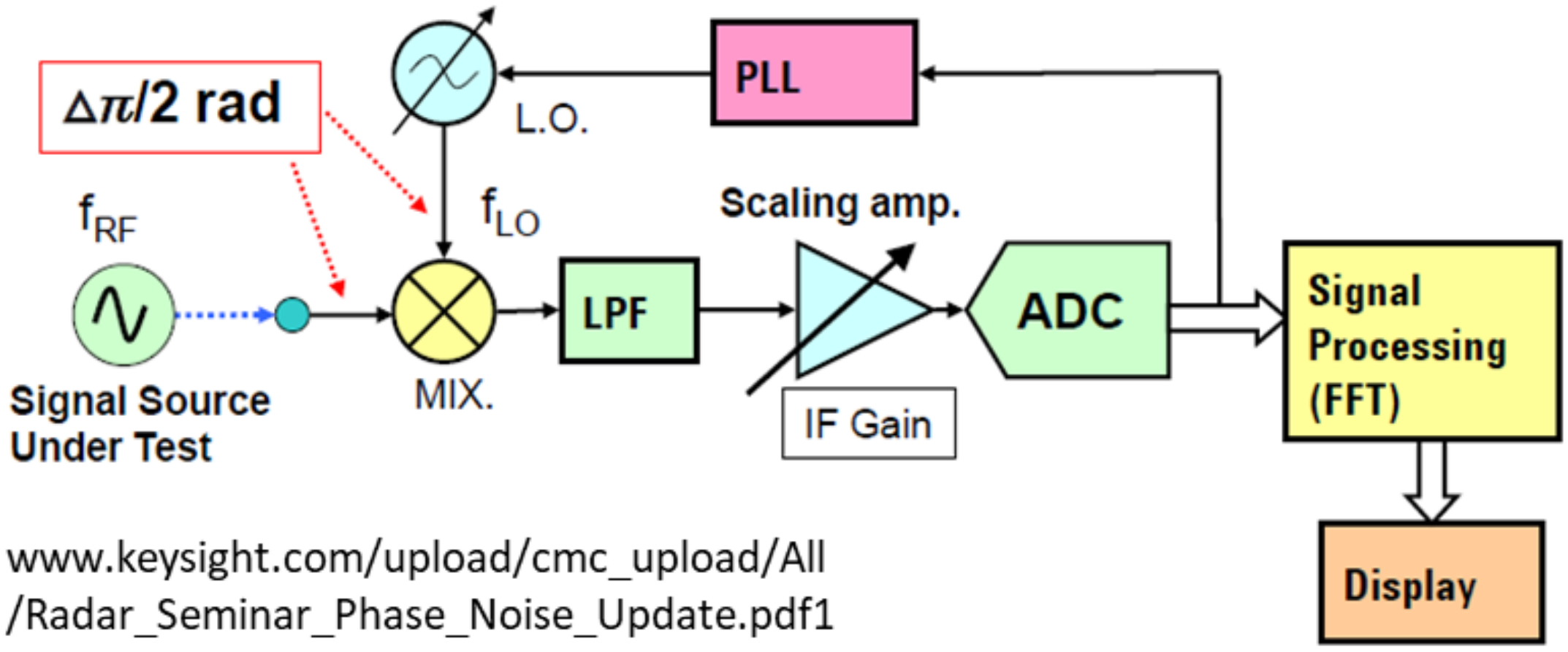

Fig. 22: Schematics of the PLL with Reference Source method to measure the SUT phase noise spectrum.

The quantity actually measured is the phase error between the SUT and the reference. The acquired sample is numerically Fourier transformed, and according to eq. (12), the squared module of the Fourier transform normalized to the time duration of the sample gives the phase noise power spectrum:

$$\Delta\varphi_{meas}(t) = \varphi_{SUT}(t) - \varphi_{LO}(t) \quad \xrightarrow{\text{source uncorrelation}} \quad S_{\varphi_{meas}}(\omega) = S_{\varphi_{SUT}}(\omega) + S_{\varphi_{LO}}(\omega) \qquad (25)$$

Since the phase noises of the two sources are uncorrelated the measured phase noise spectrum is just the sum of the SUT and reference individual spectra. Clearly, the **reference source noise** contribution to the measurement can be neglected only when it is **smaller enough** (i.e. 15÷20 dB lower) with respect to the SUT contribution. Only sources **remarkably worse** (i.e. more noisy) **than reference** can be accurately characterized.

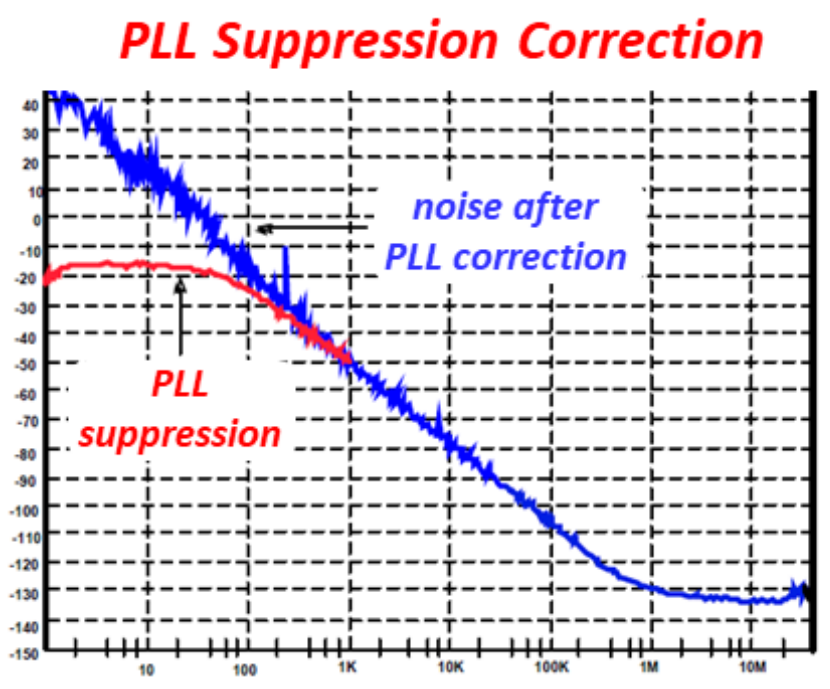

Fig. 23: phase noise measurement

Moreover, since the PLL partially suppresses the SUT noise at frequencies within the closed loop bandwidth, measured data are corrected by proper algorithms accounting for PLL frequency response to provide accurate results.

Signal Source Analyzers (SSAs) are commercially available instruments for phase noise characterization of RF oscillators [18][19]. They integrate an optimized front end including the PLL circuits, high resolution digitizers and data processing software allowing for precise phase noise measurements based on the PLL with reference source technique.

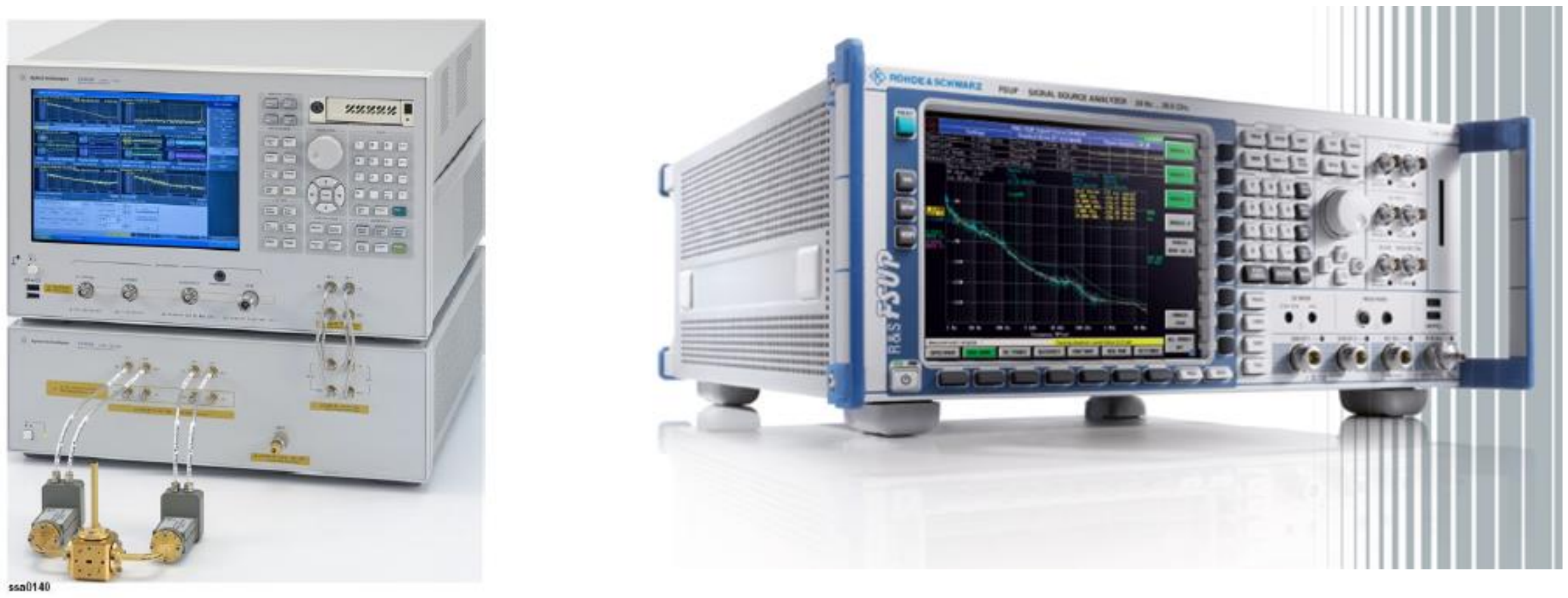

Fig. 24: Top-class commercial Source Signal Analyzers

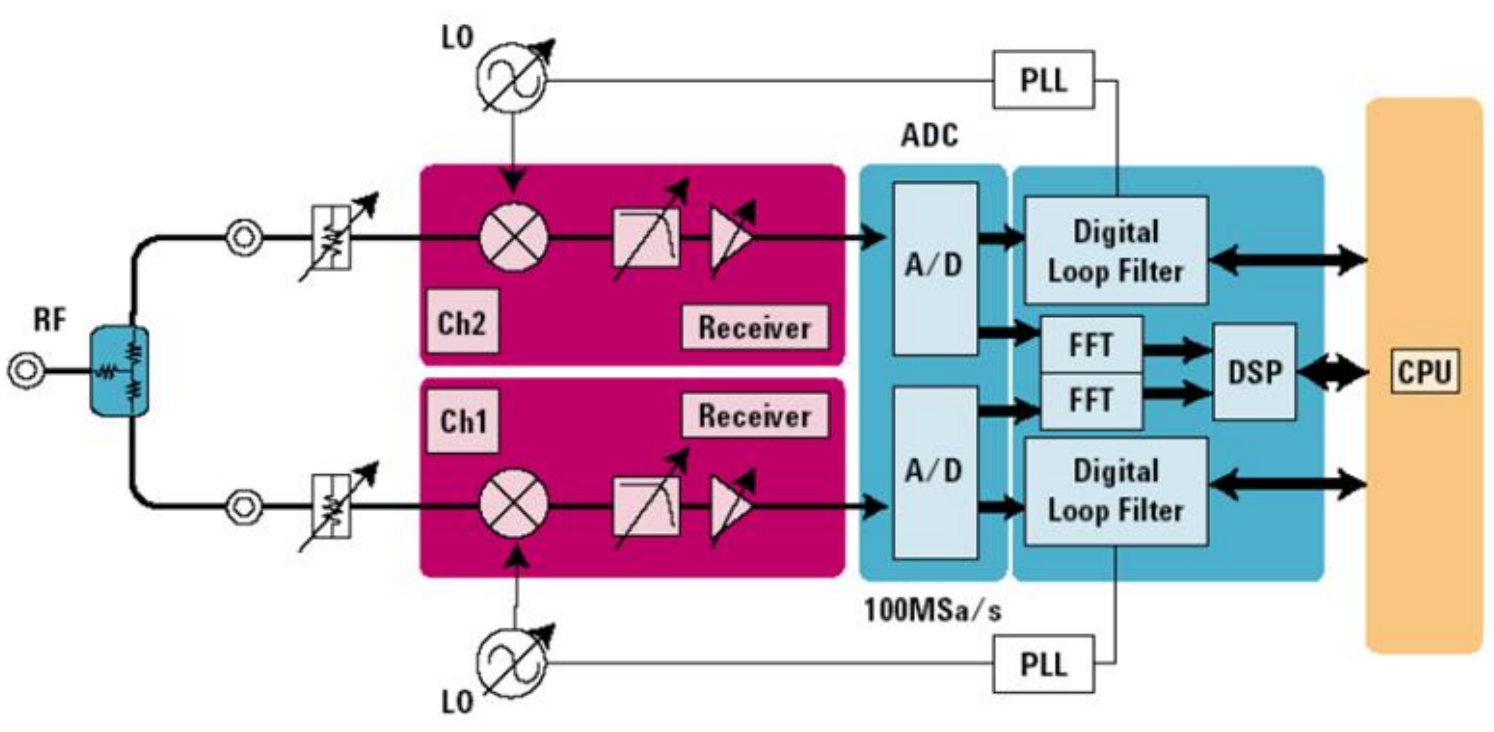

Fig. 25: PLL with two reference sources

As shown in Fig. 25 most SSAs are equipped with two low noise LO oscillators locked to the SUT signal to reduce the limitation coming from the reference contribution to the measurements. The phase noise of the source under test is measured in parallel along two independent channels. The acquired data are Fourier transformed and mathematically cross-correlated to reduce the contribution of the references to the measurement.

The cross-correlation between two independent measurements taken from two totally uncorrelated reference sources is an effective approach to characterized the phase noise spectrum of SUT of the same class of the reference, or even better. Let's start considering the SUT phase noise $\varphi_{SUT}(t)$ measured simultaneously with respect to the two SSA reference LOs, which is:

$$\Delta\varphi_{1,2}(t) = \varphi_{SUT}(t) - \varphi_{LO_{1,2}}(t) \quad \xrightarrow{FFT} \quad \Delta\Phi_{1,2}(f) = \Phi_{SUT}(f) - \Phi_{LO_{1,2}}(f) \tag{26}$$

The cross correlation function of $\Delta\varphi_1(t)$ and $\Delta\varphi_2(t)$, indicated as $r(\tau)$, and its Fourier transform $R(f)$ are given by the following mathematical expressions:

$$r(\tau) = \int_{-\infty}^{+\infty} \Delta\varphi_1(t) \cdot \Delta\varphi_2(t+\tau)\, dt \quad \xrightarrow{FFT} \quad R(f) = \Delta\Phi_1^*(f) \cdot \Delta\Phi_2(f) \tag{27}$$

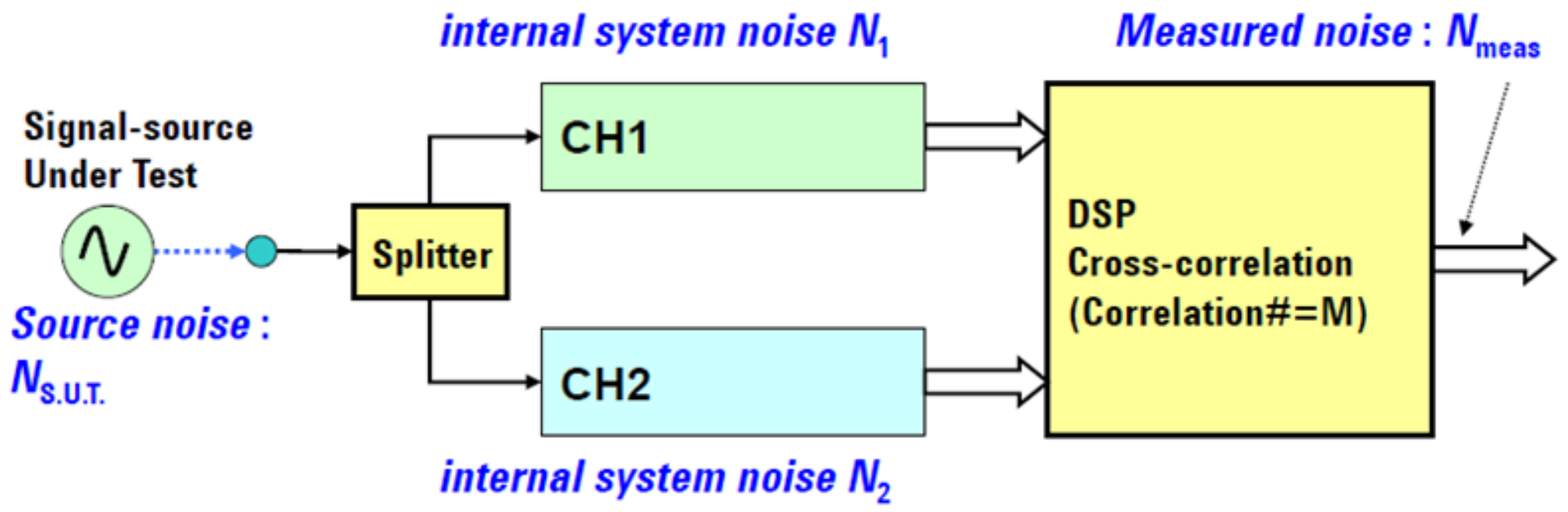

Fig. 26: Schematics of the cross-correlation measurement in PLL with two reference sources instruments

Combining eqs. (26) and (27) we got:

$$R(f) = |\Phi_{SUT}(f)|^2 - \left[\Phi_{SUT}^*(f) \cdot \Phi_{LO_2}(f) + \Phi_{SUT}(f) \cdot \Phi_{LO_1}^*(f)\right] + \Phi_{LO_1}^*(f) \cdot \Phi_{LO_2}(f) \tag{28}$$

The Fourier transform $R(f)$ of the cross correlation function $r(\tau)$ presents three terms. However, if we assume to deal with stationary random processes and suppose to perform a series of $M$ consecutive measurements, we expect that the first term will give a constant, real and positive contribution, while each of the two remain terms will gives a complex contribution of constant module and random phase. If we extract the phase noise spectrum by averaging over the $M$ different measurements, we expect to measure a $S_{\varphi_{meas}}(f)$ function given by:

$$S_{\varphi_{meas}}(f) = S_{\varphi_{SUT}}(f) + \left[\sqrt{S_{\varphi_{LO1}}(f)} + \sqrt{S_{\varphi_{LO2}}(f)}\right] \sqrt{\frac{S_{\varphi_{SUT}}(f)}{M}} + \sqrt{\frac{S_{\varphi_{LO1}}(f) S_{\varphi_{LO2}}(f)}{M}} \tag{29}$$

After M averages the magnitude of the uncorrelated contributions to the measurement (including the cross product of the phase noises of the 2 reference sources) is expected to decrease by a factor $\sqrt{M}$. **Measurement accuracy** at level of the phase noise of the **reference sources** or even lower can be achieved provided that the number of **correlations** M is **large enough**. Sources of quality comparable with references or even better can be accurately characterized, but the price to be paid is the measurement time duration which increases linearly with the number of correlations.

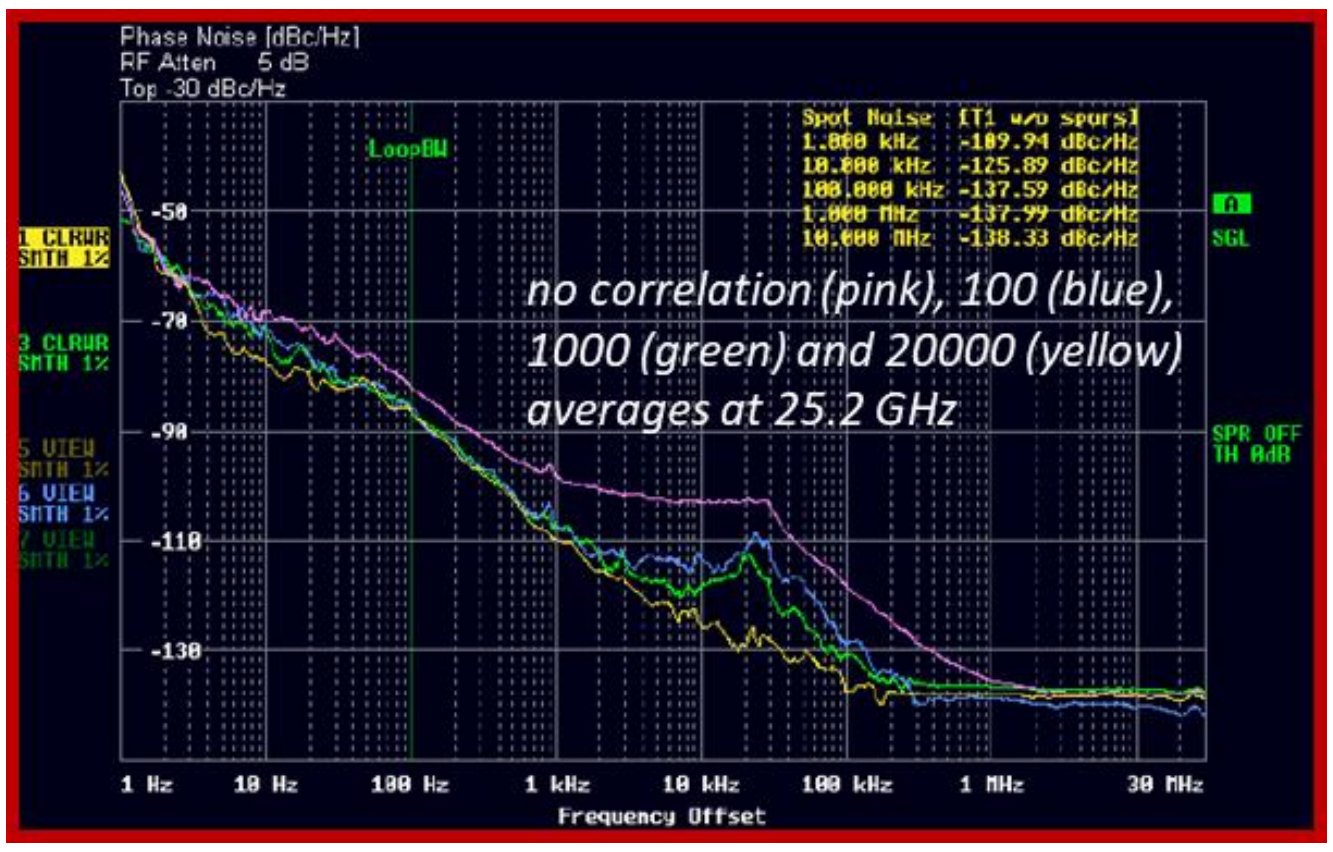

Fig. 27: Measurement resolution increase with number of cross-correlations

## 2.6 Phase detection

### *2.6.1 Balanced mixers:*

Phase detectors are the front-end devices of PLL systems and phase noise measurement instruments. When dealing with electrical signals, standard RF & microwave devices and techniques can be used. In this context **balanced mixers** are the most **widely used** devices [17][20]. A Double Balanced Mixer (DBM) is a non-linear device that can be realized by the circuit schematic reported in Fig. 28. A diode bridge is dynamically biased by a variable voltage applied to the LO input, so that the voltage appearing at the IF output is simply the signal applied to the RF input ports times a polarity function which is established by the instantaneous polarity of the LO signal, that is:

$$V_{IF}(t) = V_{RF}(t) \cdot \mathrm{sgn}[V_{LO}(t)] \tag{30}$$

where sgn(x) is a function assuming values $\pm 1$ depending on the sign of its argument x.

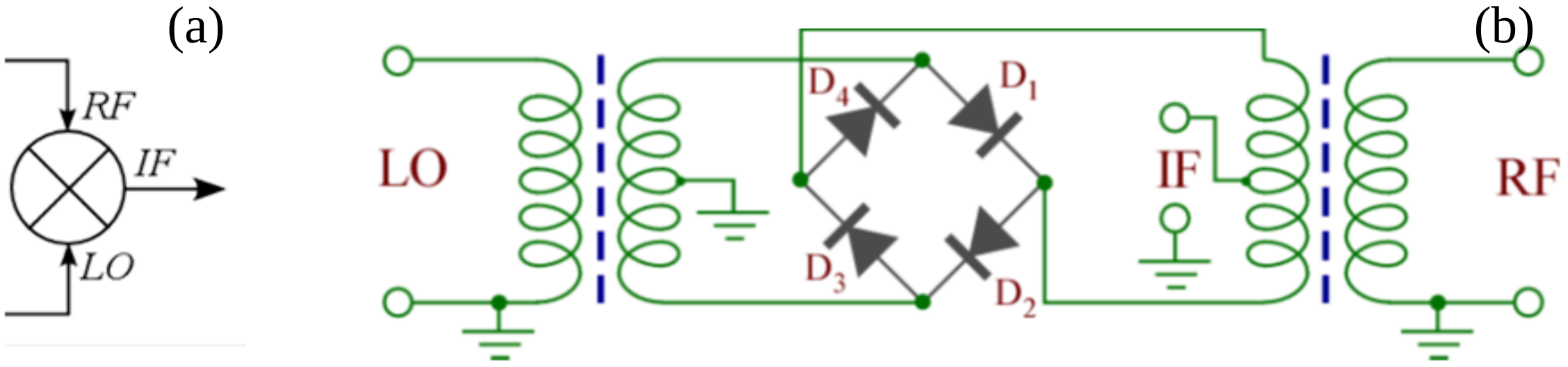

Fig. 28: Double Balanced Mixer: symbol (a) and internal circuit (b)

If both RF and LO signals are sine waves oscillations, which means $V_{RF}(t) = V_{RF}cos(\omega_{RF}t + \varphi_{RF})$ and $V_{LO}(t) = V_{LO}cos(\omega_{LO}t)$, the output signal at the IF port is given by:

$$V_{IF}(t) = V_{RF}cos(\omega_{RF}t + \varphi_{RF}) \cdot sgn[cos(\omega_{LO}t)] = V_{RF}cos(\omega_{RF}t + \varphi_{RF}) \cdot \sum_{n=odds} \frac{4}{n\pi} cos(n\omega_{LO}t) \qquad (31)$$

where the second term in the product is a square wave of frequency $\omega_{LO}$ and it has been Fourier expanded. The product generates a series of terms of frequency $n\omega_{LO} \pm \omega_{RF}$. The terms of frequency $\omega_{LO} \pm \omega_{RF}$ are the principal ones, while the others are smaller and called **intermodulation products**.

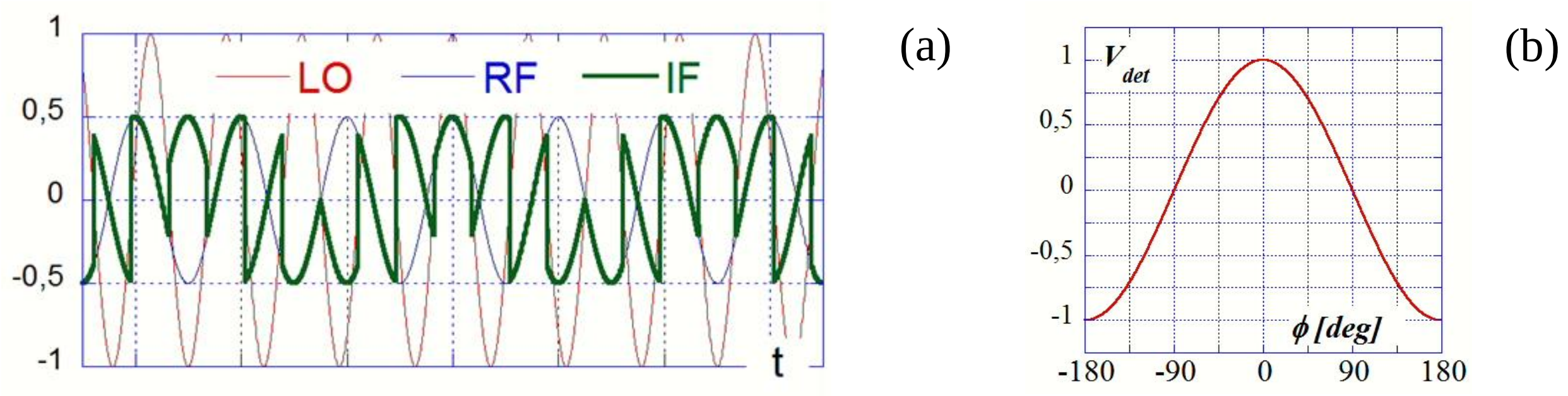

Fig. 29: IF voltage of a DBM with $\omega_{LO} \neq \omega_{RF}$ (a) and phase detection characteristics at $\omega_{LO} = \omega_{RF}$ (b)

If only the principal terms are retained, the IF voltage is given by:

$$V_{IF}(t) = \frac{2}{\pi} V_{RF}cos[(\omega_{RF} - \omega_{LO})t + \varphi_{RF}] + \frac{2}{\pi} V_{RF}cos[(\omega_{RF} + \omega_{LO})t + \varphi_{RF}] \qquad (32)$$

According to Eq. (32) the phase information of the RF signal can be translated at a lower frequency $\omega_{IF} = \omega_{RF} - \omega_{LO}$ (the second term of higher frequency can be filter out) or even in baseband if $\omega_{RF} = \omega_{LO}$. In this last case the relative phase is measured directly according to:

$$V_{IF}(t) = \frac{2}{\pi} V_{RF} \cdot cos[\varphi_{RF}(t)] \qquad (33)$$

The phase detection characteristics is shown in Fig. 29 (b), where the highest sensitivity and linearity are obtained around $\varphi_{RF} = \pm\pi/2$.

Double balanced mixers are extremely popular devices in μwave engineering. They are simple, passive, cheap and robust components suitable for a large number of applications. Used as phase detectors they show sensitivity of the order of $5 \div 10$ mV/deg. That sensitivity can be enough for synchronization applications, especially when used at high carrier frequency. However, they can suffer of AM to PM conversion, and the **sensitivity** is way **lower** when compared to **optical** and **electro-optical** devices.

### *2.6.2 Balanced mixers in conjunction with photo-detectors:*

There are cases where phase detectors need to compare a laser pulse train with respect to an RF oscillator voltage. This happens, for instance, when an RF power plant need to be synchronized with a local copy of the facility Optical Master Oscillator reference transported from the central synchronization hutch. Clearly, one obvious approach to measure such a relative phase is to convert the **light pulses** into **electric signals** by means of a **fast photodiode**. The generated electric pulses have the same repetition rate of the laser and the spectrum of the electrical signal is a comb containing all the harmonics of the laser fundamental repetition frequency up to the photodiode band limit which may extend to many GHz, while the original laser pulses can be $\ll 1$ ps long, which means that the spectrum of the optical train extends well beyond 1 THz. One selected **harmonic** of the **photodiode** signal spectrum can be extracted with a bandpass filter (BPF) to obtain a sine wave to be **phase compared** with an **RF oscillator** by means of a balanced mixer or other standard devices.

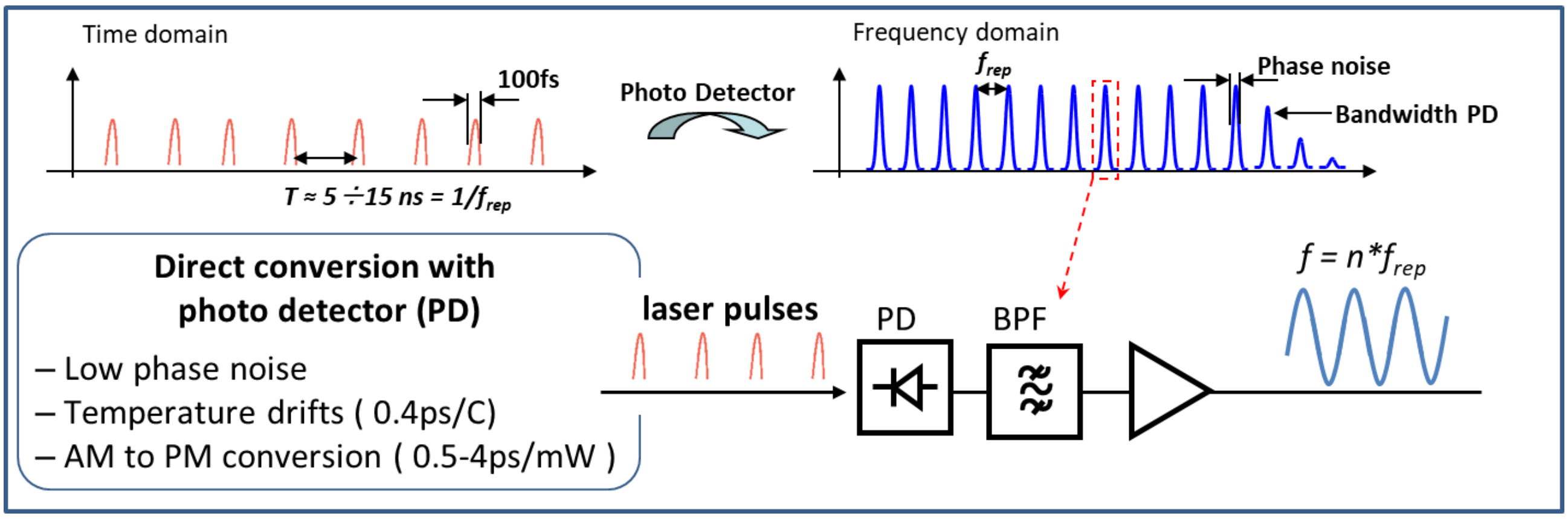

Fig. 30: Conversion of a train of optical pulses into an electrical sine-wave signal by a photodiode and a BPF

Although this is a simple and cost-effective method, it is temperature sensitive and especially prone to AM to PM conversion in the photodiode. For best performances other detection techniques to more **directly compare** the phase between **optical** pulses and **RF** oscillators are better suited.

Balanced Optical Microwave Phase Detector:

A dedicated instrument to **directly measure** the relative phase between a train of short **optical pulses** and an **RF oscillator** signal has been developed and commercialized in recent years. It is called **Balanced Optical Microwave Phase Detector** (or BOM-PD) and it is based on a Sagnac-loop interferometer ring including a directional electro-optic phase modulator [21]. The BOM-PD converts the **phase error** between a laser pulse train and a µwave oscillator into an **amplitude modulation** of the laser pulse train downstream the interferometer, which can be furtherly converted into an analog voltage by a photodiode.

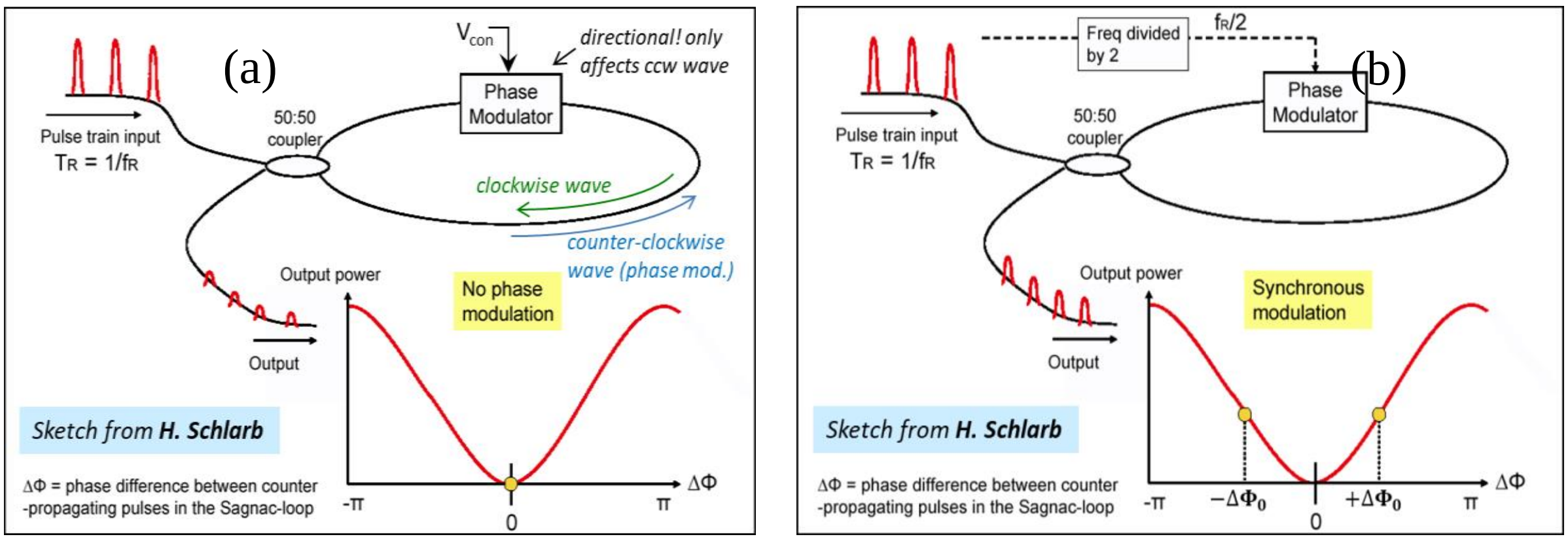

Fig. 31: BOM-PD unbiased (a) and biased at half the laser pulse repetition rate (b)

The electro-optic phase modulator produces a dephasing $\Delta\Phi$ proportional to the applied control voltage $V_{con}$ between the optical carriers of the 2 pulse trains counter-propagating along the ring. The intensity $I_{out}$ of the laser train emerging from the interferometer is then:

$$I_{out} \div I_{in}\overline{[cos(\omega_c t) - cos(\omega_c t + \Delta\Phi)]^2} \div I_{in} sin^2(\Delta\Phi/2) \quad (34)$$

If no voltage is applied at the modulator control port then $\Delta\Phi = 0$ and the 2 counter-propagating waves interfere destructively at the output combiner. The amplitude of the output pulses is nearly zero in this case. The BOM-PD requires to be biased by a sine wave voltage at frequency $f_R/2$, being $f_R$ the laser repetition frequency. The $f_R/2$ sine wave is extracted from the input pulse train with dedicated electronics, and it is phased such that the laser pulses of the counter-clockwise wave cross the modulator aligned with the sine wave maxima and minima, as shown in Fig. 32 (a).

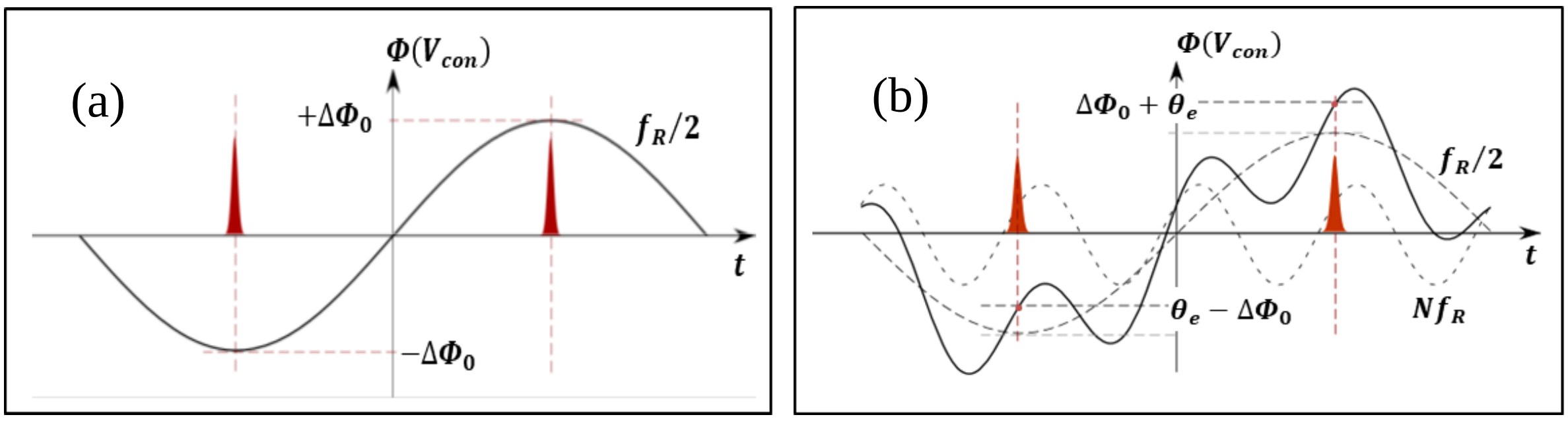

Fig. 32: BOM-PD biasing with single tone at $f_R/2$ (a) and two-tones at $f_R/2$ and $nf_R$ (b)

Under this condition the pulses experience in sequence a phase shift of $\pm\Delta\Phi_0$. The intensity of the laser output train is non-zero in this case, but it does not show amplitude modulation since all pulses are equally attenuated according to Eq. (34).

Let's consider adding a sine wave voltage of frequency $nf_R$ , an integer multiple of the laser repetition frequency, on the modulator control port. The harmonic voltage will imprint a constant contribution $\theta_e$ to the phase of all pulses, so finally the pulses will show in sequence a phase modulation equal to $\theta_e \pm \Delta\Phi_0$. According to Eq. (34) the intensities of "even" and "odds" pulses emerging from the interferometer will be different in this case, depending on the value of the relative phase $\theta_e$. The output pulse train is therefore amplitude modulated, as shown in Fig. 33 (a), with a modulating frequency $f_R/2$ and a modulation depth depending on the relative phase between the harmonic voltage and the input laser train.

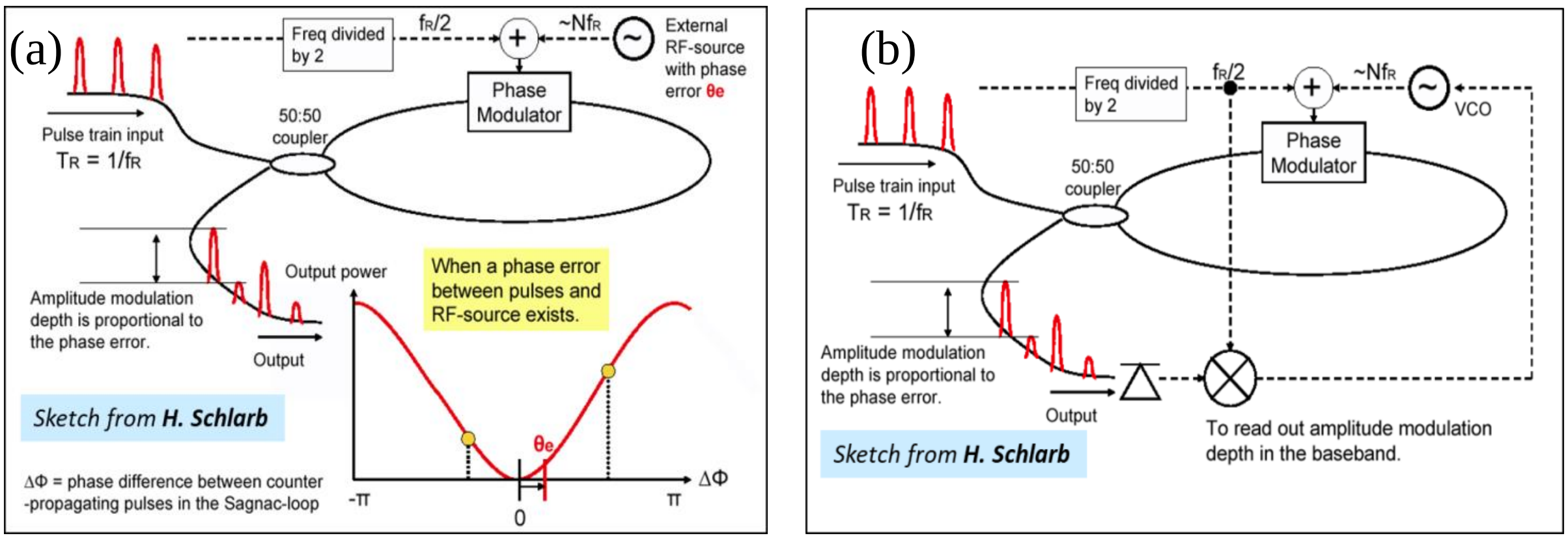

Fig. 33: BOM-PD as phase detector for an ext. RF source (a) and locking a VCO to $nf_R$ in a PLL topology (b)

The laser pulse train emerging from the interferometer is amplitude demodulated to extract the phase error information. A possible use of the phase error signal is **driving a PLL** to lock a VCO tuned around **one harmonic** $nf_R$ of the original **laser repetition rate**, as shown in Fig. 33 (b). This configuration allows extracting and converting to a µwave signal the timing information encoded in the laser repetition rate, with the best preservation of the phase purity.

BOM-PD are definitely superior with respect to balanced mixers equipped with photo-detectors in measuring relative phase between RF oscillators and optical pulse trains. In particular they are much more insensitive against laser amplitude fluctuations and temperature drifts. For RF carrier of $f = 1.3\ GHz$ it has been measured instrument jitter and drift well below $10\ fs\ rms$.

### *2.6.3 Optical Balanced Cross-correlators:*

**Optical cross-correlation** is definitely the **highest sensitivity** phase detection technique when dealing with phase measurements between two trains of short optical pulses [[22]][[23]]. This method exploits the short time duration of the pules produced by mode-locked lasers ($\sigma_t \approx 100$ fs) corresponding to extremely large bandwidths exceeding 1 THz.

Optical cross-correlation is based on the property of special **non-linear crystals**, when illuminated simultaneously by two incident beams of photons of wavelengths $\lambda_1$ and $\lambda_2$, to emit radiation at a **shorter wavelength** $\lambda_3$, with $1/\lambda_3 = 1/\lambda_1 + 1/\lambda_2$. The instantaneous **intensity** of the emitted radiation is proportional to the **product of the incident radiations**, and it is zero if the incident pulses do not overlap in time. The residuals of the incident radiations emerging from the crystal are filter out and a photo-detector converts the intensity of the cross-correlation radiation into an electrical signal, as shown in the sketch of Fig. 34. In order to provide a **balanced output**, that means to produce an output voltage to the first order only proportional to the relative time delay and not to the intensity of the incident beams, the instrument is equipped with **a second detection arm**. The two incident beams have orthogonal polarization and in the second arm is made such that the two polarizations experience a controlled differential delay whose value $\Delta T$ is chosen to be similar to the pulse duration $\sigma_t$.

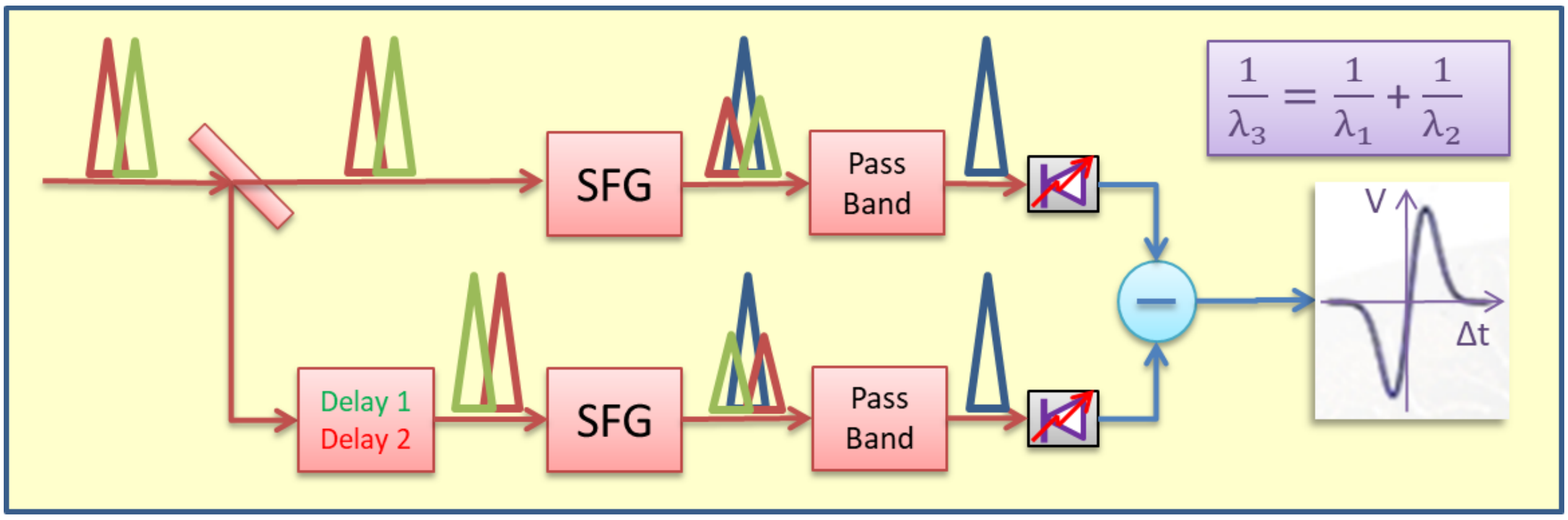

Fig. 34: Schematics of an optical balanced cross-correlator

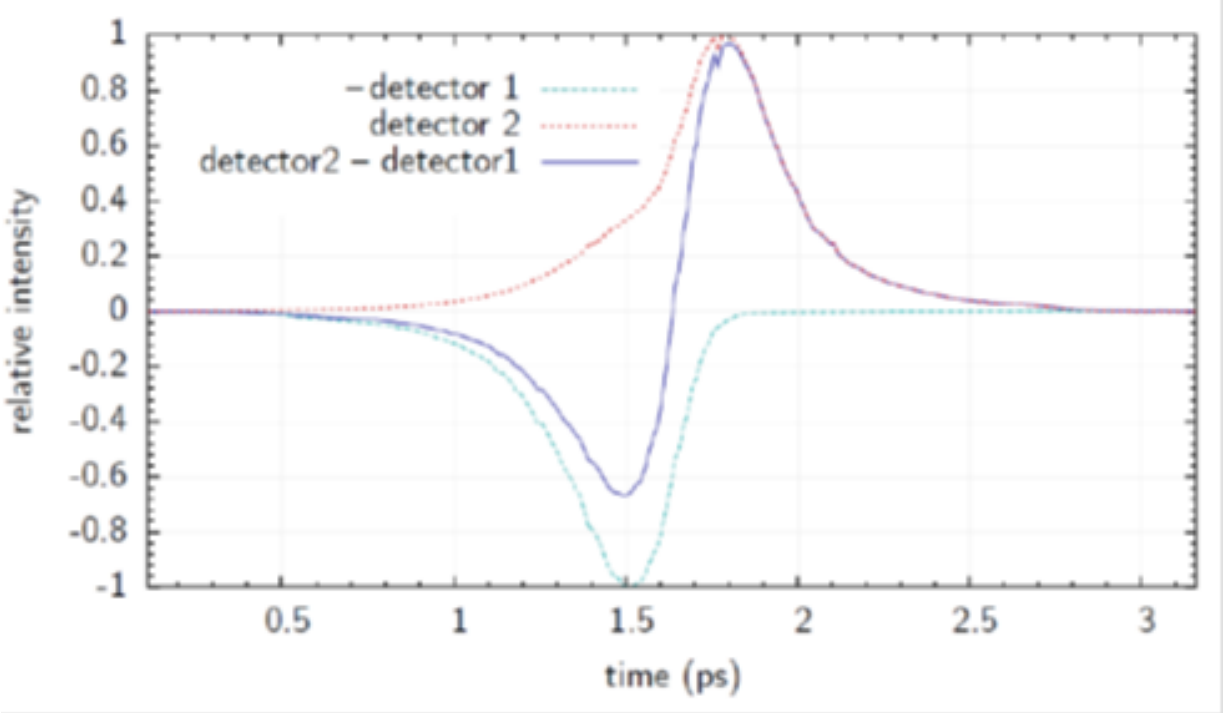

Fig. 35: Detection plot of an optical X-correlator

The detector output $V_0$ is the difference of the voltages generated by the two photo-detectors converting the cross-correlation radiation produced in each arm. If the time delay between the pulses in the upper arm is exactly $\Delta T/2$ then clearly $V_0 \approx 0$ (balanced output), while it grows rapidly as soon as the delay deviates from $\Delta T/2$ because of the short time duration of the pulses.

Detection sensitivity up to $10$ mV/fs can be achieved with short pulses, which makes this instrument by far the **highest resolution phase detector** available.

# 3 Performance of the synchronization systems

## 3.1 Client Residual Jitter

As extensively discussed in the introductory section of this paper, the synchronization system of an accelerator based facility consists in the distribution network of a common reference clock signal to any location where a client oscillator is positioned (see for instance Figs. 7a and 10), and in a series of local Phase Locked Loops whose task is to conform at the best the oscillator repetition rates to the reference one. According to Eq. (19), a client with a free-run phase noise $\Phi_{i_0}(f)$ and spectral density $S_{i_0}(f)$ once being locked to the reference with a PLL with loop gain $H_i(f)$ will show a residual phase jitter $\Phi_i(f)$ and a phase noise power spectrum $S_i(f)$ given by:

$$\Phi_i = \frac{H_i}{1+H_i}\Phi_{ref} + \frac{1}{1+H_i}\Phi_{i_0} \xrightarrow[contributions]{inchoerent} S_i = \frac{|H_i|^2}{|1+H_i|^2}S_{ref} + \frac{1}{|1+H_i|^2}S_{i_0} \qquad (35)$$

where the two residual spectral density contributions can be added being uncorrelated.

According to Eq. (18) the standard deviation $\sigma_{t_i}$ of the absolute residual temporal error of the client oscillator is given by:

$$\sigma_{t_i}^2 = \frac{1}{\omega_{ref}^2}\int_{f_{min}}^{+\infty} \frac{|H_i|^2 S_{ref}(f) + S_{i_0}(f)}{|1+H_i|^2} df \qquad (36)$$

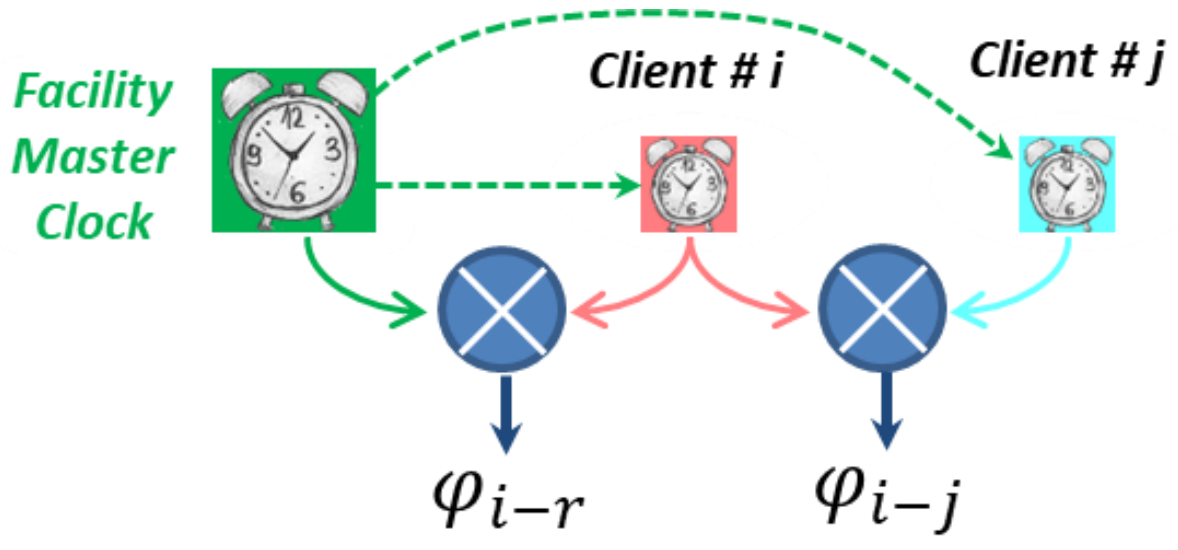

Fig. 36: time jitter between clients and reference

The quantity $\sigma_{t_i}$ measures the rms time deviation with respect to an ideal noise-free sine-wave oscillator. However, **more realistic** quantities impacting directly the beam dynamics are the relative **jitters** between **clients and reference** $\varphi_{i-r}(t) = \varphi_i(t) - \varphi_{ref}(t)$, and among **different clients** $\varphi_{i-j}(t) = \varphi_i(t) - \varphi_j(t)$. Eqs. (35) and (36) are easily modified to compute the standard deviations of the relative residual temporal errors between the client oscillators and the facility reference, according to:

$$\Phi_{i-r} = \frac{\Phi_{i_0} - \Phi_{ref}}{1+H_i} \Rightarrow S_{i-r}(f) = \frac{S_{i_0}(f) + S_{ref}(f)}{|1+H_i|^2} \Rightarrow \sigma_{t_{i-r}}^2 = \frac{1}{\omega_{ref}^2}\int_{f_{min}}^{+\infty} \frac{S_{ref}(f) + S_{i_0}(f)}{|1+H_i|^2} df \qquad (37)$$

Eq. (37) shows that the residual jitter between client and reference oscillators is obtained by adding the phase noise spectral densities of the two sources reduced by the PLL closed loop gain squared. The contribution coming from the free-run client oscillator is typically dominant. However, this is not necessarily always true, especially in the high frequency region beyond the PLL bandwidth. Laser oscillators, for instance, show very low phase noise densities beyond the audio spectrum, which eventually might be better than µ-wave reference oscillators in that range.

Concerning the relative jitter between different clients Eq. (35) leads to:

$$\Phi_{i-j} = \frac{\Phi_{i_0} - \Phi_{ref}}{1+H_i} - \frac{\Phi_{j_0} - \Phi_{ref}}{1+H_j} \rightarrow \ S_{i-j}(f) = \frac{S_{i_0}(f)}{|1+H_i|^2} + \frac{S_{j_0}(f)}{|1+H_j|^2} + \left|\frac{H_i - H_j}{(1+H_i)(1+H_j)}\right|^2 S_{ref}(f)$$

$$\sigma_{t_{i-j}}^2 = \frac{1}{\omega_{ref}^2}\int_{f_{min}}^{+\infty} S_{i-j}(f)\,df = \frac{1}{\omega_{ref}^2}\int_{f_{min}}^{+\infty}\left[\frac{S_{i_0}(f)}{|1+H_i|^2} + \frac{S_{j_0}(f)}{|1+H_j|^2} + \left|\frac{H_i - H_j}{(1+H_i)(1+H_j)}\right|^2 S_{ref}(f)\right]df \quad (38)$$

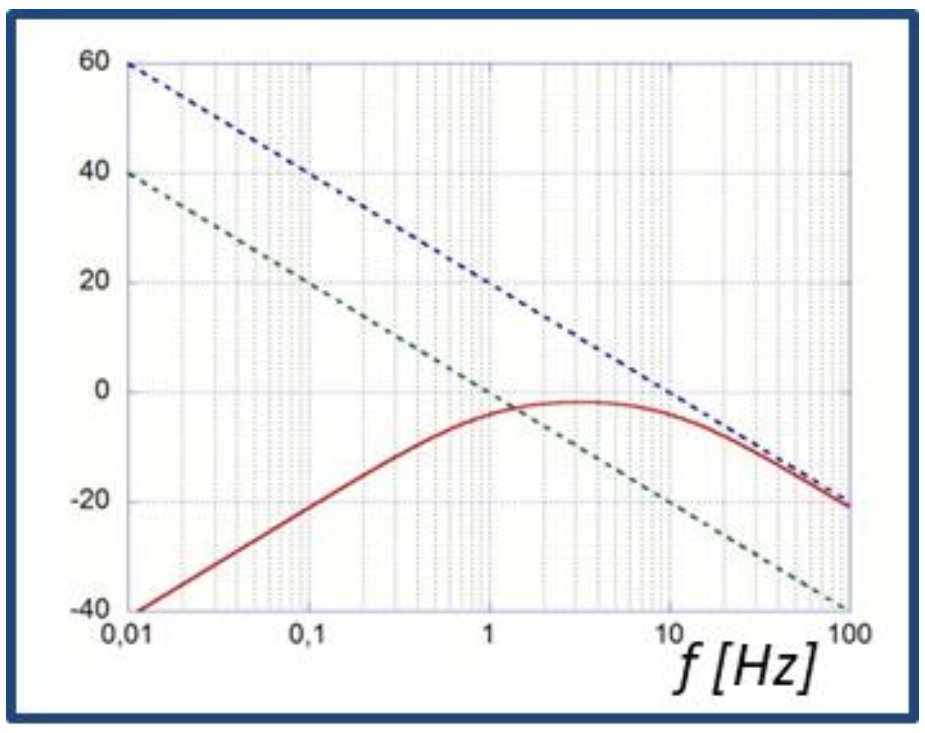

Fig. 37: Function weighting the reference contribution to the clients relative jitter

According to Eq. (38) the residual jitter between two clients locked to the same reference is obtained by adding three contributions. The first two contributions are simply the original phase noise spectral densities of the free-run clients reduced by the squared closed loop gains of the two independent PLLs locking the clients to the common reference. The third term instead depends on the **reference spectral density** itself, weighted by a function that accounts for the **mismatch** of the two **PLL transfer functions**. Clearly, if the two PLL transfer functions were equal ($H_i = H_j$) this term would vanish. As matter of fact, it is very difficult and improbable to precisely equalize the various PLLs present in the synchronization system of an accelerator facility.

As already discussed, laser clients will be locked through piezo-controlled actuators, showing a PLL bandwidth much narrower than RF VCOs locked to the same reference and driving, for instance, RF power plants. As an example, let's consider two clients with simple and similar PLL open loop transfer functions (just a single pole at $f = 0$) but different bandwidths as shown in Fig. 37 (1 kHz and 10 kHz, dashed lines). In this specific case the function $|H_i - H_j|/|(1+H_i)(1+H_j)|$ weighting the contribution of the reference phase noise spectral density to the relative jitter between the clients is represented by the solid red line in Fig. 37. This function shows a value close to unity in the frequency region 1 ÷ 10 kHz between the two PLL cut-off frequencies, which means that the reference phase noise characteristics in that frequency range is directly imprinted into the relative noise between the two locked clients. This consideration explains why **RMOs** are specified at level of **state-of-the-art low noise oscillators** in a wide spectral region including the cut-off frequencies of every client PLLs (0.1 ÷ 100 kHz typical).

### 3.2 Drift of the reference distribution

In the previous paragraphs it has been described how a local client oscillator is locked to a local copy of the facility RMO and what are the expected values for absolute and relative residual jitters. Obviously, another essential task of the facility synchronization system is to **transport the reference signal** from a central station, where the RMO (and/or the OMO) are physically located, **to all clients** requiring synchronization that are typically distributed over the whole facility. The length of the transport links can vary depending on the facility size and the position of the specific clients, ranging from few meters to some kilometers. Ideally, whatever the length is, the **signal downstream the link** should be merely a **copy of the reference**, with no added phase drift and jitter.

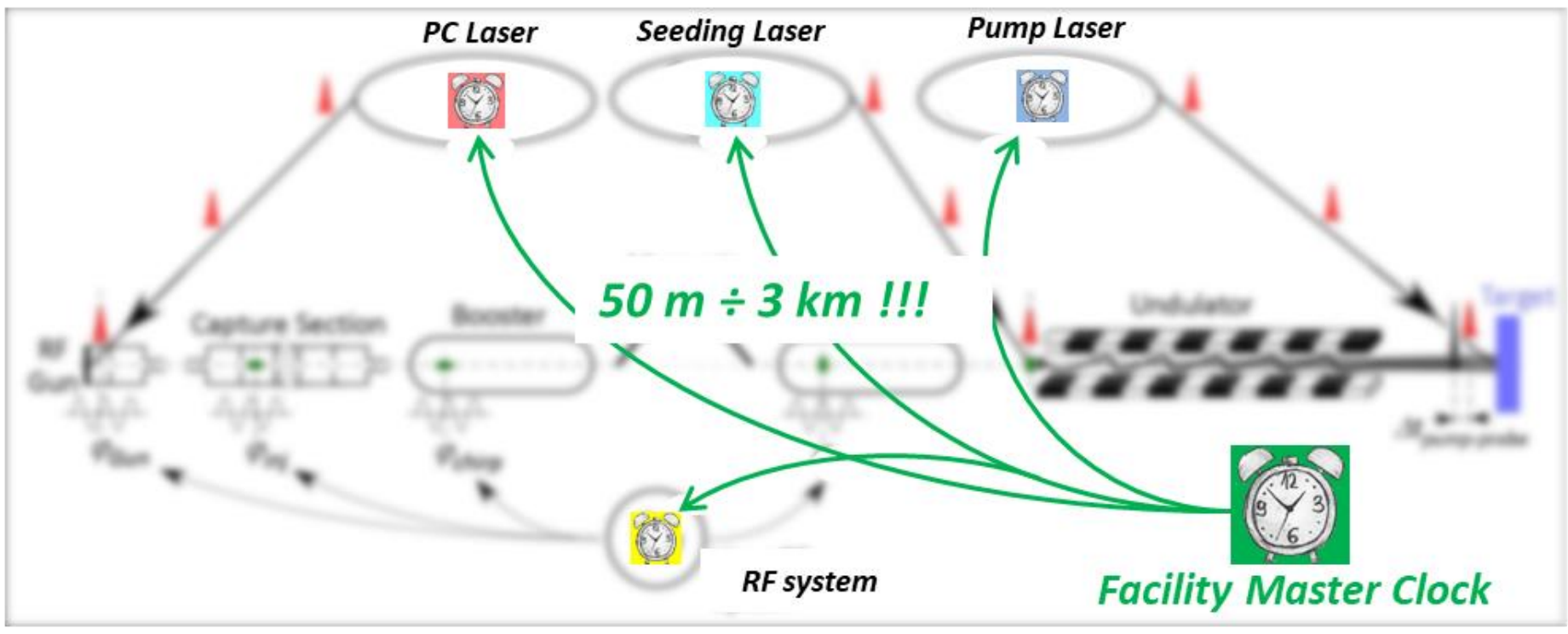

Fig. 38: Distribution of the facility master clock over the whole facility

More realistically the perturbation added by the link must be acceptable for the specific application, or at least traceable and correctable. Is this the case for slow varying fluctuations (drifts) that can be tracked and corrected in real time; links implementing an automatic correction of the length fluctuations caused by temperature and other environmental drifts are called **actively stabilized links**. In general, depending on the physical nature of the transported reference signal, the distribution can be electrical or optical, while it can be active or passive depending on the presence or absence of **automatic feedback systems** stabilizing the link total length or, better saying, the **total link delay**.

Electrical distribution is the simplest and cheapest option. The distributed reference is an RF signal taken from the RMO, possibly amplified and splitted, and transported through coaxial cables. The frequency of the RMO ranges typically from $\approx 100$ MHz to few GHz. The carrier frequency choice is a trade-off between cable attenuation, where low frequencies are preferred, and phase detection time resolution, where high frequencies provide better performances.

Coaxial cables are sensitive to temperature variations. The linear expansion coefficient for copper is $\approx 1.7 \cdot 10^{-5}/°\mathrm{C}$ . This means that a $\approx 100$ ns air dielectric cable, corresponding to a quite short link ($\approx 30$ m), will suffer a thermal elongation of $\approx 1.7$ ps/°C, a value unacceptable for most applications even assuming a tight temperature stabilization of the cable path ($\leq 0.1$ °C).

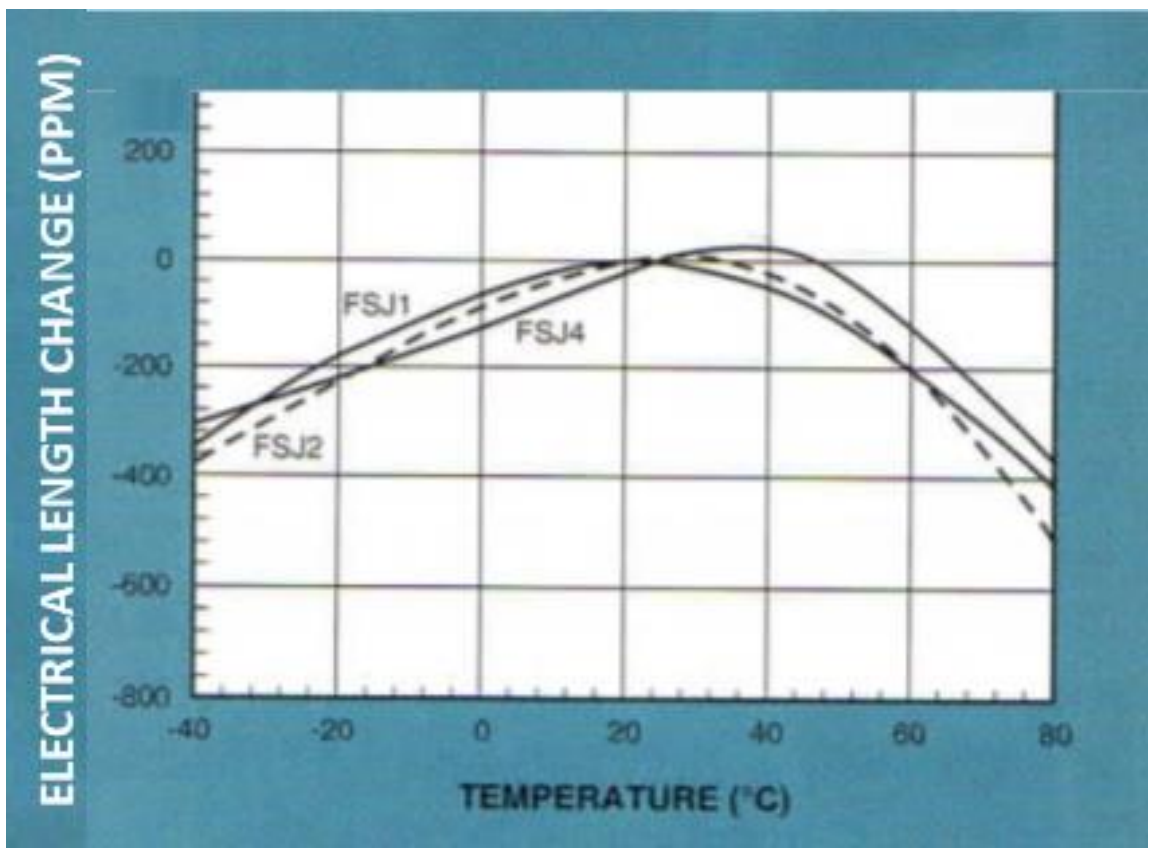

Fig. 39: Andrew Heliax FSJ delay compensated cables

This problem can be mitigated by using special **delay compensated** coaxial cables. In this cables the coaxial conductors are separated by a special dielectric whose permittivity variation with temperature $\varepsilon(T)$ compensates to the first order the delay.

The first order delay compensation depends on the temperature itself and is full at a certain optimal value $T_{opt}$ depending on the cable characteristics. As an example, for the low-loss $3/8$" FSJ2 Andrew Heliax cable [24], whose characteristics are reported in Fig. 39, this value $T_{opt} \approx 24$ °C.

If the cable is operated in a small temperature range around $T_{opt}$ then the relative delay variation $\Delta\tau/\tau$ (expressed in PPM = part per million) reported in the Fig. 39 curve can be approximated with a pure 2nd order term, according to:

$$\left.\frac{\Delta\tau}{\tau}\right|_{PPM} \approx -\left(\frac{T - T_{opt}}{T_c}\right)^2 \tag{39}$$

where $T$ is the actual cable temperature and $T_c \approx 2$ °C is a parameter obtained by fitting the FSJ2 plot. We can use Eq. (39) expression to estimate the delay variation for specific applications and compare it with synchronization specification. For instance if we assume a $1\ km$ long link ($\tau \approx 5\ \mu s$) and a link stability specification of $\Delta\tau \leq 5\ fs$, we have $\Delta\tau/\tau|_{PPM} \leq 10^{-3}$ which requires a thermal stabilization of the link as tight as $T - T_{opt} \leq \sqrt{10^{-3}} \cdot T_c \approx 0.06$ °C. Since keeping the temperature stability of $\approx 1\ km$ long link at level of 6 hundredths of °C in 24/7 operation is unrealistic, even a delay compensated cable distribution can be insufficient for most demanding applications. In this cases ***active stabilized distributions*** must be used.

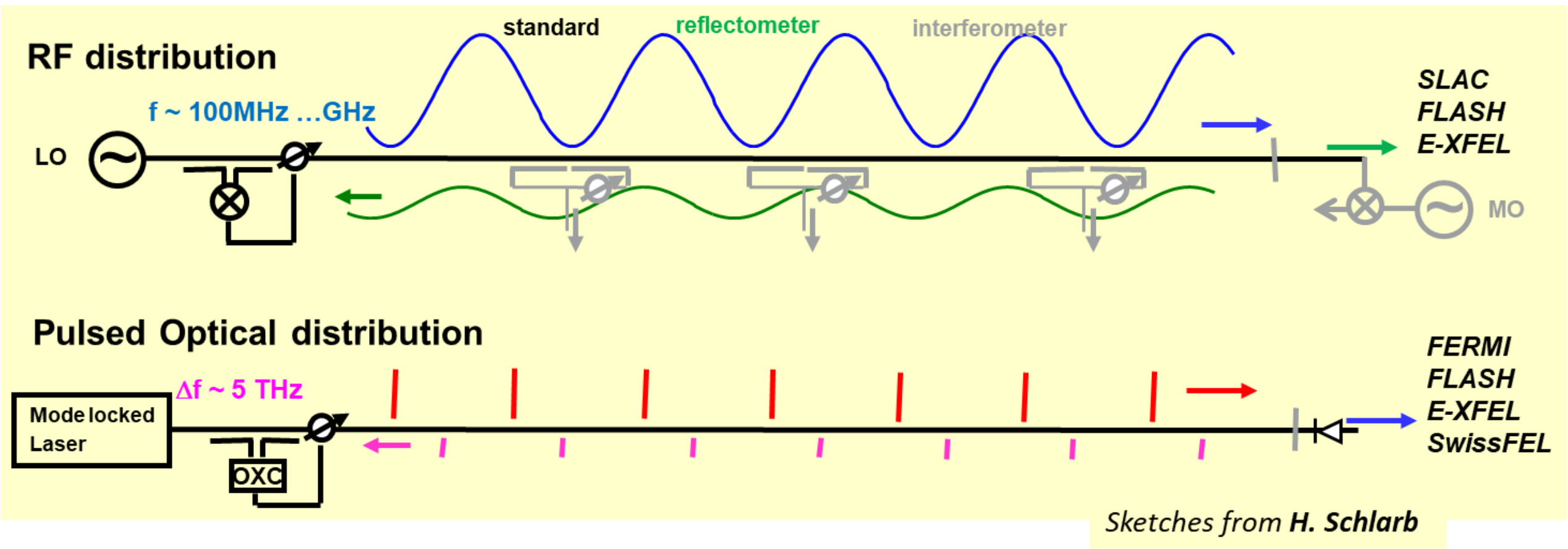

Fig. 40: Actively Stabilized Reference Distribution systems

In an ***active stabilized*** distribution system each link is provided with a ***feedback system*** capable to keep constant its ***round-trip delay***. For RF electrical distribution the link needs to be terminated with a partially reflective pad, so that a reflected wave along the cable can be captured by a directional coupler and phase compared with the forward wave. Any drift caused by an electrical elongation of the link is then detected and corrected by an actuator, in this case a variable phase shifter, in a feedback loop configuration. An RF actively stabilized link is sketched in the upper part of Fig. 40.

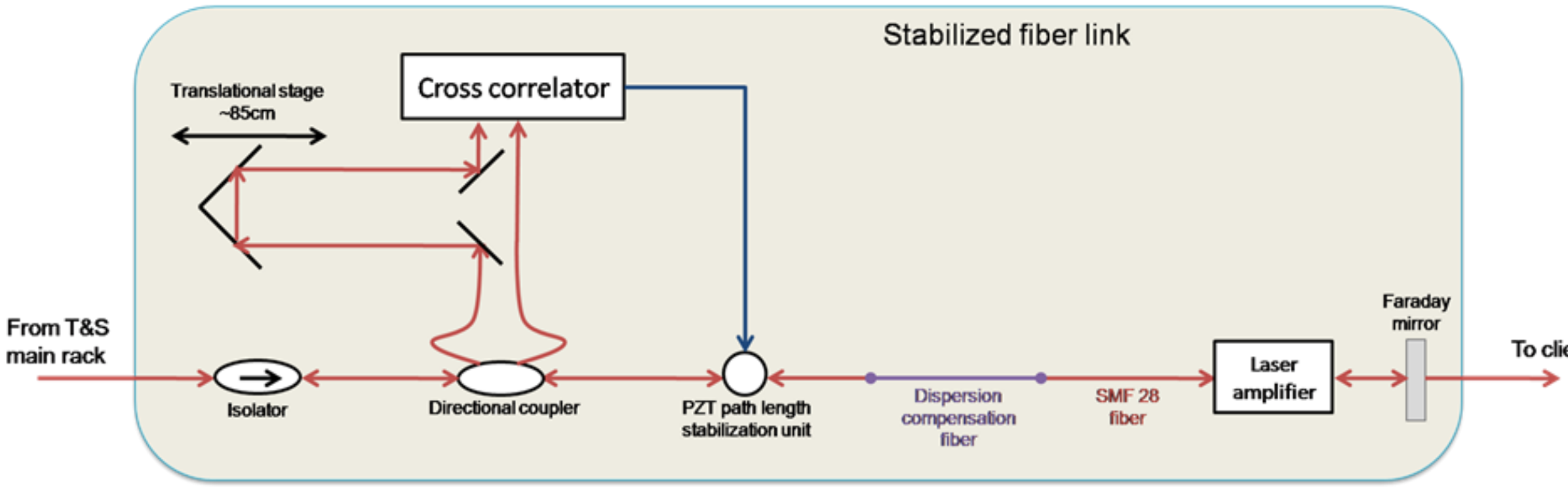

Fig. 41: Actively Stabilized Optical link

In general **RF electrical distributions** are suited for **small/medium size** facilities, while **optical distributions** are used for **medium/large** facilities. In optical distributions the RMO synchronization information is encoded in a mode-locked laser pulse repetition rate[25] or, less used, is imprinted in the amplitude modulation of a CW laser (OMO systems). The OMO synchronization information is then transported to the clients by means of glass fiber links. There are huge benefits associated with optical distribution. First of all short pulse duration ($\approx 100$ fs) of mode locked lasers provides **extremely large bandwidth** well above the THz threshold, which allow using **high sensitivity error detection** methods such as optical cross correlation and, for AM CW laser carriers, interferometry [26]. Another huge advantage is represented by the **low attenuation** of the fiberlinks, which allows transporting the reference easily over distances of many kilometers. However, optical distribution systems present also some disadvantages. They are, by far, more complicated and expensive with respect to RF distributions, and **mandatorily require link active compensation** because of the large thermal sensitivity of the glass fibers, as depicted in the lower part of Fig. 40 and, in more details, in Fig. 41. An optical actively stabilized link is sketched in the lower part of Fig. 40. In this case optical cross-correlators can be implemented to detect any tiny slippage of forward and backward pulse overlapping, and the actuator is realized with a fiber piezo-stretcher or by inserting a free-space laser propagation section equipped with piezo-controlled mirrors to adjust the path length.

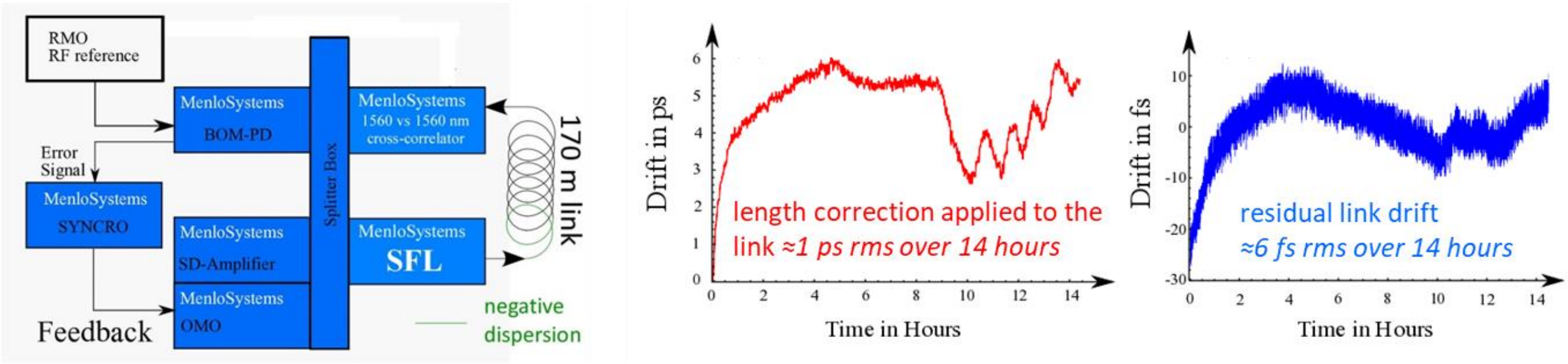

Fig. 42: Schematics and results of a factory stability test on a stabilized optical link ( MenloSystems GmbH)

Moreover, the links require dispersion compensation otherwise the large spectral content of the optical pulses would cause a substantial growth of the pulse duration well above the ps threshold, spoiling the time resolution characteristics of the reference signal emerging from the link. Dispersion compensation is obtained by adding a piece of special negative dispersion fiber whose length is precisely cut to match the total link length. This allows keeping the final pulse length close to the pulse transform limit ($\approx 100$ fs or shorter). A sketch of a 14 hours long time factory test on the stability of an optical stabilized link, together with the obtained results, is shown in Fig. 42. The pulses emerging from a $\approx 170$ m stabilized and dispersion compensated link are compared out-of-the-loop with a copy of the OMO pulses by using a 1560 vs. 1560 nm cross correlator. The result (blue curve) in an rms fluctuation of $\approx 6$ fs, obtaind by applying an rms correction of $\approx 1$ ps corresponding to the natural uncorrected link drift over the measurement time duration and under the test environmental conditions.

### 3.3 Arrival time diagnostics - Bunch Arrival Monitors

Beam arrival time diagnostics is a crucial topic to ultimate check the performances of the synchronization system of a particle accelerator based facility, and to measure residual errors to be corrected by automated feedback processes. Various kinds of bunch arrival monitors are available, based on different physics processes, with different characteristics in terms of time resolution, refresh rate, beam perturbation, and possibility to provide additional beam diagnostics. They are placed at some target positions along the accelerator vacuum chamber, to measure the arrival time of the particles (or sometimes only of the bunch longitudinal distribution centroids) at that specific locations. As any physical time measurement, particle **arrival times** are measured with **respect to some reference clock**. Depending on type and nature of the monitor the reference clock can be **either the facility master clock** (RMO or OMO) **or one of the client oscillators** of the facility synchronization system, such as the RF signal in a particular power plant or the optical pulse train taken from one of the mode-locked laser cavities included in the facility. In this respect, the information can be post-processed or manipulated to be referred also to clocks different from that physically used in the device. Criteria and formulae relating beam arrival time to different clocks will be given in the final section. The most used devices and techniques to measure the beam arrival time are described in this paragraph.

### *3.3.1 RF deflectors:*

For some special applications RF fields are used ***to deflect*** a charged beam ***more than to accelerate*** it. Structures called RF deflectors are designed for this task, mostly based on circular waveguide dipole modes $TM_{1m}$ and $TE_{1m}$, i.e. modes showing an azimuthal periodicity of order 1 properly iris-loaded (for $TW$ deflectors) or short-circuited (for $SW$ deflecting cavities). The figure of merit qualifying the efficiency of a $SW$ RF deflecting structure of length $L$ is the transverse shunt impedance $R_{\perp}$ defined as:

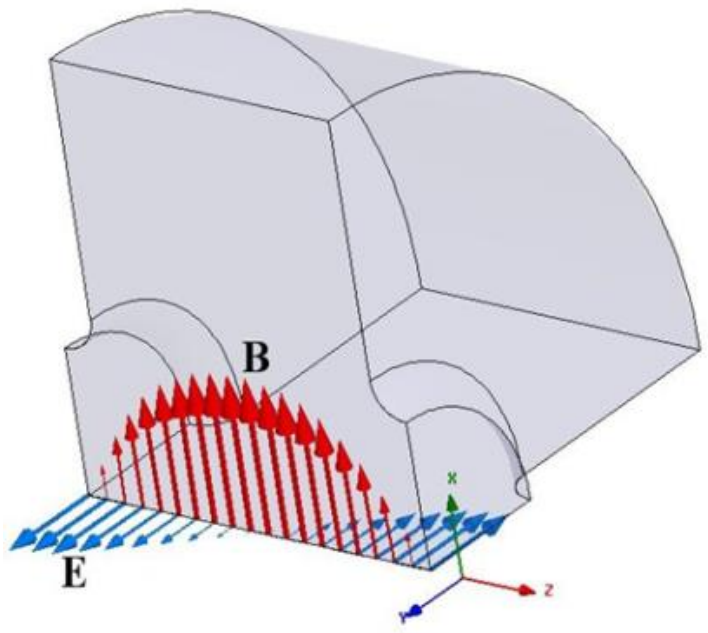

Fig. 43: RF deflecting fields in a deflector

$$R_{\perp} = \frac{V_{\perp}^2}{2P} \qquad with \quad V_{\perp} = \left| \int_{-L/2}^{L/2} \left[ E_y(z) + \beta c \, B_x(z) \right] e^{j\omega z/(\beta c)} dz \right| = \frac{\beta c}{q} \Delta p_{\perp} \tag{40}$$

where a deflection in the $y$-direction for a charge $q$ moving along the $z$-direction with a velocity $\beta c$ has been considered, $\omega$ is the angular frequency of the RF field, $P$ is the RF power dissipated in the structure and $\Delta p_{\perp}$ is the variation of the particle transverse momentum caused by the transverse RF fields in the deflector. The variation of charge propagation direction with respect to the nominal longitudinal accelerator trajectory $\Delta\phi_{def}$ is therefore:

$$\Delta\phi_{def} = \frac{\Delta p_{\perp}}{p} = \frac{qV_{\perp}}{\beta^2 W} \tag{41}$$

where $p$ is the particle total momentum and $W$ is the particle energy.

There are many different applications requiring RF deflecting structures. They are used, for instance, as crab crossing cavities to obtain head-on bunch collisions in colliders where beam trajectories cross with an angle, or as RF injection kickers to create closed orbit bumps rapidly varying with time in rings requiring complex stacking configuration, or as RF separators in ion sources to separate and collect different ion species. In our context RF deflectors are **beam diagnostics devices**, for intra-bunch tomography of the longitudinal phase space [27][28]. Some RF deflecting structures used for different tasks are shown in Fig. 44.

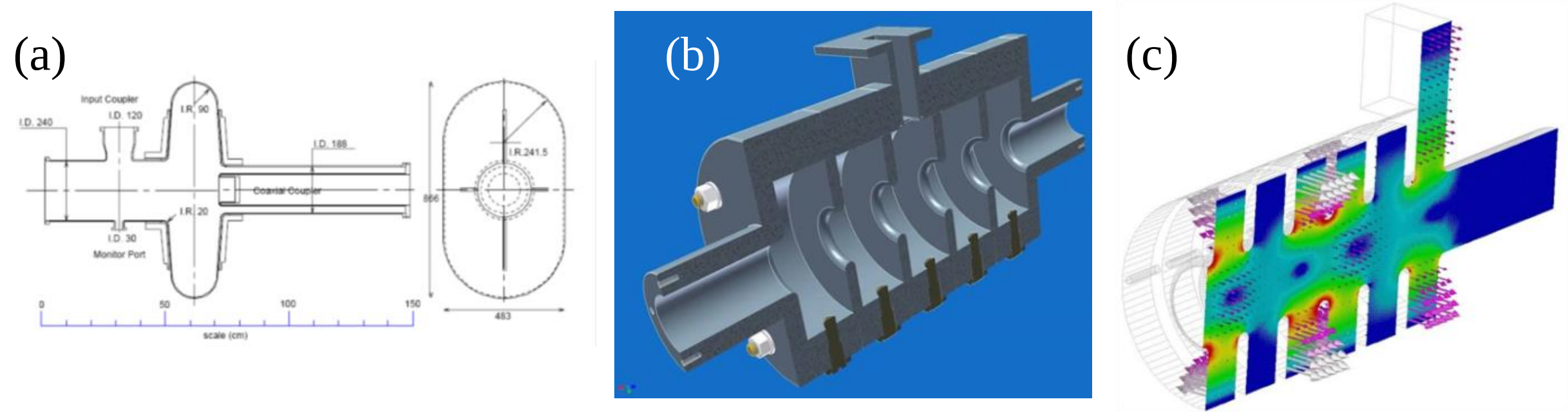

Fig. 44: RF deflecting structures: KEK-B SC crab cavity (a), 5-cells SW RF Deflector for beam diagnostics at INFN Frascati SPARC_Lab (b), TW RF Deflector for CTF3 combiner ring at CERN (c).

RF deflectors are used for beam longitudinal phase space diagnostics by simply streaking the bunch on a fluorescent screen placed at a distance L applying a time dependent transverse kick to establish a correlation between the arrival time $t_p$ of a particle at the deflector and its final transverse position $y_s$ on the screen. For bunch much shorter than the RF wavelength injected near the zero-crossing of the deflecting field integral the correlation is pretty linear. The beam image on the screen is captured by a camera so that longitudinal charge distribution and centroid longitudinal position can be measured.

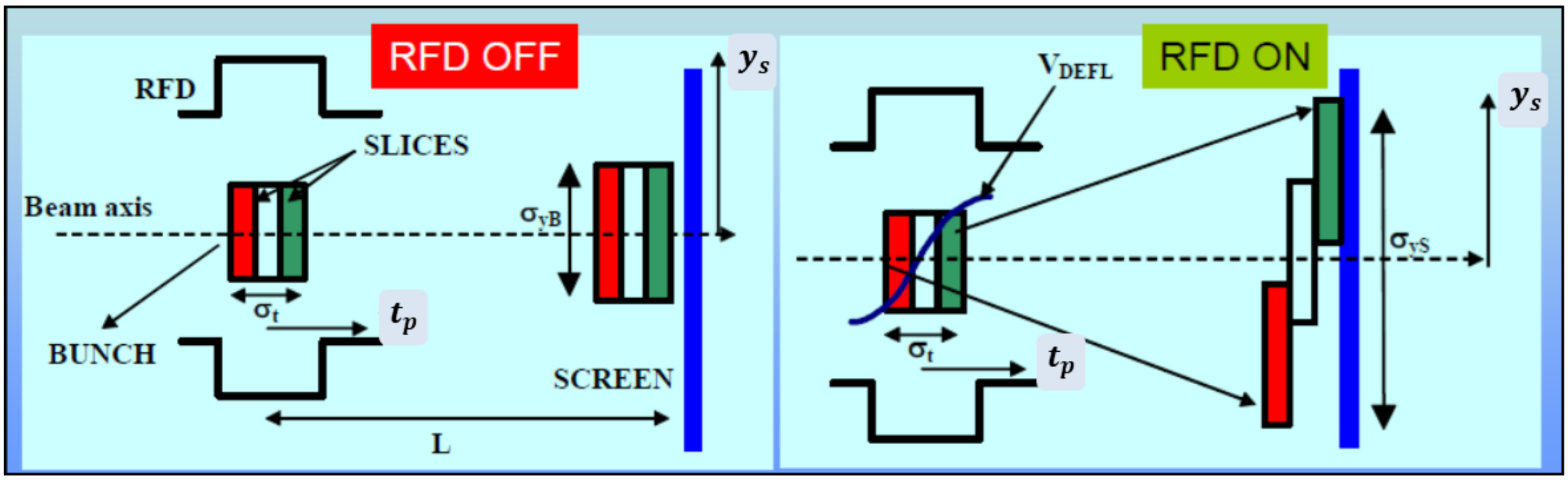

Fig. 45: RF deflector for beam longitudinal phase space diagnostics

Assuming a free-space beam propagation between the deflector and the screen, elementary cinematics gives the final position on the screen $y_s$ of a relativistic particle entering the deflector at time $t$ with transverse coordinates $(y_{def}, y'_{def})$:

$$y_s = \frac{V_\perp L}{W/q} sin[\omega_{RF}(t - t_{RF})] + y'_{def}L + y_{def} \approx K_\perp(t - t_{RF}) + y'_{def}L + y_{def}$$

$$with \quad K_\perp \stackrel{\text{def}}{=} \frac{V_\perp \omega_{RF} L}{W/q} \tag{42}$$

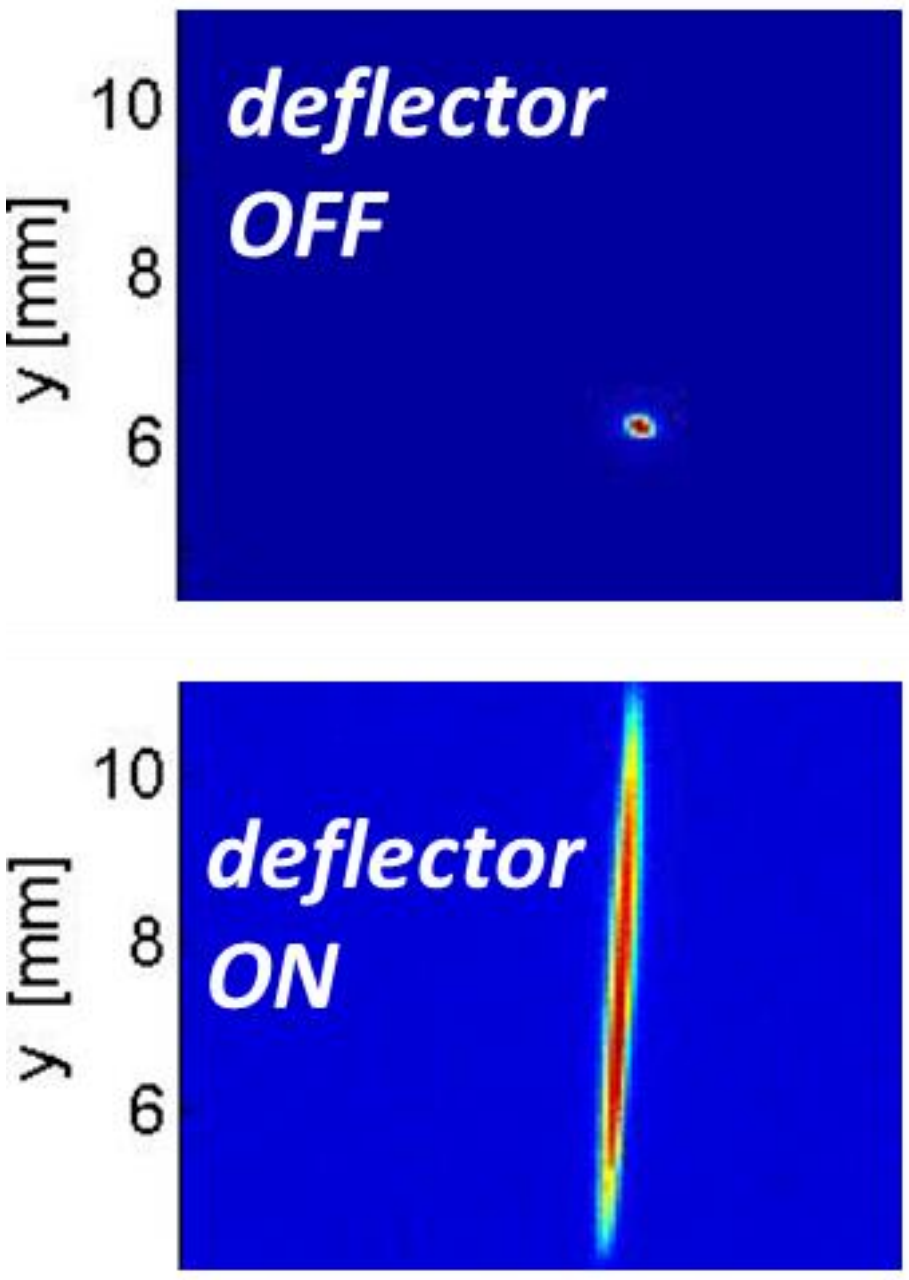

Fig. 46: Beam spots at screen

The **time-resolution** $\tau_{res}$ provided by this set up is defined as the **minimum arrival time deviation** for a particle or a distribution centroid corresponding to a **transverse displacement** on the screen equal to the **natural beam spot-size** $\sigma_{y_{so}}$, i.e. the vertical width of the beam spot on the screen observed in absence of RF deflecting fields (RF deflector "off" that means $V_\perp, K_\perp = 0$ ).

To compute $\sigma_{y_{so}}$ let's assume $K_\perp = 0$ in Eq. (42). The variance $\sigma^2_{y_{so}}$ is obtained by averaging over the bunch particle population the squared vertical displacement at the screen $y^2_{s0}$. We get:

$$\sigma^2_{y_{so}} = \langle y^2_{s0} \rangle = \langle y^2_{def} \rangle + 2L\langle y_{def}\, y'_{def} \rangle + L^2 \langle (y'_{def})^2 \rangle \qquad (43)$$

Clearly, the natural spot size at the screen depends on the characteristics of the particle distribution at the deflector. According to the Twiss linear optics theory of particle beams we have:

$$\langle y^2_{def} \rangle = \beta_\perp^{defl} \varepsilon_\perp; \quad \langle y_{def}\, y'_{def} \rangle = -\alpha_\perp^{defl} \varepsilon_\perp; \quad \left\langle \left( y'_{def} \right)^2 \right\rangle = \gamma_\perp^{defl} \varepsilon_\perp \qquad (44)$$

where $\varepsilon_\perp$ is the beam transverse emittance, and $\beta_\perp^{defl}$, $\alpha_\perp^{defl}$, $\gamma_\perp^{defl}$ are the values of the Twiss functions at the deflector position. Taking into account the expression mutually relating the Twiss functions $\gamma = (1+\alpha^2)/\beta$ Eq. (43) becomes:

$$\sigma^2_{y_{so}} = \frac{\varepsilon_\perp}{\beta_\perp^{defl}} L^2 \left[ 1 + \left( \frac{\beta_\perp^{defl}}{L} - \alpha_\perp^{defl} \right)^2 \right] \qquad (45)$$

The time-resolution provided by the deflector $\tau_{res}$ can be estimated dividing the width of the natural spot $\sigma_{y_{so}}$ by the deflecting constant $K_\perp$ obtaining:

$$\tau_{res} = \frac{\sigma_{y_{so}}}{K_\perp} = \frac{W/q}{\omega_{RF} V_\perp} \sqrt{\frac{\varepsilon_\perp}{\beta_\perp^{defl}}} \sqrt{1 + \left( \frac{\beta_\perp^{defl}}{L} - \alpha_\perp^{defl} \right)^2} \geq \frac{W/q}{\omega_{RF} V_\perp} \sqrt{\frac{\varepsilon_\perp}{\beta_\perp^{defl}}} \qquad (46)$$

According to Eq. (46) the finest time resolution is obtained under the optimal condition $\beta_\perp^{defl} = \alpha_\perp^{defl} \cdot L$ which minimizes the term under the square root. Therefore the right-hand side of Eq. (46) represents the best time resolution provided by a measurement based on an RF deflector. Typical numbers for an electron linac such as $W/q = 500\, MeV$ ($\gamma = 1000$at the deflector position), $f_{RF} = 3\, GHz$, $V_\perp = 10\, MV$, $\varepsilon_\perp = 10^{-6}/\gamma = 10^{-9}\, m$, $\beta_\perp^{defl} = 10\, m$ already gives $\tau_{res} \approx 25\, fs$. Resolutions down to $< 10\, fs$ can be reached by pushing further the parameters. The $\tau_{res}$ value estimates the longitudinal extension of the intra-bunch beamlet (or slice) which can be experimentally resolved. It also gives an estimation of the shot-to-shot bunch centroid arrival time measurement resolution, provided that the bunch total time duration is not orders of magnitude longer.

A typical shot-to-shot arrival time measurement performed at the SPARC_LAB facility of the INFN Frascati Labs is shown in Fig. 47.

As a summary, beam diagnostics based on **RF deflector is intercepting** since a target screen has to be inserted on the beam trajectory. It works typically on **single bunch** or at low repetition rates. Bunch trains can be possibly resolved by using fast gated cameras and short persistence fluorescent screens. Particle arrival time is measured with **respect to the phase of the RF fields in the deflector**, which is in general affected by the phase noise present in the associated RF power plant. If used in combination with a spectrometer, the deflector allows for longitudinal phase space imaging by mapping the particle longitudinal coordinates $(z, \epsilon)$ into the screen transverse coordinates $(y, x)$.

Even though the fields along the deflector beam axis are purely transverse, according to Panofsky-Wenzel theorem particles travelling off-axis experience accelerating E-field proportional to their transverse displacement. RF deflectors therefore introduce an **energy spread** correlated with the beam **transverse dimension**. The induced relative energy spread is also proportional to the peak deflecting voltage $V_{\perp}$.

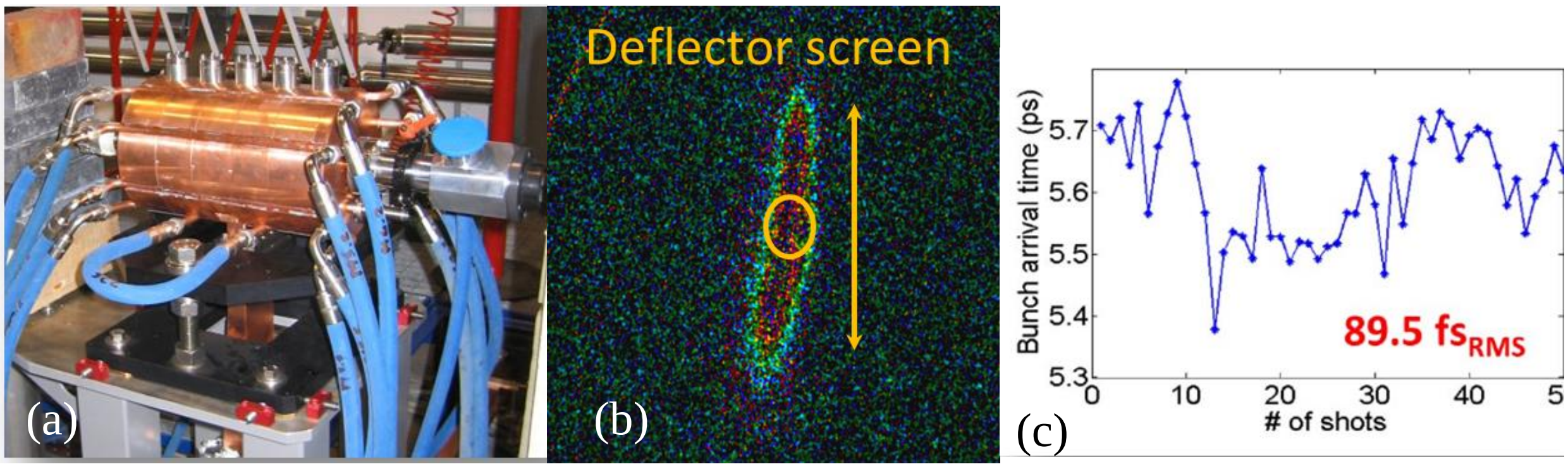

Fig. 47: Bunch Arrival Time measurements performed at INFN SPARC_LAB using the RF deflectorDeflector (a), streaked beam screen image (b), shot-to-shot arrival time measurements and statistics (c)

### *3.3.2 Electro-Optical Bunch Arrival Monitor:*

In superconducting (SC) linacs the beam temporal structure is more complex than in normal conducting (NC) ones. RF and beam pulses are much longer, typically in the ms scale, and allow hosting many bunches arranged in trains. The number of bunches in the train, their charge and time spacing depend on the applications but in principle can be arranged in a quite flexible way, and in general the carried average currents are much larger compared to NC linacs capability. Monitoring the **arrival time of each bunch in a train** is very important, especially for applications requiring tight control of the synchronization of the beam with other external system, such as pump-probe FEL experiments. The electro-optical Bunch Arrival Monitor is a device developed at the FLASH facility (DESY Hamburg) to measure the arrival time of each electron bunches in a long train with a **non-intercepting technique** [29][30]. The working principle of the device exploits the large slew rate, of the order of $1\ \mathrm{V/ps}$, of the voltage delivered by a button beam position Monitor when excited by the passage of a short bunch, as shown in Fig. 48. The BPM signal is used to amplitude modulate a reference laser train (a copy of the facility OMO) by means of a **Mach-Zehnder Electro-Optical modulator (EOM)**. The EOM is simply a Mach-Zehnder interferometer where one of the two arms includes an optical phase shifter driven by an external voltage [31]. If no voltage is applied, the two arms are in-phase and the intensity of the emerging radiation is maximum, while it decreases when the two paths are de-phased down to a minimum which is ideally zero when the relative phase reaches 180°. What is peculiar of this device is its **large bandwidth** which can overall exceed $20\ GHz$ from the input modulation port (electrical signal) to the AM of the emerging light. This characteristic is exploited to reach the required **time resolution** of the arrival time monitor.

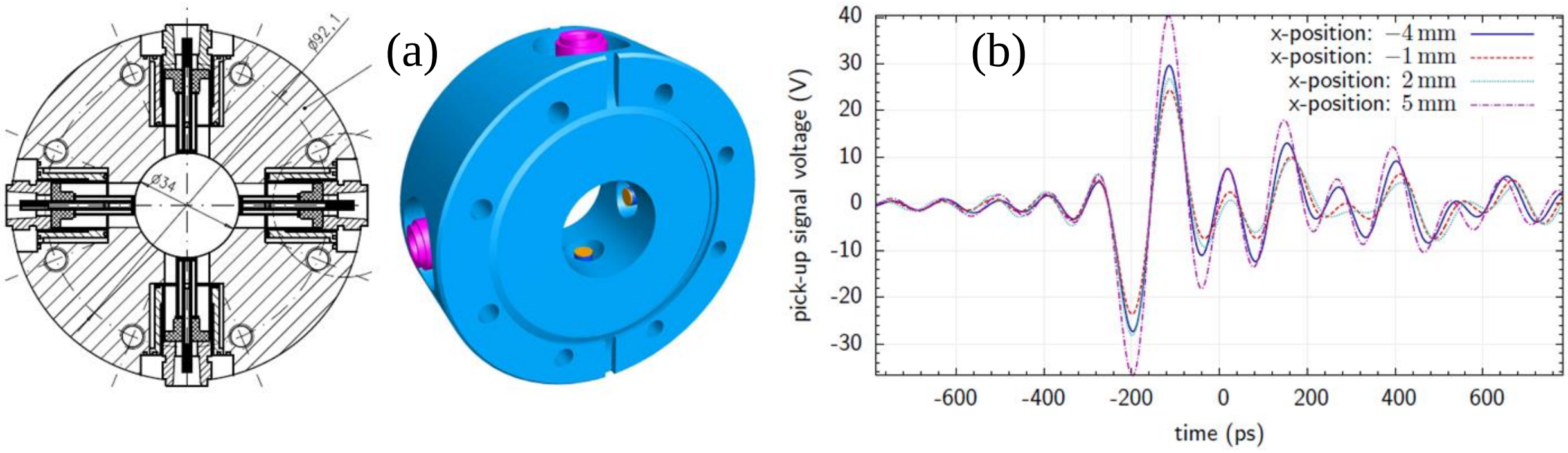

Fig. 48: Bipolar signal (b) delivered by a button BPM (a) at the passage of a short bunch

The bunch arrival time information is encoded in the time position of the zero-crossing of the fastest front of the BPM delivered signal shown in Fig 48 b).

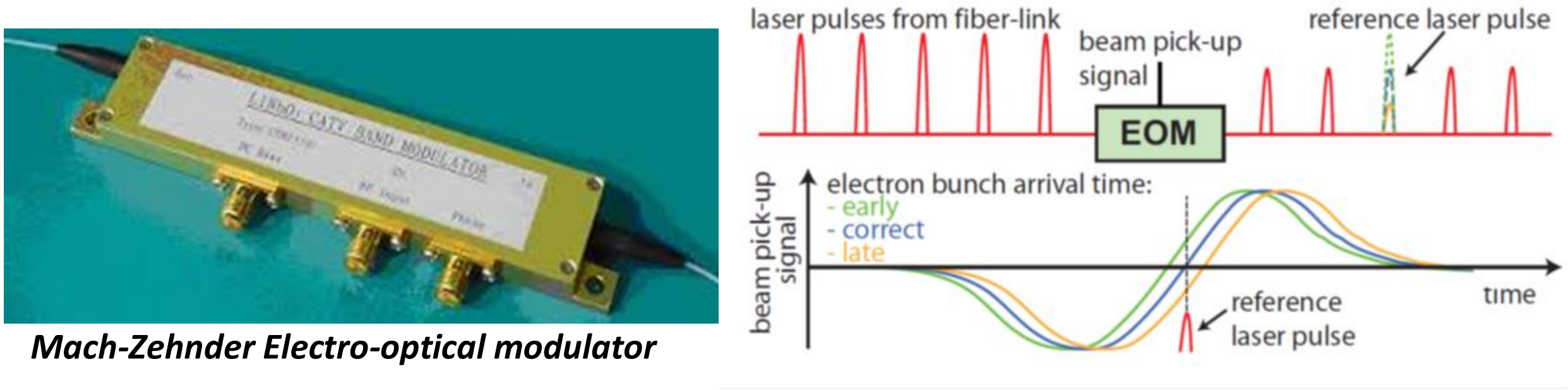

Fig. 49: Reference laser pulses modulated by a Mach-Zender EOM driven by the BPM beam induced signal

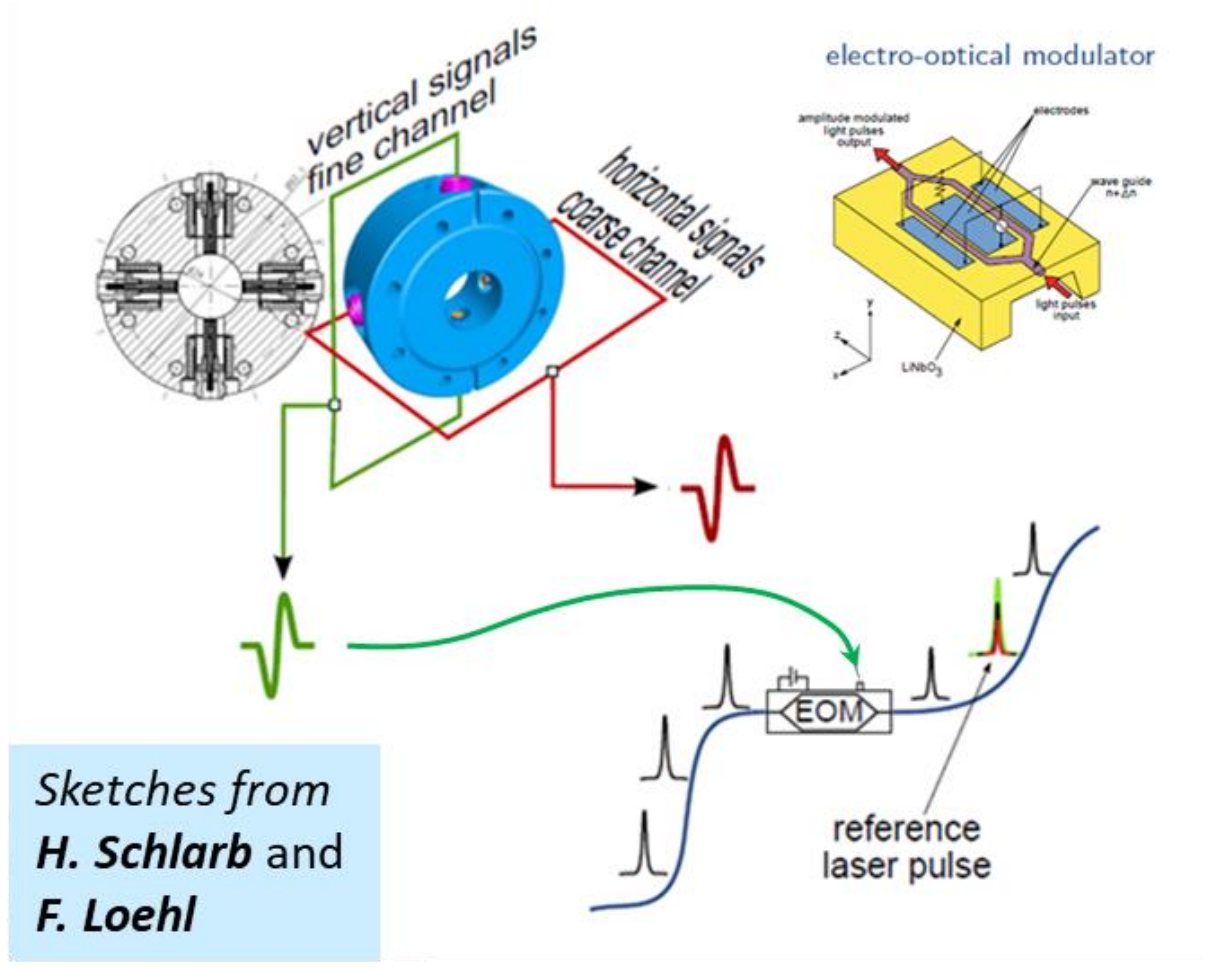

Fig. 50: Electro-Optical Bunch Arrival Monitor sketch

The EOM optical input is fed with a laser pulse train derived from the facility master oscillator which is aligned approximately with the **zero-crossing of the BPM beam induced signals**. The electrical input of the EOM is biased by a proper dc voltage and connected with the BPM signals. The bunch repetition rate is a in general a fraction (typically of the order of $1/10 \div 1/100$) of the OMO pulse repetition rate, so that most of the OMO pulses do not interact with the beam induced signal and emerge from the EOM with a constant attenuation depending on the dc bias. The OMO pulse interacting with the BPM signal will show the same attenuation only when perfect aligned with the signal zero crossing.

Bunches arriving **early or late** will add a non-zero voltage (positive or negative, respectively) causing an **intensity variation** of the interacting OMO pulse. The bunch arrival time fluctuations are therefore converted in amplitude modulation of the reference laser pulses that can be measured and analyzed by using proper photodiodes and a dedicated acquisition system.

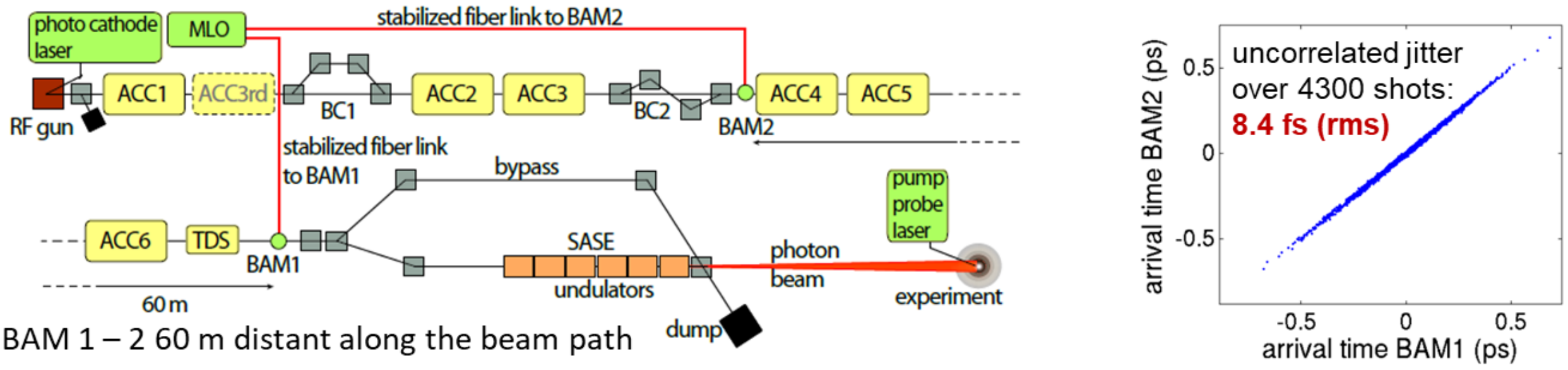

Fig. 51: Layout and results of a resolution test performed with 2 BAMs placed 60 m apart

As a summary, the Electro-Optical Bunch Arrival Monitor is perfectly suited to resolve **bunch trains with narrow spacing**. It is **not-interceptive**, which means that it can run continuously during operation providing a real-time diagnostics, and refers the bunch arrival time **directly to the Optical Master Oscillator**, which is in principle the best clock available in the facility. The experimentally demonstrated resolution is $< 10$ fs, as shown in Fig. 51 where the measurements of two BAMs placed 60 m apart along the beam path have been acquired and compared, showing a high correlation degree with a residual uncorrelated jitter of only 8.4 fs rms. Contrarily to the RF deflector and Electro Optical Sampling based monitors, electro-optical BAMs do not provide additional information on the bunch longitudinal phase space other than the centroid arrival time.

### *3.3.3 Electro Optical Sampling:*

***Electro-optical sampling*** *(EOS) is a beam diagnostics technique providing information on bunch* ***longitudinal charge distribution*** *and* ***arrival time*** [32] [33].

The technique is based on the property of some optically active crystals (such as Zinc Telluride ZnTe, Gallium Phosphide GaP, Gallium Selenide GaSe) to become **birefringent** when exposed to an **intense electric field** (in the MV/m range).

The birefringence is a physical effect proper of anisotropic crystals showing different refraction indexes for different polarization (i.e. different orientations of the electric field vector) of an electromagnetic wave propagating through them.

As a consequence, a linearly polarized electro-magnetic wave (possibly confined in a laser pulse) impinging on a birefringent

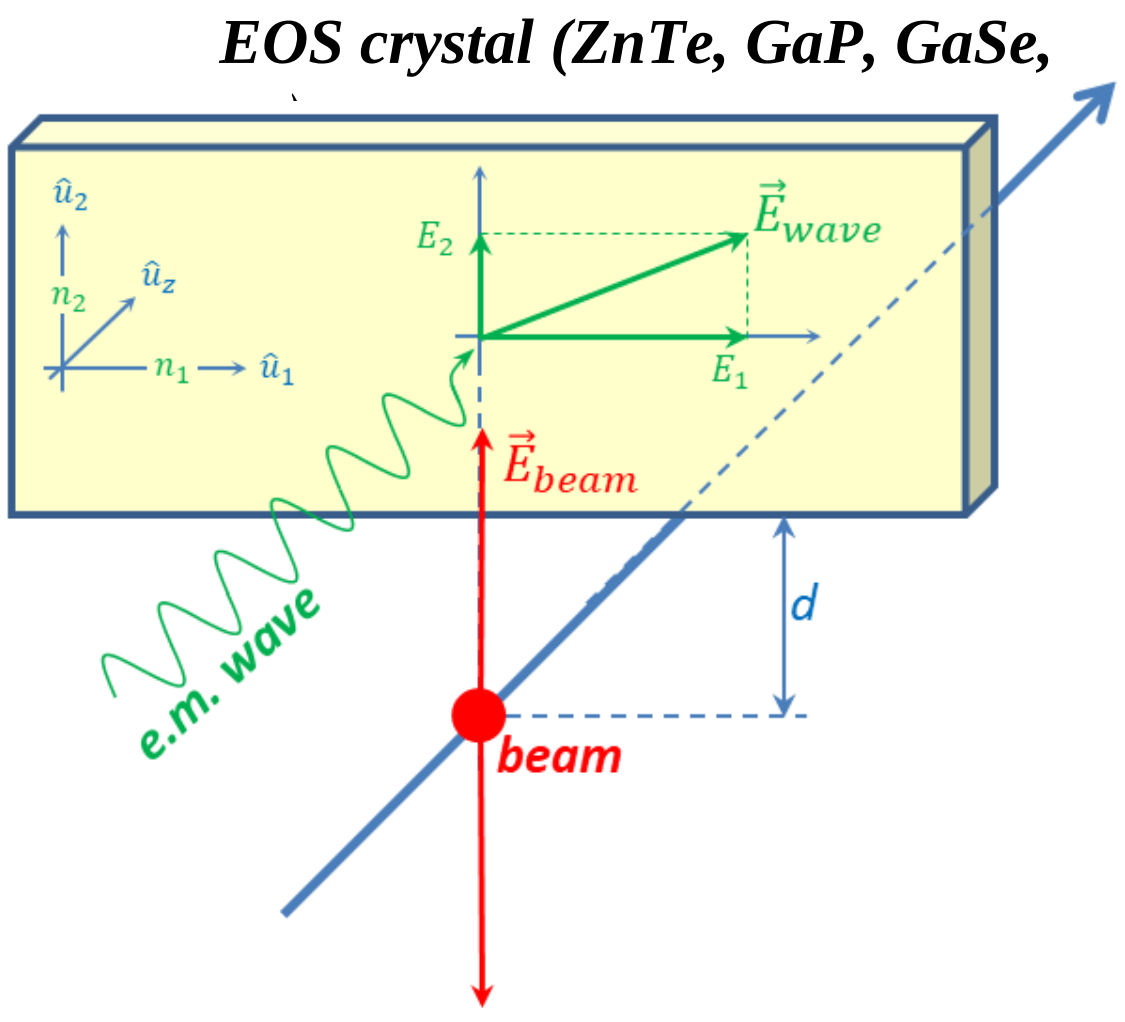

Fig. 52: Birefringence induced on an optically active crystal by the bunch electric field

crystal will gain elliptical polarization while propagating along it, as sketched in Fig. 52, according to:

$$\vec{E}_{wave}(z,t) = E_1 \cos[\omega(t - n_1\, z/c)]\, \hat{u}_1 + E_2 \cos[\omega(t - n_2\, z/c)]\, \hat{u}_2 \qquad (47)$$

To exploit the induced birefringence for beam diagnostics purposes an **optically active crystal** is placed in the vicinity of the beam trajectory such that birefringence effects will show up only **during the bunch passage** because of the Lorentz contracted Coulombian electric field travelling with the bunch.

A basic set-up of an EOS based beam diagnostics station is sketched in Fig. 53. A laser pulse synchronous with the beam illuminates an optically active crystal placed in the vicinity ($d < 1$ mm) of the beam trajectory. A polarizer $P$ and an analyzer $A$ orthogonally oriented (but not aligned with the crystal principal axes $\hat{u}_1, \hat{u}_2$) are placed upstream and downstream the crystal. In **absence of beam** there is no induced birefringence, the laser polarization remains unchanged while travelling through the crystal and **no radiation can propagate** beyond the analyzer $A$. On the contrary, in presence of a beam the **laser intensity downstream** the transport line is modulated by the intensity of the **bunch transverse electric field**.

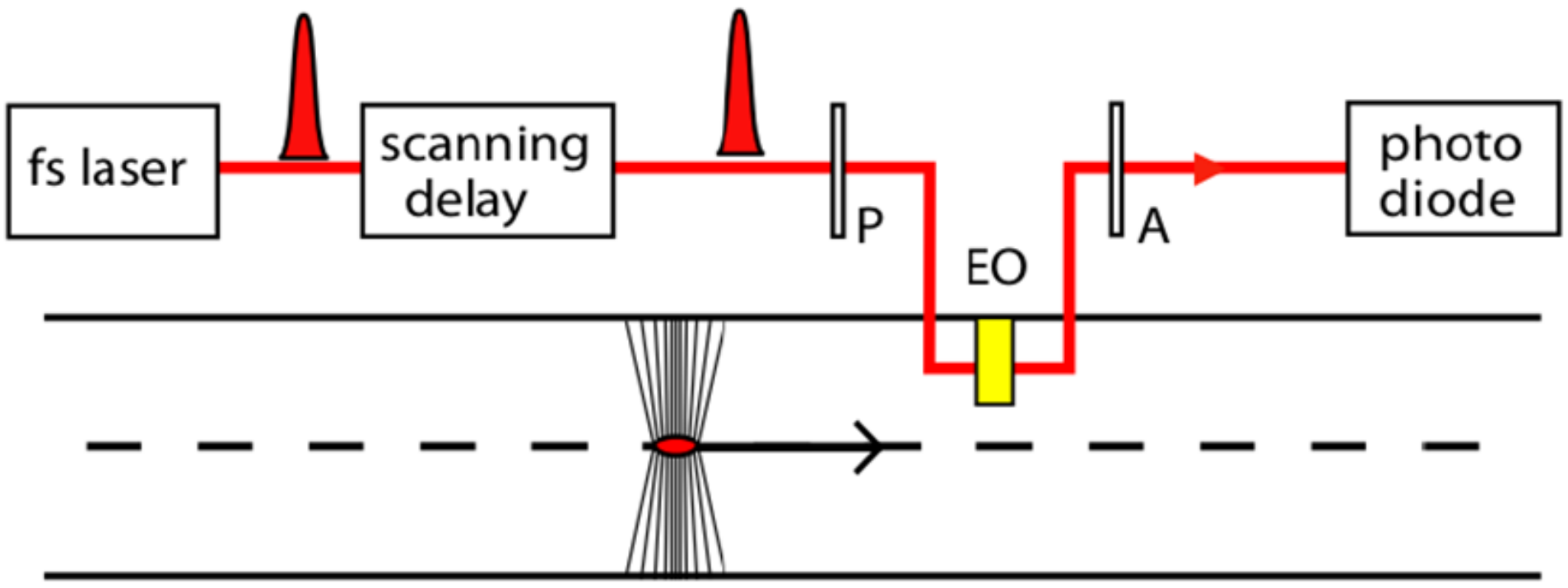

Fig. 53: EOS beam diagnostics basic set-up

In this configuration the **EOS** set-up imprints an **intensity modulation** on a laser pulse **proportional** to the instantaneous bunch current, i.e. proportional to the **bunch charge longitudinal distribution**. For short bunches (in the sub picosecond regime) the imprinted modulation is too fast to be directly time-resolved in single shot. The bandwidth of the laser envelope extends beyond the THz region, so that no electronics devices can be used.

This kind of EOS set-up can only be used in combination with laser pulses much shorter than bunch duration, with a scanning delay line to sample portions of the bunch distribution in a shot-by-shot acquisition regime. Clearly this approach is barely usable for bunch arrival time monitoring since it mixes-up bunch arrival time and charge fluctuations. However, **single shot EOS diagnostics** is possible with different set-ups that convert the **time dependency** of the laser intensity into a **displacement** of the laser beam image footprint on a screen that can be captured by a CCD camera.

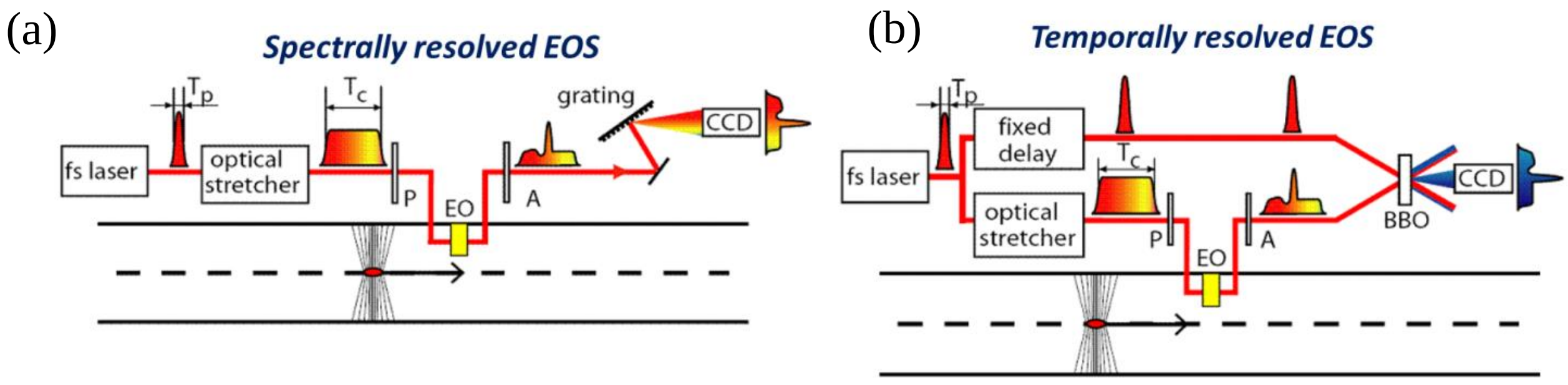

Fig. 54: Spectrally (a) and temporally (b) resolved EOS set-up for single shot beam diagnostics

Two different EOS set-ups allowing single shot beam diagnostics are shown in Fig. 54. In the **spectrally resolved EOS** the laser pulse is stretched and linearly chirped before interaction to produce a time-wavelength correlation. Different bunch beamlets modulate the amplitude of different carrier wavelengths, that can be finally resolved by means of a dispersive grating. The main limitation of this technique comes from the **frequency mixing** between the THz field of the particle beam and the carrier of the optical waves which **broadens** the instantaneous **carrier spectrum**, limiting the resolution to $\approx 200$ fs.

In the **temporally resolved EOS** the stretched and intensity modulated EOS pulse is scanned by a copy of the unstretched pulse by performing a large-angle cross-correlation on a non-linear BBO crystal. Because of the crossing angle and the transverse dimensions of the two beams, the EOS signal beamlets interact at different transverse positions of the crystal corresponding to a different transverse position of the cross-correlation radiation source. This technique provides a quite **better resolution** of $\approx 50$ fs but it requires a substantially **larger laser intensity** which is necessary to drive the cross-correlation process.

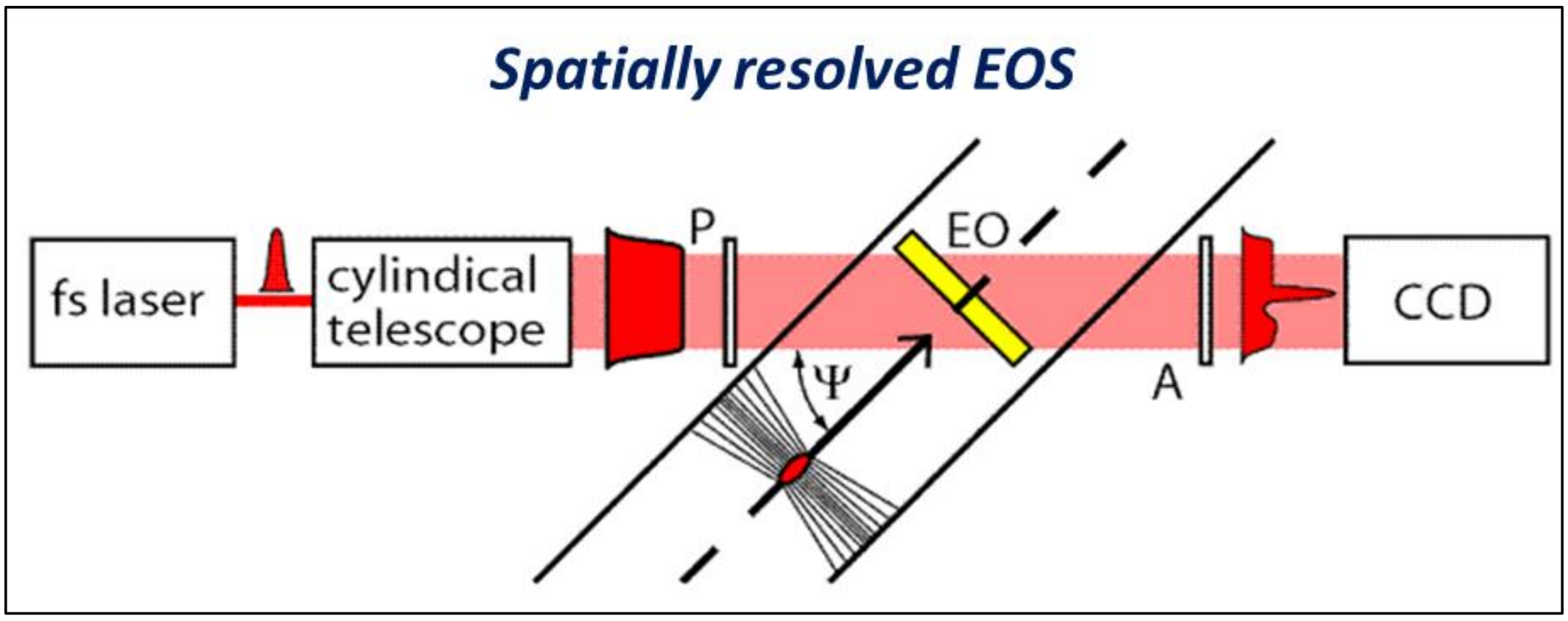

Fig. 55: Spatially resolved EOS set-up for single shot beam diagnostics

Another EOS set-up allowing single shot beam diagnostics is sketched in Fig. 55, the so-called **spatially resolved EOS**. In this configuration the laser pulse is **transversally stretched** and impinges the EOS crystal at a large angle. **One side** of the crystal is reached earlier by the laser wavefront and **samples the bunch head field**, while **the opposite side** is reached later and probes the **bunch tail field**. The image of the laser intensity profile on the screen captured by the CCD camera is directly correlated to the charge longitudinal distribution.

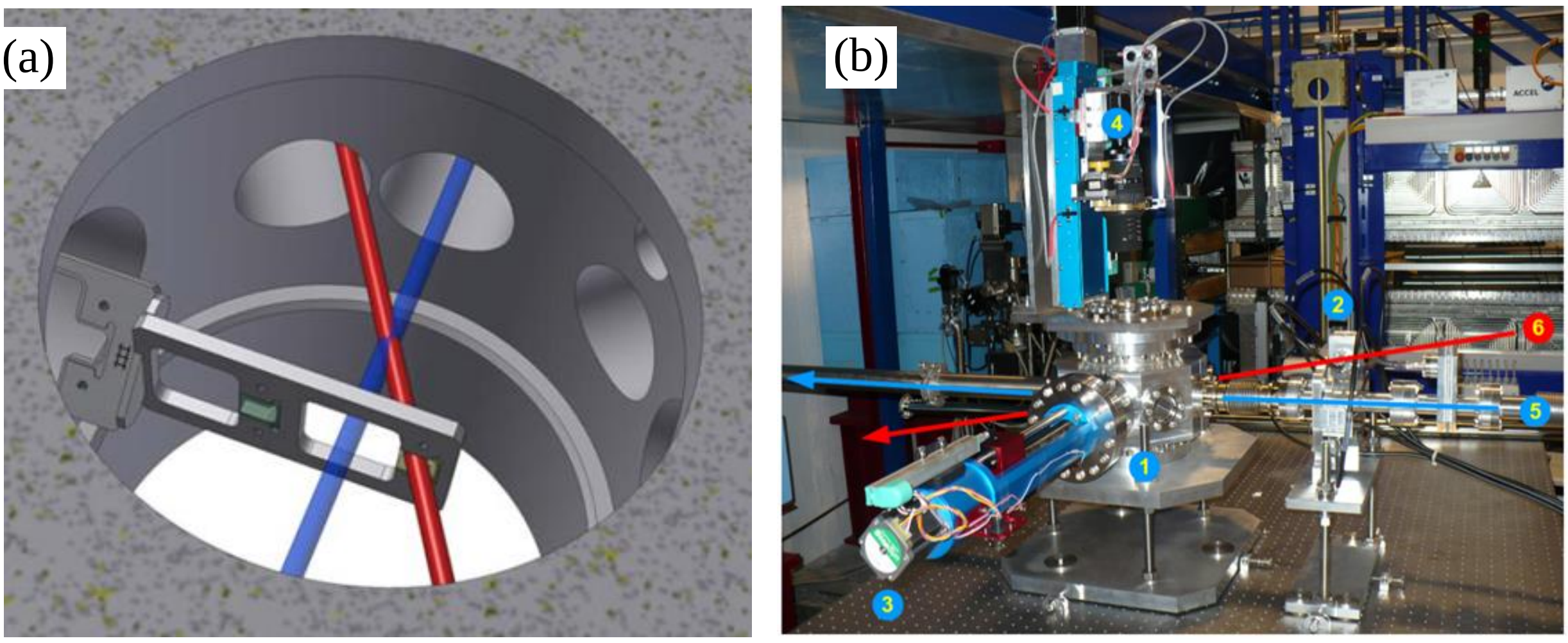

Fig. 56: Spatially resolved EOS model (a) and experimental set-up (b) at SPARC_LAB (INFN Frascati)

**Spatially resolved** configuration is the simplest **EOS** experimental set-up providing single-shot beam diagnostics. Bunch length measurement resolution is $\geq 50$ fs, limited by the **dispersion** of the optically active material that tends to **enlarge the duration** of the **beam THz pulse** while travelling across the crystal. This **only partially affects** the bunch arrival time $\mathrm{T_{arr}}$ measurement, i.e. the shot-to-shot variation of the position of the distribution centroid. In this case the estimated resolution can be reduced down to $\sigma_{\mathrm{T_{arr}}} \approx 10$ fs.

A $3D$ model of a spatially resolved EOS crystal with movable support is shown in Fig. 56 a, while a picture of set-up installed at the INFN Frascati SPARC_LAB facility is shown in Fig. 56 b.

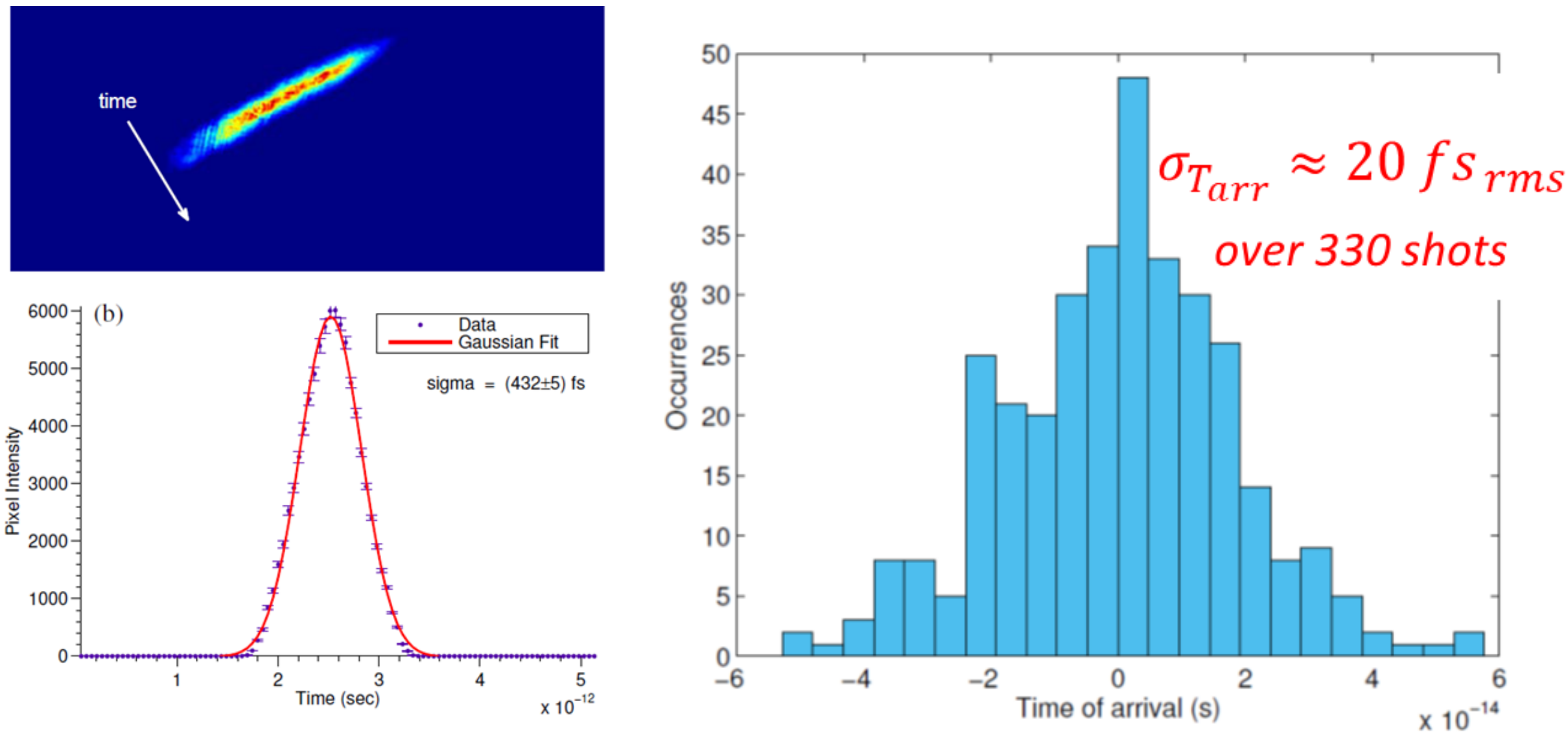

Fig. 57: EOS based arrival time jitter measurements and statistics performed at SPARC_LAB (INFN Frascati)

The best experimental result obtained by using this set-up is reported in Fig. 57, where a jitter of the bunch arrival time of $\approx 20$ fs has been measured [34]with respect to the **injector photo-cathode laser** system that has been split and guided to the EOS station on the crystal as **timing reference**.

As summary, Electro-Optical Sampling is a **not-interceptive diagnostics** technique that can be used for **single shot measurements** provided that the THz intensity modulation imprinted in the sampling laser profile is converted into a displacement of the beam image on a screen by proper set-ups. Similarly to RF deflector based diagnostics it works mainly at **low repetition rates**, while bunch trains need fast gated cameras and short persistence fluorescent screens to be resolved, and provides a direct measurement of the **longitudinal distribution** of the bunch charge. Information on the **bunch centroid arrival time** is extracted by averaging the longitudinal distribution of each acquired shot. **Arrival times** are referred to the **laser system** used to **drive the EOS** diagnostic station, which could be either the Optical Master Oscillator or a sample of any other laser system available in the facility.

# 4 Beam Synchronization

In this section a **linear model** to compute the expected fluctuations of the beam arrival time as consequence of the facility sub-systems residual synchronization errors will be introduced and discussed. As an introductory example, the case of a **magnetic compressor** will be presented, to study the combined effects of errors in the **bunch injection time** and **RF amplitude and phase noise** on the bunch arrival time downstream a magnetic chicane. The **results will be generalized** to describe the dependence of the beam timing on the other accelerator sub-systems for generic machines and selected working-points. In conclusion some practical example will be illustrated.

## 4.1 Bunch magnetic compressor

**Magnetic compressors** are widely used in electron linac injectors to **reduce the bunch length**[5]. The working principle described in Fig. 58 is very well-known and elementary. The bunch coming from a source is accelerated off-crest in a portion of the linac. If the bunch is placed on the positive slope of the accelerating field, the tail (which arrives at larger t) gains more energy than the bunch head. A time-correlated energy chirp is therefore imprinted on the bunch which is then injected in a magnetic chicane where low energy particles travels longer paths with respect to high energy ones. This causes a bunch compression which might be **more or less pronounced** depending on the **balance** between the bunch **energy chirp** and the **path elongation** with energy provided by the chicane. For an **optimal balance** between the two factors the bunch will be **totally compressed**, which means that the tail will exactly catch the head. Imperfect balance will result in under-compressed (tail always behind the head) or even over-compressed (tail beyond the head) bunches. The energy chirp adds up to the bunch natural energy spread, but the post-compression acceleration will be much less sensitive to the RF curvature because of shorter time duration of the bunch. Compressors can be also staged in linacs to when ultra-short bunches are required and beam quality needs to be preserved.

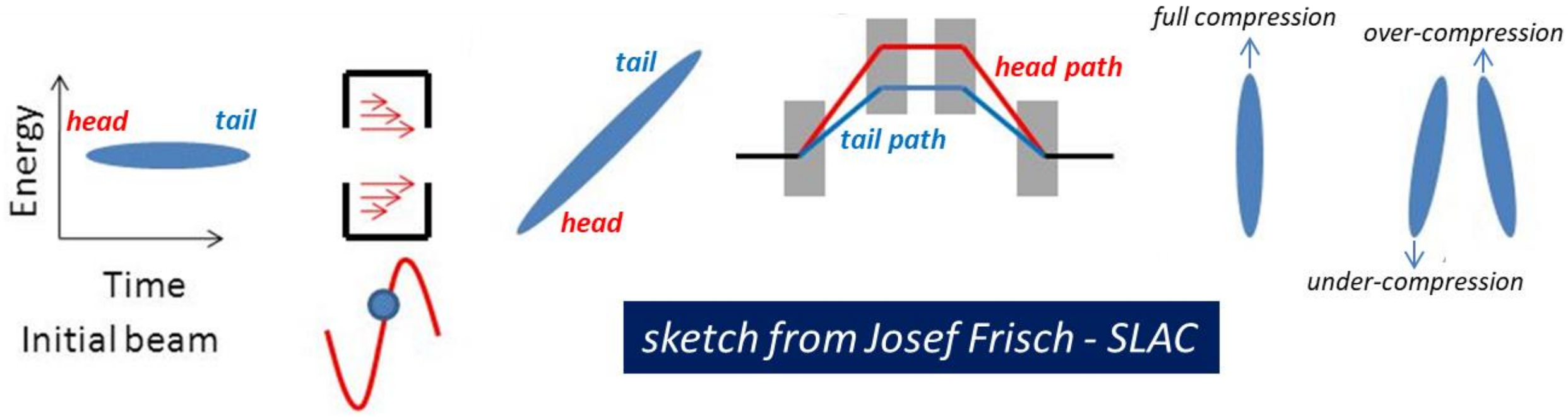

Fig. 58: Sketch of the working principle of a bunch compressor based on a magnetic chicane

In the following a set of linear equations describing the particle longitudinal dynamics in a magnetic compressor will be derived. This aims at introducing a simple unidimensional **linear model** to study the bunch **arrival time dependence** on the **machine parameters**, while it is worth advising the reader that it is too simple and rough to correctly describe the intra-bunch particle dynamics where many physical processes that are completely neglected here (such as space charge effects, coherent synchrotron radiation emission, RF non linearities, …) play a very important role.

Let's consider a particle entering in a magnetic compressor complex with proper injection energy $W_{in}$ and design phase $\varphi_0 = \omega_{RF}(t_{in} - t_{RF})$ with respect to the chirping RF field. The particle will then leave the compressor with the design energy $W_0$ given by:

$$W_0 = W_{in} + qV_{RF}\cos(\varphi_0) \tag{48}$$

Errors of the injection energy and phase $\Delta W_{in}$ and $\Delta\varphi$ will result in an output energy error $\Delta W_o$ which can be calculated by differentiating Eq. (48) according to:

$$\Delta W_o = \Delta W_{in} - qV_{RF}\sin(\varphi_0)\,\Delta\varphi = \Delta W_{in} + h\frac{c}{\omega_{RF}}W_0\,\Delta\varphi \tag{49}$$

where the introduced chirp coefficient h is the relative energy deviation $\Delta W/W_0$ normalized to the particle longitudinal displacement $\Delta z$, corresponding to:

$$h \overset{\text{def}}{=} \frac{\Delta W/W_0}{\Delta z} = \frac{\omega_{RF}}{c}\frac{\Delta W/W_0}{\Delta\varphi} = \frac{-qV_{RF}\sin(\varphi_0)}{W_{in} + qV_{RF}\cos(\varphi_0)}\frac{\omega_{RF}}{c} \tag{50}$$

To complete the bunch compression process the **chirped beam** travels along the magnetic chicane which is a **non-isochronous transfer line**. Particles with different energies travel along paths of different lengths according to:

$$\Delta L = R_{56}(\Delta W_0/W_0) \tag{51}$$

where $R_{56}$ is the coefficient expressing the first order dependence of the path elongation on the relative energy error. By combining Eqs. (49) and (51), a particle entering the magnetic compressor with a time error $\Delta t_{in}$ and a relative energy error $\Delta W_{in}/W_{in}$ will leave it with time and relative energy errors $\Delta W_0/W_0$ and $\Delta t_o$ given by:

$$\begin{aligned}
\Delta W_0/W_0 &= hc\,\Delta t_{in} + \frac{W_{in}}{W_0}\Delta W_{in}/W_{in} \\
\Delta t_o &= \Delta t_{in} + \frac{\Delta L}{c} = \Delta t_{in} + \frac{R_{56}}{c}(\Delta W_0/W_0) \\
&= (1 + h\,R_{56})\Delta t_{in} + \frac{R_{56}}{c}\frac{W_{in}}{W_0}(\Delta W_{in}/W_{in})
\end{aligned} \tag{52}$$

To produce a nearly full bunch compression the chirp and path elongation coefficient have to be tuned such that $h \cdot R_{56} \approx -1$. Under this condition, according to Eq. (52), the exit time of a particle is almost independent on the entering time. This mechanism describes to the first order the deformation (compression) of the longitudinal distribution of the particles in a bunch, **but also the multi-shot dynamics of the bunch center of mass**. **The bunch arrival time downstream the compressor is weakly related to the upstream arrival time**.

Eq. (52) can be also expressed in a matrix notation, according to:

$$\begin{pmatrix} \Delta t \\ \Delta W/W \end{pmatrix}_o = \begin{bmatrix} 1 & \frac{R_{56}}{c} \\ 0 & 1 \end{bmatrix} \begin{bmatrix} 1 & 0 \\ hc & \frac{W_{in}}{W_0} \end{bmatrix} \begin{pmatrix} \Delta t \\ \Delta W/W \end{pmatrix}_{in} = \begin{bmatrix} 1 + hR_{56} & \frac{R_{56}}{c}\frac{W_{in}}{W_0} \\ hc & \frac{W_{in}}{W_0} \end{bmatrix} \begin{pmatrix} \Delta t \\ \Delta W/W \end{pmatrix}_{in} \quad (53)$$

where the matrices $\hat{A}$ and $\hat{B}$ defined as:

$$\hat{A} = \begin{bmatrix} 1 & 0 \\ hc & \frac{W_{in}}{W_0} \end{bmatrix}; \quad \hat{B} = \begin{bmatrix} 1 & \frac{R_{56}}{c} \\ 0 & 1 \end{bmatrix}; \quad \hat{B} \cdot \hat{A} = \hat{C} = \begin{bmatrix} 1 + hR_{56} & \frac{R_{56}}{c}\frac{W_{in}}{W_0} \\ hc & \frac{W_{in}}{W_0} \end{bmatrix} \quad (54)$$

represent the chirping acceleration and the propagation through the non-isochronous drift separately, while the matrix $\hat{C} = \hat{B} \cdot \hat{A}$ describes globally the compressor stage.

Let's now consider the presence of phase ($\Delta\varphi_o = -\omega_{RF}\Delta t_{RF}$) and amplitude ($\Delta V_{RF}/V_{RF}$) errors in the RF section of the compressor. The resulting energy error of the beam entering in the non-isochronous drift can be calculated again by differentiating Eq. (48) with respect to the RF errors according to:

$$\Delta W_o = q\Delta V_{RF} cos(\varphi_o) - qV_{RF} sin(\varphi_o)\Delta\varphi_o \quad \Longrightarrow \quad \frac{\Delta W_o}{W_o} = -hc\Delta t_{RF} + \frac{W_o - W_{in}}{W_o}\frac{\Delta V_{RF}}{V_{RF}} \quad (55)$$

which can be again expressed in matrix notation, where the effect of the RF noise are described by the matrix $\hat{N}$:

$$\begin{pmatrix} \Delta t \\ \Delta W/W \end{pmatrix}_{\substack{chicane \\ input}} = \begin{bmatrix} 0 & 0 \\ -hc & \frac{W_0 - W_{in}}{W_0} \end{bmatrix} \begin{pmatrix} \Delta t_{RF} \\ \Delta V_{RF}/V_{RF} \end{pmatrix} = \hat{N} \begin{pmatrix} \Delta t_{RF} \\ \Delta V_{RF}/V_{RF} \end{pmatrix} \quad (56)$$

To the first order the **RF noise does not affect the bunch internal distribution**, since RF amplitude and phase does not change significantly over a bunch time duration. It will more affect the **bunch as a whole**, resulting in bunch-to-bunch energy and arrival time deviations.

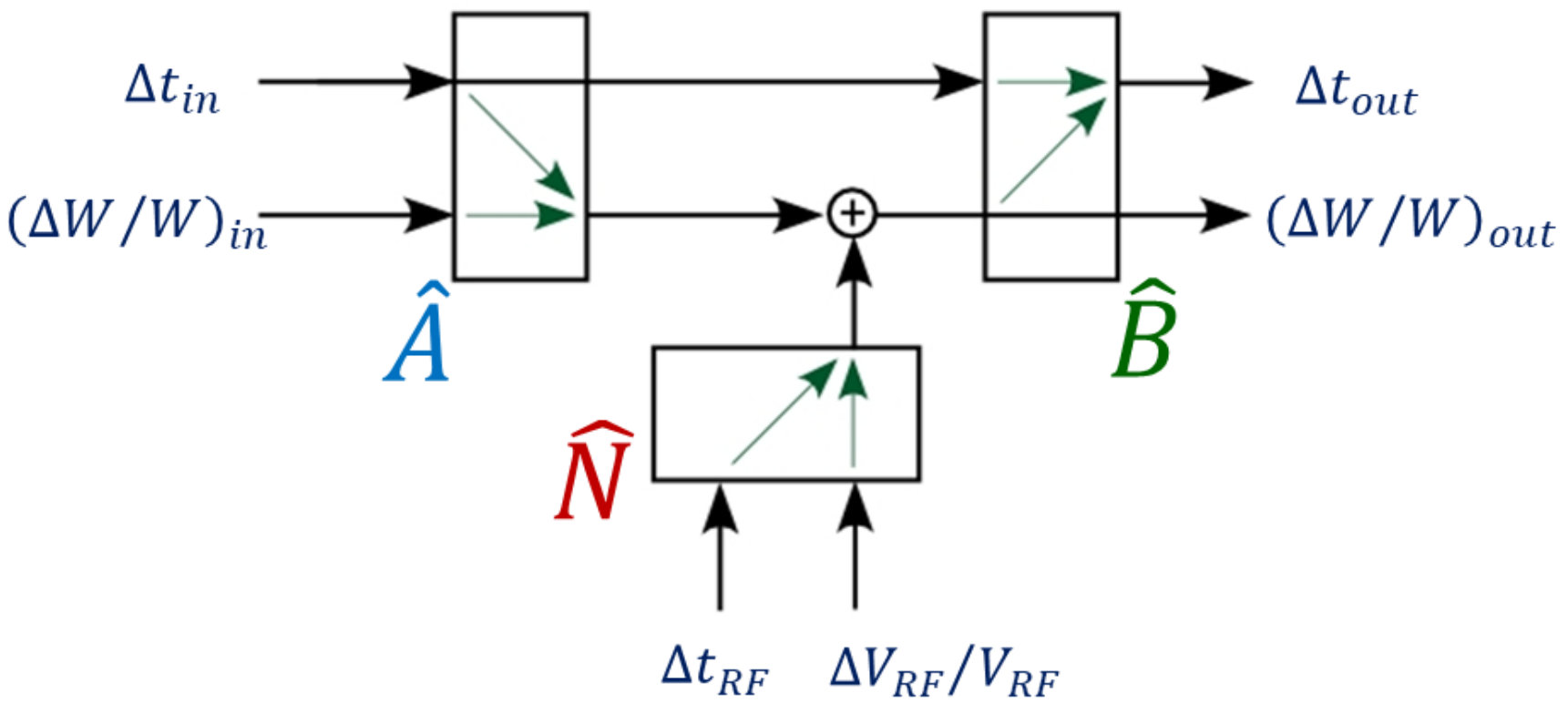

Fig. 59: Effects of injection and RF errors on the particle output energy and timing - Block diagram

The global error of the particle timing and energy downstream the compressor caused by injection errors and RF noise can be evaluated by combining the effects according to the block diagram of Fig. 59. The timing error at the chicane input will be just equal to the initial timing error, while the energy error at the chicane input will depend on all injection and RF errors and will remain unchanged downstream the chicane. Using matrix notation the Fig. 59 block diagram is described by the following equation:

$$\begin{pmatrix} \Delta t \\ \Delta W/W \end{pmatrix}_0 = \hat{B}\left[\hat{A}\begin{pmatrix} \Delta t \\ \Delta W/W \end{pmatrix}_{in} + \hat{N}\begin{pmatrix} \Delta t_{RF} \\ \Delta V_{RF}/V_{RF} \end{pmatrix}\right] \qquad (56)$$

The timing error downstream the chicane $\Delta t_0$ is a linear combination of the timing injection error $\Delta t_{in}$ and the final energy error $\Delta W_0/W_0$, which in turn depends on all injection and RF errors, and can be calculated by using Eq. 56 according to:

$$\Delta t_0 = (1 + hR_{56})\Delta t_{in} - hR_{56}\Delta t_{RF} + \frac{R_{56}}{c}\frac{W_{in}}{W_0}(\Delta W/W)_{in} + \frac{R_{56}}{c}\frac{W_0 - W_{in}}{W_0}(\Delta V_{RF}/V_{RF}) \qquad (57)$$

Looking at the first two terms in Eq. (57), since bunch compression requires $hR_{56} \approx -1$ it clearly appears that **the bunch arrival time downstream the compressor is strongly related to the phase of the chirping RF** and **almost independent on the injection time**. According to this model, in this configuration the beam timing jitter will follow exactly the RF phase noise, plus other contributions coming from the RF amplitude noise and energy injection errors (the 4$^{th}$ and 3$^{rd}$ terms in Eq. 57, respectively) while it will be independent on the injection time jitters. However, for a compressor stage not tuned near the full compression we have $1 + hR_{56} \neq 0$ and the beam final arrival time will be a linear combination of the timing errors of both beam injection and RF accelerating field, with **different weights**. In the following we will generalize this result.

## 4.2 General model

Let's now consider a generic accelerator based facility, such as that sketched in Fig. 6, where a tight control of the beam arrival time at some specific target positions is a vital requirement. The machine will be designed to operate at a nominal (and optimal) working-point, requiring that the time (or phase) of all sub-systems are properly set and kept to the design values $T_i$ to provide the required beam characteristics at the Linac end (or at some intermediate target positions), where the bunch centroid arrives at time $T_b$. This condition represents the **nominal synchronization** of the facility, which depends on the specific implemented working-point. When the **same facility is tuned differently**, to serve for instance another experiment, then **another nominal synchronization working-point** has to be configured.

Given a nominal synchronization working point, the bunch arrival time $T_b$ is a function of the timing sets $T_i$ of all the facility sub-systems impacting on the beam longitudinal dynamics, i.e. $T_b = T_b(T_i)$.

To evaluate how beam arrival time is affected by the synchronization errors of the sub-systems is convenient to **expand** the generic relation $T_b = T_b(T_i)$ to the **first order**. Perturbations of sub-system phasing $\Delta t_i$ will produce a change $\Delta t_b$ of the beam arrival time given by:

$$\Delta t_b = \sum_i a_i \Delta t_i = \sum_i \frac{\Delta t_i}{c_i} \quad \text{with} \quad \sum_i a_i = 1 \qquad (58)$$

The introduced $a_i$ coefficients express the **weights** of the various clients in **determining** the **beam arrival time**. The sum of all weights $a_i$ must be equal to 1 following the simple logic argument that if all the sub-systems are shifted by exactly the same time offset $\Delta t_{off}$ than all the accelerator internal processes will remain unchanged, and the only consequence will be an identical variation $\Delta t_{off}$ of the beam arrival time.

The reciprocals $c_i = 1/a_i$ of the weighting coefficients can be used, representing the compression coefficients, since a low $a_i$ value indicates that the influence of the i-th sub-system on the beam arrival time in the specific considered working-point is modest, and its perturbation contribution is "compressed". Values of $a_i$ can be computed analytically, by simulations or even measured experimentally by shifting the sub-systems one at a time and measuring the correlated variation of the beam arrival time.

Again it is worth stressing that the $a_i$ values depend very much on the **machine working-point** and on the beam **target position** (observation point). As an example we may consider a simplified version of the Fig. 6 facility, consisting in a linac driven by a photo-injector and including a magnetic compressor stage, as depicted in Fig. 60. In this simplified configuration we can identify 5 main sub-systems (or clients of the facility synchronization system) affecting the beam arrival time at the linac end, namely: the photo-cathode laser system, and the RF fields in the RF Gun, in the Capture Section (the very first accelerating section downstream the RF Gun), in the Booster (upstream the chicane) and in the Main Linac (downstream the chicane). Let's then consider and discuss 3 different working-points.

I. No compression: Beam captured by the GUN and accelerated on-crest in the Capture Section (CS), in the Booster and in the main Linac.

In this case, according to beam dynamics studies tracking the bunch particles travelling through the RF Gun cavity, the beam arrival time is mostly correlated with the photo-cathode (PC) laser system, and with the RF phase of the Gun accelerating field, while it almost independent on the phase of the other RF clients. Typical values of the $a_i$ coefficients are:

$$a_{PC} \approx 0.6 \div 0.7 \,; \quad a_{RF_{GUN}} \approx 0.4 \div 0.3 \,; \qquad \text{others } a_i \approx 0 \qquad (59)$$

II. Magnetic compression: an energy-time chirp is imprinted by an off-crest acceleration in the booster and exploited in magnetic chicane to compress the bunch, while the Capture Section and the Main Linac are operate on-crest. As demonstrated in the previous paragraph, in this case the bunch arrival time is tightly defined by the timing of the Booster RF accelerating field, especially when operating in full compression. A residual loose dependence on the PC and RF Gun timing has to be considered, whose relevance is inversely proportional to the operational compression factor, while the other clients play a negligible role. Typical values of the $a_i$ coefficients are then:

$$a_{RF_{boost}} \approx 1 \,; \qquad |a_{PC}|, |a_{RF_{GUN}}| \ll 1 \,; \qquad \text{others } a_i \approx 0 \qquad (60)$$

If the bunch is over-compressed the head and tail downstream the chicane are reversed, which means $a_{RF_{boost}} > 1$ and $a_{PC} < 0$.

III. RF compression: a not-fully relativistic bunch ($E_0 \approx$ few MEV at the RF Gun exit) injected ahead the crest in the RF Capture Section slips back toward an equilibrium phase closer to the crest during acceleration, being also compressed in this process. This linear compression technique can be used instead of standard magnetic compression, or in combination to it in a two-stages arrangement. If we consider a working-point based only on RF compression, with both Booster and Main Linac operated on crest, expected values of the $a_i$ coefficients are:

$$a_{RF_{CS}} \approx 1\,; \qquad |a_{PC}|, |a_{RF_{GUN}}| \ll 1; \qquad others\ a_i \approx 0 \tag{61}$$

The values of the CS RF phase at the beam injection define the compression factor. According to the linear theory for a travelling wave capture section, the full compression is obtained for $\varphi_{RF_{CS}} \approx -\pi/2$ (at the RF zero-crossing ahead the crest), while over-compression is obtained at larger distances from the crest ($\varphi_{RF_{CS}} < -\pi/2$).

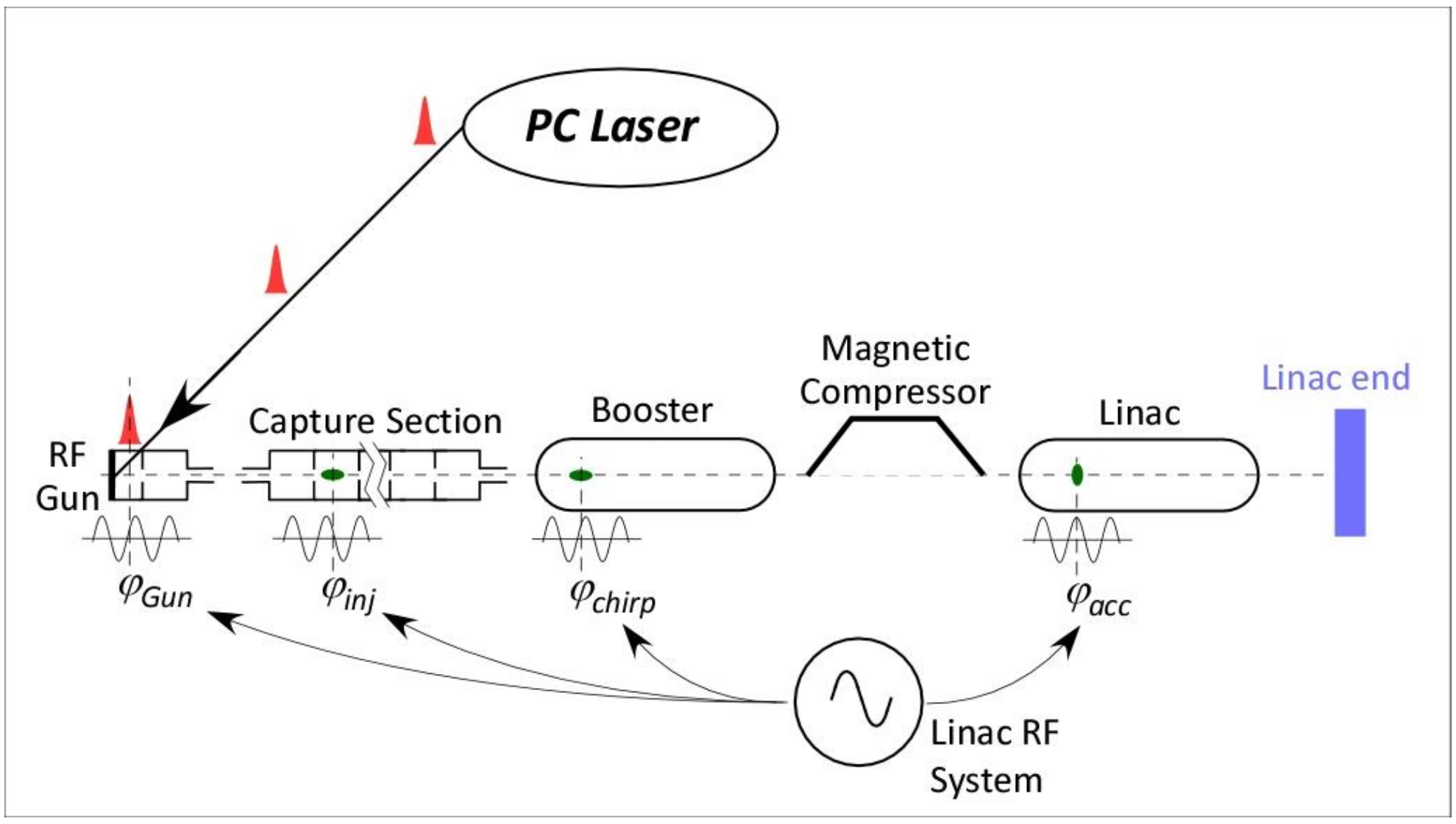

Fig. 60: A photo-injector driven linac equipped with a magnetic compression stage

According to Eqs. (59), (60) and (61) the **same hardware** tuned in **different ways** provides very **different values of the timing coefficients** $a_i$. Particle distributions within the bunch and shot-to-shot centroid distributions show a similar dependence on the sub-systems timing errors, but the actual values of the coefficients $a_i$ might be different since the intra-bunch longitudinal dynamics is heavily affected by space charge forces [[34]].

### 4.3 Beam Jitter estimation

The knowledge of the $a_i$ values **allows estimating the expected bunch arrival time jitter** $\sigma_{t_b}$ on the base of the **residual jitter** characteristics $\sigma_{t_i}$ of the facility **clients**. Starting from Eq. (58) the calculation is quite straightforward if the random temporal variables $\Delta t_i$ (measured with respect to the facility reference clock) can be considered fully uncorrelated, which is a reasonable assumption whenever the residual phase noise of all sub-systems are essentially dominated by the contribution of their free-run spectral density, as expressed in Eq. (37). In this case the variance of the bunch arrival time is simply given by a linear combination of the clients timing errors, according to:

$$\sigma_{t_b}^2 = \sum_i a_i^2\, \sigma_{t_i}^2 \tag{62}$$

where the variances $\sigma_{t_i}^2$ of the sub-system time jitters are weighted by the timing coefficient squared $a_i^2$.

According to Eq. (58) the relative time error between the beam and the j-th facility subsystem is given by:

$$\Delta t_{b-j} = \Delta t_b - \Delta t_j = (a_j - 1)\Delta t_j + \sum_{i\neq j} a_i \, \Delta t_i \tag{63}$$

that leads to an estimate of the beam jitter variance measured with respect to a specific facility sub-system (such as the PC laser or the RF accelerating voltage of a certain group of cavities) given by:

$$\sigma_{t_{b-j}}^2 = (a_j - 1)^2 \sigma_{t_j}^2 + \sum_{i\neq j} a_i^2 \, \sigma_{t_i}^2 \tag{64}$$

For a more rigorous analysis we have to take into consideration that the **clients are all correlated to the reference source, but not exactly in the same way**, because of the different efficiency of the local PLLs corresponding to different loop gains $H_i(s)$. By combining Eqs. (37) and (58), in the Laplace domain the beam-to-reference phase $\Phi_{br}(s)$ is related to the inherent client phase noise $\Phi_{i_0}(s)$ and to the reference $\Phi_{ref}(s)$ according to:

$$\Phi_{br} = \sum_i a_i \, \phi_i = \sum_i \frac{a_i \Phi_{i_0}}{1 + H_i} - \left[\sum_i \frac{a_i}{1 + H_i}\right] \Phi_{ref} \tag{65}$$

The phase noise spectral density of the random variable $\Delta t_{b-r}$, i.e. the beam arrival time error referred to the facility reference clock, is proportional to $|\Phi_{br}|^2$. To compute it the right-hand side of Eq. (65) need to be squared, which leads to:

$$|\Phi_{br}|^2 = \sum_i \frac{a_i^2 \left(\left|\Phi_{i_0}\right|^2 + |\Phi_{ref}|^2\right)}{|1 + H_i|^2} + \left[\sum_{i\neq j} \frac{a_i a_j}{(1 + H_i)(1 + H_j)^*}\right] |\Phi_{ref}|^2 \tag{66}$$

where all terms containing the products $\Phi_{i_0} \cdot \Phi_{j_0}$ and $\Phi_{i_0} \cdot \Phi_{ref}$ have been neglected since refer to uncorrelated random variables and average out when evaluating the statistical properties of $\Delta t_{b-r}$. With the definition of $S_{i-r}(f)$ given in Eq. (37), the phase noise spectral density of the $\Delta t_{b-r}$ variable is obtained from Eq. (66) and it is equal to:

$$S_{b-r}(f) = \sum_i a_i^2 \, S_{i-r}(f) + \left[\sum_{i\neq j} \frac{a_i a_j}{(1 + H_i)(1 + H_j)^*}\right] S_{ref}(f) = a_r^2(f) \, S_{ref}(f) + \sum_i a_i^2 \, S_{i-r}(f) \tag{67}$$

where the function $a_r^2(f)$ is just the result of the sum in the square brackets. The variance of the $\Delta t_{b-r}$ variable is obtained by frequency integration of the noise spectral density and gives:

$$\sigma_{t_{br}}^2 = \sum_i a_i^2 \, \sigma_{t_{ir}}^2 + \sigma_{t_{res}}^2 \quad \text{with } \sigma_{t_{res}}^2 = \frac{1}{\omega_{ref}^2} \int a_r^2(f) \, S_{ref}(f) df \tag{68}$$

where the term $\sigma_{t_{res}}^2$ is a residual coming from the reference clock phase noise spectrum weighted by the cross products of the PLLs closed loop frequency responses. It is worth noticing that if $S_{ref}(f) \ll S_{i_0}(f)$ for all clients, then the term $\sigma_{t_{res}}^2$ can be neglected, and the locked clients can be considered uncorrelated just as in Eqs (62) and (64).

Moreover, if the PLL loop transfer functions were all similar ($H_i \approx H_0$) then according to Eq. (5) a common residual phase $\Phi_0 = \Phi_{ref} H_0/(1 + H_0)$ would be imprinted on all clients, which could be neglected when dealing with relative phases.

Let' conclude the section by giving some numerical example of Eqs. (62) and (64) applications. Still referring to the Fig. 60 linac let's assume that the PC laser timing jitter has a standard deviation of $\sigma_{t_{PC}} \approx 70$ fs, while all the RF devices have a common timing jitter whose standard deviation is $\sigma_{t_{RF}} \approx 30$ fs. The standard deviation of the beam arrival time will depend on the implemented machine working-point. Let's consider two cases: (I) no compression and (II) magnetic over-compression.

I. For the no compression working point let us assume the following weighting coefficient values:

$$a_{PC} \approx 0.65; \quad a_{RF_{GUN}} \approx 0.35; \qquad \text{others } a_i \approx 0$$

By applying Eqs. (62) and (64) we get:

$$\sigma_{t_b} \approx 47 \text{ fs}; \ \sigma_{t_{b-PC}} \approx 27 \text{ fs}; \ \sigma_{t_{b-RF}} \approx 50\text{fs}$$

Because of the uncomplete correlation the beam has a lower absolute jitter than the PC laser, but the 65% correlation is sufficient to reduce down to $\approx 27$ fs the relative jitter between the two.

II. For the magnetic over-compression working point let's assume the following weighting coefficient values:

$$a_{PC} \approx -0.13; \quad a_{RF_{GUN}} \approx -0.07; \qquad a_{RF_{boost}} \approx 1.2; \quad \text{others } a_i \approx 0$$

By applying Eqs. (62) and (64) and taking into account that the Booster and the RF Gun are driven by the same RF signal we get:

$$\sigma_{t_b} \approx 35 \text{ fs}; \ \sigma_{t_{b-PC}} \approx 86 \text{ fs}; \ \sigma_{t_{b-RF}} \approx 10\text{fs}$$

that shows very little residual jitter between the beam and the RF because of a correlation factor close to unity (113%, since $a_{RF_{boost}} + a_{RF_{GUN}} = 1.2 - 0.07 = 1.13$).

The numerical example illustrates how **different working points** provide **different beam arrival time** performances even though the jitter characteristics of the sub-systems involved (PC laser and RF) remain the same.

## 4.4 CONCLUSIONS

Timing and Synchronization is a quite recent branch of the particle accelerator physics which started developing in the early 2000 mainly as a consequence of the advent of a new generation of machines such as the FEL radiation sources, and it has grown rapidly and considerably in the last years under the pushing demands of the facility users as well as of newly and more advanced particle acceleration concepts and experiments such as beam external injection in laser driven plasma waves.

The basic architecture of synchronization systems consists in a star connection from a central hutch hosting the facility master clock and all the peripheral machine sub-systems (clients) requiring temporal alignment. The link is typically optical, dispersion compensated and actively stabilized to provide short and constant delay pulses as timing reference. Clients are typically controlled oscillators, either optical or RF, which are locked to the reference using proper phase detectors and optimized PLL feedback loops. Design and handling of such systems require knowledge and expertise from various fields such as Electronics, RF, Laser, Optics, Control, Diagnostics, Beam dynamics, … In this respect timing and synchronization can be considered a multi-disciplinary branch of particle accelerator physics.

Understanding the real synchronization needs of a facility to draw proper specifications for the clients on the base of physical models is a crucial step for successful design and efficient operation. This is also necessary to avoid over-specification that leads to extra-costs and unnecessary complexity.

Synchronization diagnostics, based on high resolution bunch arrival time monitors, is fundamental to understand beam behavior, to identify the main sources of timing jitter and fluctuations, and to provide input data for beam-based feedback systems correcting synchronization residual errors.

Although stability down to the femto-second scale has been already demonstrated and in the best cases routinely obtained, new challenges arise as synchronization requirements get more and more tight following the evolution of the particle accelerator concepts and technology. The atto-second frontier is the new horizon [9], and there are studies and proposals under elaboration to move towards this new territory.